\documentclass[reprint,aps,rmp]{revtex4-1}

\usepackage{graphicx}
\usepackage{amsmath}
\usepackage{amssymb}
\usepackage{dcolumn}
\usepackage{txfonts}
\usepackage{bm}

\begin{document}

\title{Controlling mass and energy diffusion with metamaterials}

\author{Fubao Yang}\thanks{These authors contributed equally to this work.}
\affiliation{Department of Physics, Key Laboratory of Micro and Nano Photonic Structures (MOE), and State Key Laboratory of Surface Physics, Fudan University, Shanghai 200438, China}

\author{Zeren Zhang}\thanks{These authors contributed equally to this work.}
\affiliation{Department of Physics, Key Laboratory of Micro and Nano Photonic Structures (MOE), and State Key Laboratory of Surface Physics, Fudan University, Shanghai 200438, China}

\author{Liujun Xu}\thanks{These authors contributed equally to this work.}
\affiliation{Graduate School of China Academy of Engineering Physics, Beijing 100193, China}

\author{Zhoufei Liu}\thanks{These authors contributed equally to this work.}
\affiliation{State Key Laboratory of Surface Physics, Department of Physics, and Key Laboratory of Micro and Nano Photonic Structures (MOE), Fudan University, Shanghai 200438, China} 

\author{Peng Jin}\thanks{These authors contributed equally to this work.}
\affiliation{State Key Laboratory of Surface Physics, Department of Physics, and Key Laboratory of Micro and Nano Photonic Structures (MOE), Fudan University, Shanghai 200438, China}

\author{Pengfei Zhuang}
\affiliation{State Key Laboratory of Surface Physics, Department of Physics, and Key Laboratory of Micro and Nano Photonic Structures (MOE), Fudan University, Shanghai 200438, China}

\author{Min Lei}
\affiliation{State Key Laboratory of Surface Physics, Department of Physics, and Key Laboratory of Micro and Nano Photonic Structures (MOE), Fudan University, Shanghai 200438, China}

\author{Jinrong Liu}
\affiliation{State Key Laboratory of Surface Physics, Department of Physics, and Key Laboratory of Micro and Nano Photonic Structures (MOE), Fudan University, Shanghai 200438, China}

\author{Jian-Hua Jiang}
\affiliation{Suzhou Institute for advanced research, University of Science and Technology of China, Suzhou 215123, China}

\author{Xiaoping Ouyang}
\affiliation{School of Materials Science and Engineering, Xiangtan University, Xiangtan 411105, China}

\author{Fabio Marchesoni}
\affiliation{MOE Key Laboratory of Advanced Micro-Structured Materials and Shanghai Key Laboratory of Special Artificial Microstructure Materials and Technology, School of Physics Science and Engineering, Tongji University, Shanghai 200092, China}
\affiliation{Department of Physics, University of Camerino, 62032 Camerino, Italy}

\author{Jiping Huang}\email{jphuang@fudan.edu.cn}
\affiliation{Department of Physics, State Key Laboratory of Surface Physics, and Key Laboratory of Micro and Nano Photonic Structures (MOE), Fudan University, Shanghai 200438, China}

\date{\today}

\begin{abstract}
Diffusion driven by temperature or concentration gradients is a fundamental mechanism of energy and mass transport that inherently differs from wave propagation in both physical foundations and application prospects. Compared with conventional schemes, metamaterials provide an unprecedented potential for governing diffusion processes, based on emerging theories like the transformation and the scattering-cancellation theory that expanded the original concepts and suggested innovative metamaterial-based devices. The term ``diffusionics'' is used in the review to generalize these noteworthy achievements in various energy and mass diffusion systems. Examples include heat diffusion systems and particle and plasma diffusion systems. For clarity the numerous studies published over the past decade are categorized by diffusion field (i.e., heat, particles, and plasmas) and discussed from three different perspectives: the theoretical perspective, to detail how the transformation principle is applied to each diffusion field; the application perspective, to introduce various intriguing metamaterial-based devices, such as cloaks and radiative coolers; and the physics perspective, to connect with concepts of recent concern, such as non-Hermitian topology, nonreciprocal transport, and spatiotemporal modulation. The possibility of controlling diffusion processes beyond metamaterials is also discussed. Finally, several future directions for diffusion metamaterial research, including the integration of metamaterials with artificial intelligence and topology concepts, are examined.
\end{abstract}


\maketitle

\tableofcontents

\section{Introduction}
\subsection{Historical review}
\begin{figure*}[!ht]
	\includegraphics[width=1.0\linewidth]{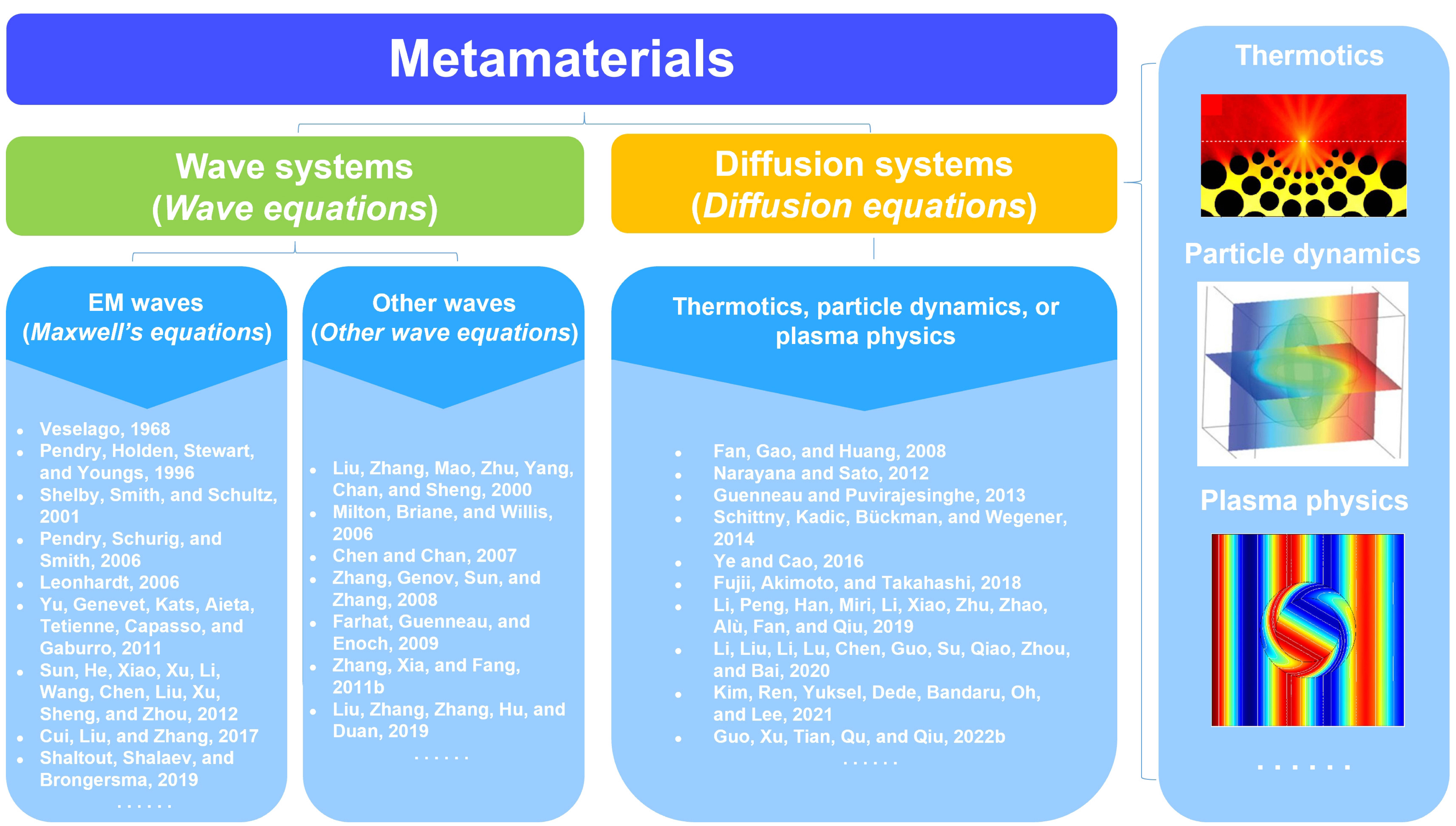}
	\caption{\label{f1}  Overview of diverse metamaterials, which can be divided into three main branches based on the type of governing equations: Maxwell's equations for the first branch, other wave equations for the second branch, and diffusion equations for the third branch~\cite{WegenerScience13,KadicRPP13,HuangSpringer20,Zhang22}. Electromagnetic (EM) waves encompass optical waves. For each branch seminal references are listed. The right column illustrates applications such as heat focusing, particle cloaking, and plasma rotation. Adapted from~\citealp{GuenneauJRSI13,AnufrievNC17,ZhangCPL22}. 	}
\end{figure*}

Cloaking is a recurrent motif in literature and art. On physical grounds, optical cloaking was first proposed in 2006 based on the transformation optics theory~\cite{LeonhardtScience06,PendryScience06}, whereby the space transformation is realized by an {\it ad hoc} spatial definition of material parameters. Contrary to natural materials and thanks to their artificial structure~\cite{SchurigScience06}, metamaterials may exhibit unusual material parameters, like negative refractive index~\cite{VeselagoSPU1968,PendryPRL1996,PendryIEEE1999,SchultzScience01}. As a result,
transformation optics and metamaterials have emerged among the most active
research topics of the last
decade~\cite{KildishevOL08,DondericiIEEE08,ChenNM10,XuNRM16,ZhangJO18,GaoSR19,FanAM20,GaburroScience11,SunNM12},
with ramifications that extend well beyond physical
optics~\cite{WegenerScience13,XuAM21,MartinezMTP22}, such as to
acoustic waves~\cite{LiuScience00,ZhangPRL11,MisseroniSR16,GeNSR18}, elastic waves~\cite{BogEPJ19,MiltonNJP06,StengerPRL12,FarhatPRL09,TianIJHMT22,LiuAS19}, water waves~\cite{XuSR15,LiPRL18,ZouPRL19}, matter waves~\cite{FleuryPRB13,ZhangPRL08,LinPRA09}
and
thermotics~\cite{FanAPL08,ChenAPL08,MaldovanNat13,HuangSpringer20,YangPR21,LiNRM21,ZhangNRP23,WangiSci20,XuPRE18,ShenAPL16-2}.
This effort led to the development of innovative
analysis techniques, such as the scattering-cancellation
scheme~\cite{CummerPRL08,AluPRL09,HanPRL14,FarhatPRAp19,FarhatPRB20} and
related numerical
algorithms~\cite{DedeSMO14,FujiiAPL18,ShaNC21,ZongCPL23}.

Besides the rapid development of wave metamaterials, other intriguing aspects have emerged of late. Firstly, research on metamaterials has been expanded to
a variety of diffusion systems, which intrinsically differ from wave systems
for both the physical mechanisms involved and their fields of application. The transformation thermotics theory~\cite{FanAPL08} was proposed in 2008 as
a thermal counterpart of transformation optics. Not surprisingly, after the first experimental
demonstrations of thermal cloaking, concentration, and rotation in
2012~\cite{NarayanaPRL12}, thermal metamaterials have become the focus of growing
interest~\cite{ChenAPL08,MaldovanPRL13,DaiPRE18,HuangES19-1,ZhouIJHMT21,LiangFoP21,TuFoP21,YangPR21,LiNRM21,LiuAM23,LiuCC23,QiCPL23}. As particle diffusion cloaking under low diffusivity condition was first
proposed in 2013~\cite{GuenneauJRSI13}, metamaterial design was further extended to particle diffusion systems~\cite{Khodayi-mehrIEEE20,AvanziniPRE20,RestrepoSR16,GuenneauAIPAdv15,LiATS18,SchittnyOpt15,YangCPL20,YinCPL21}. Light propagation in a material can be treated as photon diffusion by multiple scattering, so that light diffusion cloaking was designed accordingly~\cite{SchittnyScience14}. Recently, plasma diffusion~\cite{LiAST21,ZhouJPDAP20,TamuraIIEEE20,ReuterJPDAP18,LiangAEM18} has been manipulated by the transformation theory~\cite{ZhangCPL22}, and metamaterial-based devices (metadevices) have been designed for plasma control, thus further expanding the scope of metamaterial science in diffusion systems. Secondly, diffusion metamaterials possess distinctive dissipative characteristics that offer a promising platform for exploring novel physics, such as non-Hermitian physics, nonreciprocal transport, and spatiotemporal modulation. This explains the growing interest for this novel class of metamaterials. 

\subsection{Evolution of metamaterial physics}

Wave metamaterials are artificial structures engineered by suitably assembling
units smaller in size than their respective characteristic wavelengths. These materials do not occur naturally and offer great design and fabrication flexibility based on effective material principles arising from sub-characteristic length structure. This kind of structure sets them apart from other artificial materials, such as photonic and phononic crystals. As members of the metamaterial family, diffusion
metamaterials, too, are engineered with reference to characteristic lengths that distinguish them from
other thermal artificial materials~\cite{XuSpringer23}. For instance, in the case of thermal metamaterials, these can be the thermal radiation wavelength, the thermal diffusion length, or the thermal convection length. Notably, the latest two thermal lengths are time dependent, which explains why many thermal metamaterials operate best in a steady state. The field of metamaterial physics has yielded numerous fundamental breakthroughs and enabled a wide range of novel applications in everyday life. As illustrated in Fig.~\ref{f1}, the various fundamental governing equations divide the field of metamaterial physics into three most developed branches: Maxwell's equations for the first branch, other wave equations for the second branch, and diffusion equations for the third branch, to form a comprehensive field overview~\cite{WegenerScience13,KadicRPP13,LiNRM21,ZhangNRP23,HuangSpringer20,YangPR21}.

Branch 1: Electromagnetic and Optical wave metamaterials. Twenty-eight years after V. G. Veselago's initial introduction~\cite{VeselagoSPU1968}, Pendry's group made pioneering and substantial contributions to control transverse electromagnetic waves~\cite{PendryPRL1996,PendryIEEE1999}, leading to a research boom that continues to this day. In particular, they designed the first metal wire array to realize negative permittivity~\cite{PendryPRL1996} and the first split ring structure to achieve negative permeability~\cite{PendryIEEE1999}. They continued since to lead the research progress in this branch~\cite{PendryScience06}.

Branch 2: Other wave metamaterials. Sheng's group made pioneering contributions to acoustic (sonic) metamaterials in 2000~\cite{LiuScience00}, prompting new research into metamaterials with waves beyond electromagnetic/optical waves, such as acoustic waves, mechanical waves, elastic waves, seismic waves, and matter waves. In particular, they explored longitudinal acoustic waves and first revealed the local resonance mechanism for spectral gaps based on acoustic metamaterials~\cite{LiuScience00}.

Branch 3: Diffusion metamaterials. In 2008, Huang's group published a pioneering study on thermal conduction metamaterials~\cite{FanAPL08}. Their approach was later extended to different diffusion mechanisms~\cite{ZhangNRP23,MaldovanNat13,Yeung22}, including thermal transfer, particle dynamics, and plasma transport. In particular, they proposed transformation thermotics to control steady heat diffusion in thermal metamaterials and first predicted thermal cloaking~\cite{FanAPL08}. Current research on these three classes of metamaterials~\cite{VeselagoSPU1968,PendryPRL1996,PendryIEEE1999,PendryScience06,LiuScience00,FanAPL08} is leading to remarkable accomplishments in the science, technology and engineering of metamaterials.

\subsection{Application perspective}
\begin{figure}[!ht]
	\includegraphics[width=1.0\linewidth]{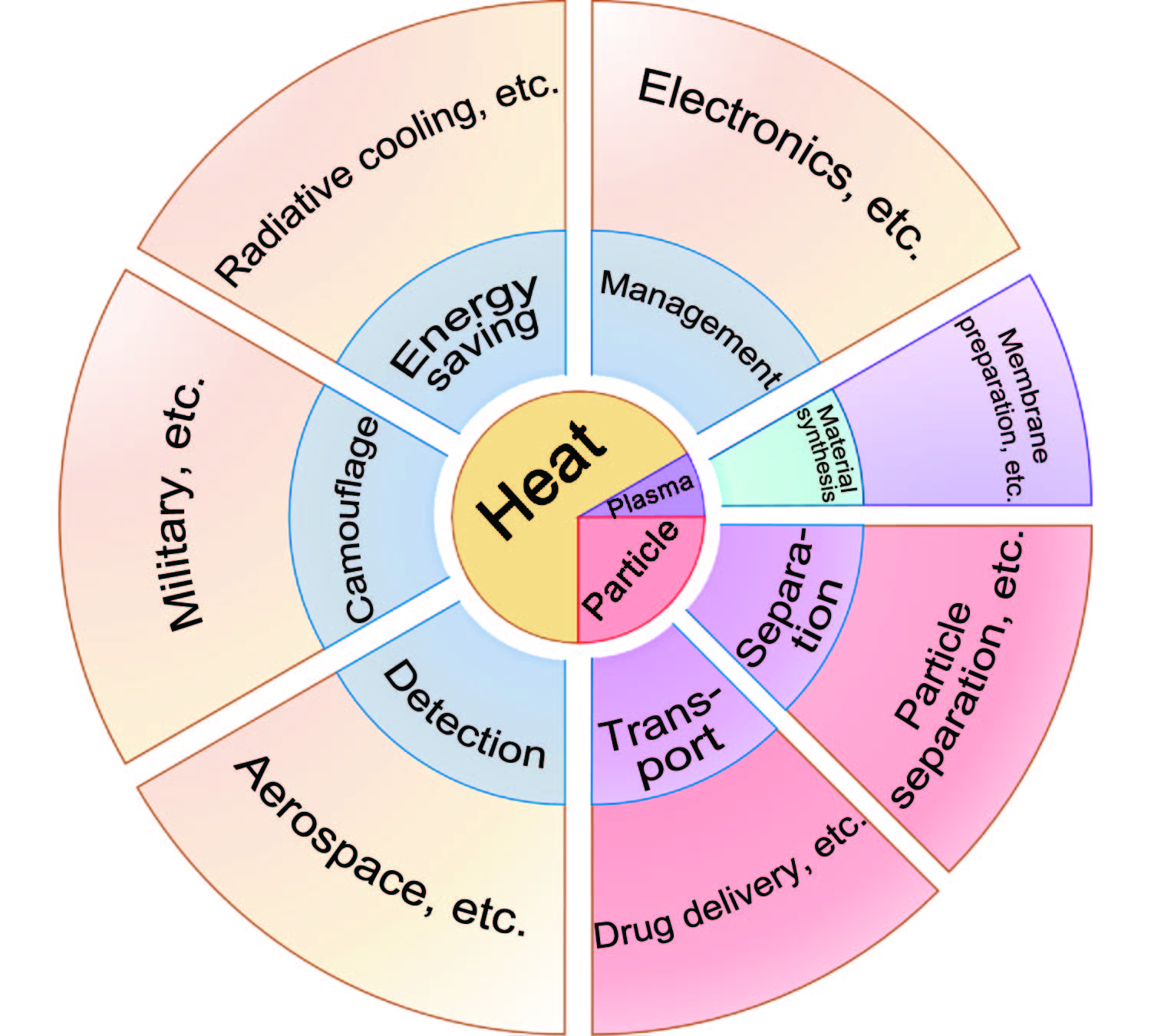}
	\caption{\label{f2} Proposed applications of diffusion metamaterials. Adapted from~\citealp{GuenneauJRSI13,HanAM14,DouEES16,YangAM15,DedeIEEE15,RamanNat14}.}
\end{figure}

Diffusion metamaterials have already inspired intriguing applications in technological fields like thermal engineering, particle manipulation, and plasma transport. In this review, we will discuss a few among the most promising applications, like heat management, energy conservation, heat camouflage, and thermal detection by heat diffusion; mass transport and separation by mass diffusion; plasma-assisted material synthesis by plasma diffusion (Fig.~\ref{f2}).

Thermal camouflage, or illusion~\cite{ZhuAIP15,DaiPRE18,YangES19,LiuNanop20,HuMT21,ZhangCPL21,FengISC22,LiuEPL21,LiM23,XingAEM23} is by now a well-established effect. In addition to hiding targets from detection, thermal illusion devices can produce fake signals to mislead the observer. A cloak plus the expected objects were constructed to realize thermal camouflage~\cite{HanAM14}. Later, a compartmentalized transformation method was proposed to achieve spatially controlled thermal illusion and realize encrypted thermal printing~\cite{HuAM18}. A 3D illusion thermotics~\cite{PengESEE19} was also established to overcome the drawbacks of imperfect 2D illusion. Camouflage in the presence of multiple physical fields has also been investigated. A multispectral camouflage device was designed for infrared, visible, laser, and microwave bands~\cite{ZhuNC21}. Device versatility and miniaturization are the future goals of thermal camouflage.

Another primary application of metamaterials is heat management in electronics and biology. For example, with the rapid development of nanoelectronics, conventional thermal management approaches like through-silicon-via optimization and thermal pipes face many challenges. Thermal metamaterials have been designed to guide heat flow in electronic packages~\cite{KimJEP21,DedeJEP18}. Compared to conventional methods, thermal metamaterials can help dissipate heat at will and avoid thermal crosstalk and local hot spots. Thermal memory and computing can be implemented by defining binary states of the heat flux as digits 1 (concentrating) and 0 (cloaking), thus suggesting a new potential direction for information technology~\cite{LokeACSAMI16,HuPRAp18}. Thermal encoding protocol has been proposed as an alternative to optical or electronic information encoding~\cite{HuPRAp18}. Personal thermal management is another frontier presenting excellent prospects. A Janus layered textile was designed to work as a heating (cooling) wrapping, while allowing integration with additional thermoelectric modules; numerical simulation suggests that such a textile could increase (decrease) the skin temperature under sunlight by 8.1 (6.0)$^\circ$C~\cite{LuoNanoL21}.

Heat energy harvesting and transport are longstanding topics in sustainable energy management. For instance, daytime radiative cooling~\cite{YuES19,LiuCPL21-1,FanNP22,XiaJQSRT23,YinCPL23}, a passive process that requires no energy input, is being widely pursued in renewable energy research. The concept of radiative cooling was first demonstrated by constructing a commercial TiO$_2$ white paint~\cite{HarrisonSE1978}. A more refined photonic approach was adopted to fabricate a radiative cooler~\cite{RamanNat14} capable of lowering the ambient air temperature by 4.9$^\circ$C. However, the multilayered structure of the membrane cooler demanded a high fabrication standard. Hybrid metamaterials were then developed for daytime radiative cooling by randomly embedding $\rm{SiO_2}$ microspheres in a polymeric matrix~\cite{ZhaiScience17}, a faster and less expensive fabrication process more suitable for industrial production. Thermal rectification~\cite{HuangES19-2,WangES20,ToyinCPL21,LiCPL21}, achieved through nonlinearity~\cite{ShenPRL16,OrdPRAP20,DaiFoP21,ZhouNC23-2} or spatiotemporal modulation~\cite{TorrentPRL18,ZhaoIJHMT22,YangPRAp22,XuPRL22-1,LiNC22,XuPRE21,XingCPL21,Ordonez-MirandaPRAp21,CamachoNC20,WangFoP22}, enables controlled heat pumping out of a heat reservoir~\cite{LiPRL15}. Black hole thermodynamics, pioneered by Danielsson \emph{et al}.~\cite{DanielssonJHEP10,DanielssonPRD21}, provides a physical understanding of the underlying transformation theory. Inspired by black holes, which can trap light within their event horizons, graded heat-conduction metadevices have been proposed as heat traps~\cite{XuNSR23}. Besides its obvious applications to waste heat
recovery~\cite{HeES19}, this solution suggests an intriguing link between
transformation thermotics and cosmology. Finally, controlling heat transport at the nanoscales is clearly essential in many areas of nanotechnology~\cite{MaldovanNat13,AnufrievNC17,LinES20,LiES20,LuES20,JiaES20,LiuCPL21-2,DongCPL21,ZhuCPL21,YuCPL21,WangCPL21-1,YuFoP22,YuFoP21,XiangFoP22,GeimNature16,Geim2DM16,GeimACSN10,GeimJMMM99,NovoselovACSP19,NovoselovNL14,NovoselovNL14-1,JiehCPL23}.

Thermal detection is another promising field of application for thermal
metamaterials. For example, in the aerospace industry new nondestructive testing methods
are being developed, which aim at detecting thermal signal distortions caused by fabrication
defects and fatigue deformations.
In this context, transient thermal waveguides can be designed to guide time-varying thermal
signals along arbitrary paths without loss of phase
information~\cite{ZhangTSEP21}. Moreover, accurate temperature measurements in medical
and other industrial applications are often perturbed by the measurement
apparatus itself. Metamaterial technology can then be employed to realize invisible
sensors capable of exploring a temperature field without
altering it~\cite{YangAM15,JinIJHMT20,JinIJHMT21}.

Applications of mass diffusion metamaterials include drug transport and
particle separation. Particle diffusion cloaking was proposed to protect
sensitive tissue areas, thus providing a new avenue to smart drug
delivery~\cite{GuenneauJRSI13}. Since the
transport properties of a metamaterial may depend on the nature of the diffusing particles, a particle separator for steady
diffusion was constructed accordingly~\cite{RestrepoSR16}. Recently, a transient
particle separator was proposed to filter out contaminated substances~\cite{ZhangATS22}.
Metamaterials can also be employed to regulate plasma transport. For example,
introducing NH$_3$-plasma can simultaneously dope and etch the surface of
graphene in oxygen reduction and oxygen evolution reactions, thereby improving
their efficiency~\cite{DouEES16}.

Overall, this review offers comprehensive and detailed coverage of the physical mechanisms and principles of diffusion metamaterials, a primer for readers interested in this fast growing field. Contrary to a recent technical report~\cite{ZhangNRP23}, we focus here more on the physics of metamaterials.

\section{Basic theoretical and experimental methods}

\begin{figure}[!ht]
	\includegraphics[width=1.0\linewidth]{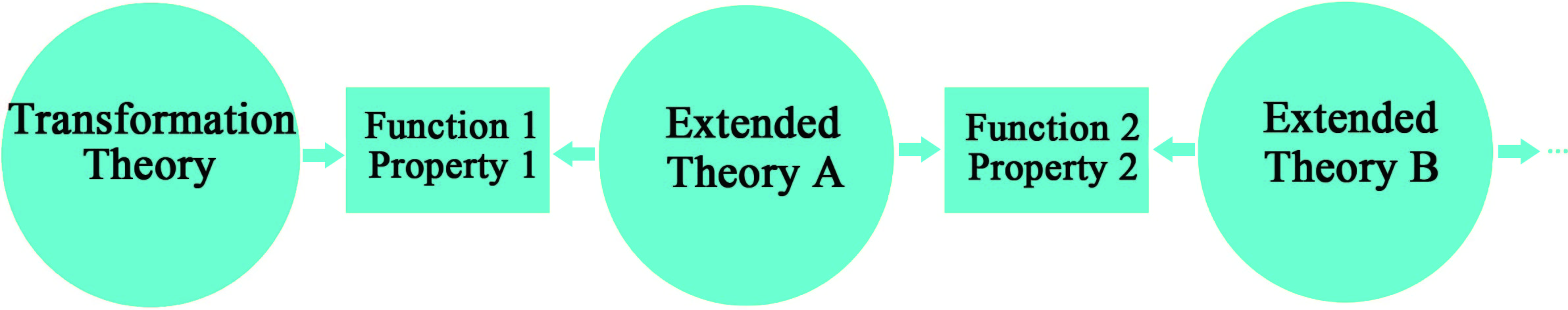}
	\caption{\label{diffusionics} The rapidly growing field of diffusionics encompasses transformation theory and its various extensions at different levels. When we take thermal diffusion as an example. we have ``transformation theory for steady state''~\cite{FanAPL08,ChenAPL08} to ``thermal cloaking (function 1)''~\cite{FanAPL08,ChenAPL08} from ``scattering-cancellation (extended theory A)''~\cite{XuPRL14,HanPRL14} to ``thermal camouflage (function 2)''~\cite{HanAM14} from ``effective medium theory (extended theory B)''~\cite{DaiPRE18}$\Rightarrow \cdots$. Adapted from~\citealp{XuSpringer23}. }
\end{figure}

In the present context the term ``diffusionics'' refers to the remarkable
advances in the use of metamaterials to control energy and mass diffusion. This new discipline comprises transformation theory and
its extended theories (see schematic of Fig.~\ref{diffusionics}). Diffusionics is based on characteristic lengths associated with diffusion metamaterials, thus distinguishing it from phononics. Phononics has two primary challenges: one is the regulation of heat energy flow on the nanoscale, and the other is the goal of processing information through phonons~\cite{LiNRMP2012}. This section provides a comprehensive overview of the pertinent theoretical models and experimental techniques for diffusionics and diffusion metamaterials.

\subsection{Transformation theory}

General relativity predicts that in curved spaces light does not travel in straight lines. However, the principle of general relativity suggests that the equation governing light propagation is form invariant in different coordinate systems. This suggests the possibility of reproducing the optical properties of a curved space using a physical material with artificially designed optical parameters.

\begin{figure}[!ht]
	\includegraphics[width=1.0\linewidth]{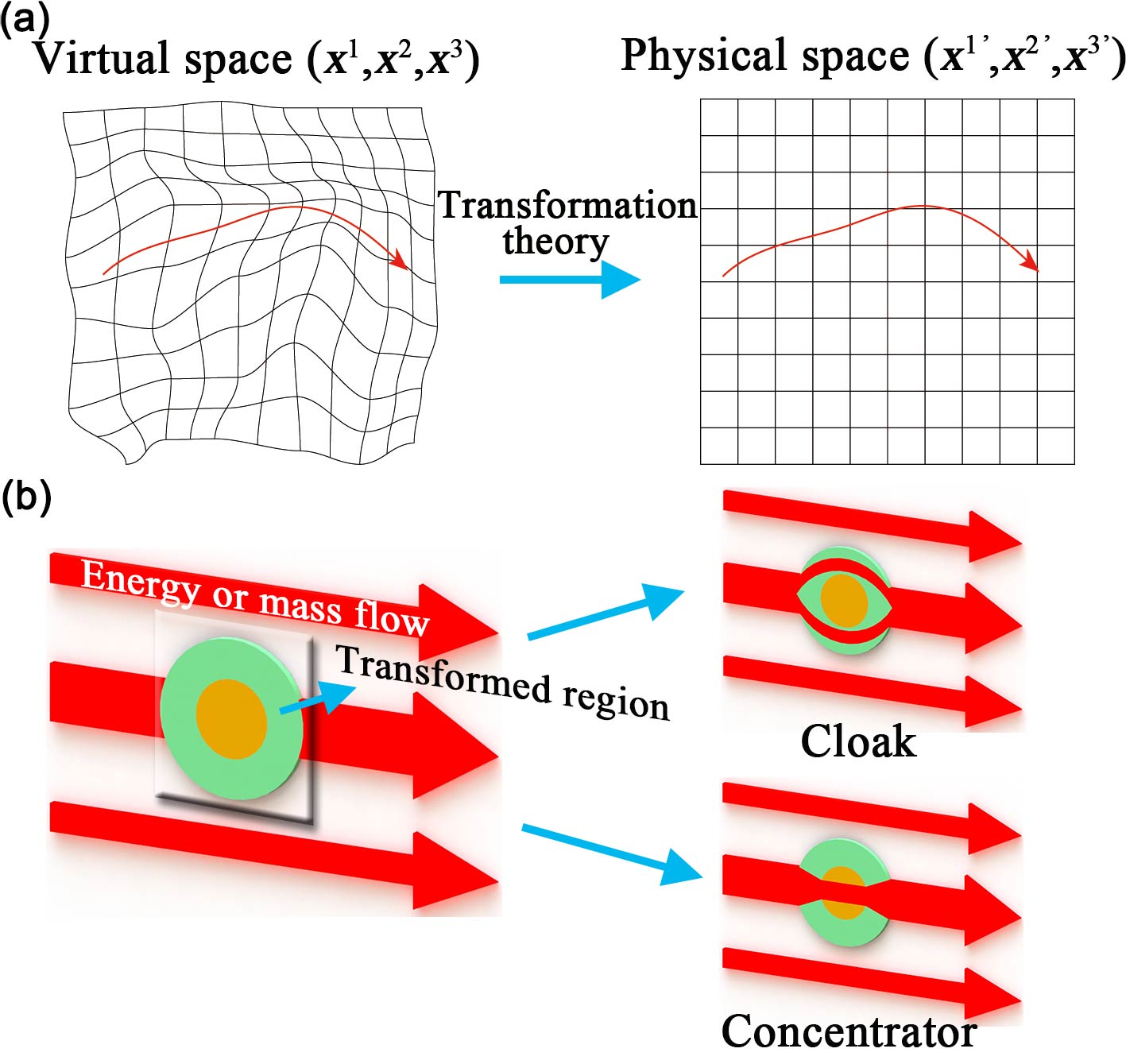}
	\caption{\label{f3} Transformation principle and its applications. (a) Schematic of the transformation theory. The arrows indicate diffusion flows, and the grids represent coordinate systems. (b) The annular (outer shell) areas delimits the transformed regions. A cloak shields the flow away from the circular (inner core) region; a concentrator focuses it inside.}
\end{figure}

Transformation theory originates from transformation optics~\cite{LeonhardtScience06,PendryScience06} and has been extended to many other physical fields, such as acoustics~\cite{CummerNJP07,CummerNRM16,ChenAPL07}, elastodynamics~\cite{FarhatPRL09,BrulePRL2014}, thermotics~\cite{FanAPL08,ChenAPL08}, particle dynamics~\cite{GuenneauJRSI13}, and plasma physics~\cite{ZhangCPL22}. This is a powerful tool to manipulate physical processes under the condition that their governing equations are form-invariant under coordinate transformation. The transformation theory realizes the same effect of a space transformation by transforming the material parameters (Fig.~\ref{f3}).

\subsubsection{General mapping}

A physical field (e.g., temperature field) can be described in different coordinate systems, e.g.,  $T(x^{1^{\prime}},x^{2^{\prime}},x^{3^{\prime}})$ or $T(x^{1},x^{2},x^{3})$, where $x^{u^{\prime}}$ (or $x^{u}$) are the Cartesian (or curvilinear) coordinates [\ref{f3}(a)]. The curvilinear coordinates are defined by $x^{u^{\prime}}=J^{u^{\prime}}_{u}x^{u}$, where $J^{u^{\prime}}_{u}=\partial x^{u^{\prime}}/\partial x^{u}$ is the relevant transformation Jacobian matrix. The transformation theory aims at reproducing in Cartesian space phenomena occurring in the curvilinear space by suitably transforming the material parameters, under the key condition that the governing equations are form invariant upon the coordinate transformation.

Transient heat conduction in macroscopic solids is governed by
\begin{equation}\label{govern.1}
 \rho c \frac{\partial T}{\partial t}-\nabla\cdot\left(\kappa\nabla T\right)=Q,
\end{equation}
where $\rho$, $c$, $\kappa$, and $Q$ are the mass density, heat capacity, thermal conductivity, and heat power density, respectively. Though the thermal conductivity of natural materials may depend on temperature, for now we treat it as a temperature-independent tensor. We start from the known temperature field $T(x^{1^{\prime}},x^{2^{\prime}},x^{3^{\prime}})$ in the Cartesian coordinates, and rewrite Eq.~(\ref{govern.1}) in the component form as
\begin{equation}\label{CarteCompo}
 \rho c\partial_{t}T- \partial_{x^{u^{\prime}}}\left(\kappa\partial_{x^{l^{\prime}}} T\right)=Q,
\end{equation}
with $\partial_t=\partial/\partial t$ and $\partial_{x^{u^{\prime}}}=\partial/\partial x^{u^{\prime}}$. Then a space transformation is applied to regulate the temperature field on demand [left panel of Fig.~\ref{f3}(a)]. Since the space transformation only affects the temperature field, the component form of the governing equation becomes
\begin{equation}\label{CurviCompo}
 \rho c\partial_{t}T- \partial_{x^{u}}\left(\kappa\partial_{x^{l}} T\right)=Q.
\end{equation}
The difference between Eqs.~(\ref{CarteCompo}) and~(\ref{CurviCompo}) is that $x^{u^{\prime}}$ becomes $x^{u}$ due to the effect of the coordinate
transformation. The temperature field in Eq.~(\ref{CurviCompo}) is described in the curvilinear coordinate system, and we need to perform a coordinate transformation to reformulate it Cartesian coordinates [right panel of Fig.~\ref{f3}(a)], namely
\begin{equation}\label{Comp.2}
 \rho c\partial_{t}T-\dfrac{1}{\sqrt{g}}\partial_{x^{u^{\prime}}}\left(\sqrt{g} \kappa g^{u^{\prime}l^{\prime}} \partial_{x^{l^{\prime}}} T\right)=Q,
\end{equation}
where $g^{u^{\prime}l^{\prime}}$ are the components of the matrix product $\bm{J}^{\dag}\bm{J}$, $\bm{J}^{\dag}$ denoting the transpose of the Jacobian matrix $\bm{J}$. $g$ is the determinant of the matrix $g_{u^{\prime}l^{\prime}}$ (i.e., the inverse of $g^{u^{\prime}l^{\prime}}$), so that $\sqrt{g}=\text{det}^{-1}\boldsymbol{J}$. Since a coordinate system transformation does not change the physical field, Eqs.~(\ref{CurviCompo}) and~(\ref{Comp.2}) describe the same phenomenon but in different coordinate systems. The question now is how to realize the temperature field described by Eq.~(\ref{Comp.2}) in the physical (Cartesian) space. The effect of $g^{u^{\prime}l^{\prime}}$ and $\sqrt{g}$ can be reproduced by having recourse to transformed material parameters. To this end, Eq.~(\ref{Comp.2}) can be rewritten as
\begin{equation}\label{Comp.f}
	\sqrt{g}\rho c\partial_{t}T-\partial_{x^{u^{\prime}}}\left(\sqrt{g} \kappa g^{u^{\prime}l^{\prime}} \partial_{x^{l^{\prime}}} T\right)=\sqrt{g}Q,
\end{equation}
and further simplified as
\begin{equation}\label{Comp.f1}
	\dfrac{\rho c}{\det\bm{J}}\partial_{t}T-\partial_{x^{u^{\prime}}}\left(\dfrac{J^{u^{\prime}}_{u}\kappa J^{l^{\prime}}_{l}}{\det\bm{J}} \partial_{x^{l^{\prime}}} T\right)=\dfrac{Q}{\det\bm{J}}.
\end{equation}
Therefore, the same temperature fields can be obtained from Eq.~(\ref{govern.1}) with transformed material parameters
\begin{subequations}\label{relation}
	\begin{align}
		\rho' c'&=\frac{\rho c}{{\det}\bm{J}},\\
		\bm{\kappa}'&=\frac{\boldsymbol{J} \kappa \bm{J}^{\dag}}{{\det}\bm{J}},\\
		Q'&=\frac{Q}{{\det}\bm{J}},
	\end{align}
\end{subequations}
where the prime symbol $'$ denotes the transformed parameters. This is how heat manipulation (like thermal cloaking or concentration) can be achieved through an appropriate choice of the thermal parameters of a conducting material [Fig.~\ref{f3}(b)].

In practice, thermal fields often appear together with other physical fields. A device controlling a single physical field can hardly meet the increasing demands for functional diversity and adaptability. The ability to manipulate multiple physical fields on a single device has become a trend. Thermal metamaterials based on transformation thermotics can adjust a diversity of thermal functions. Using the transformation principle to combine the desired effects involving different fields can greatly improve the technological impact of metamaterials, which can be called transformation multiphysics. For example, manipulating thermal and electric fields together is essential in industry and everyday life. In the presence of temperature and potential gradients, decoupled thermal and electric fields can be controlled effectively. The governing equation of electric fields and thermal fields are similar, so that both satisfy the requirement of form invariance. Therefore, appropriate tailoring the material parameters, such as the electric and thermal conductivity, can realize the simultaneous control of thermal and electric fields~\cite{LiJAP10}. Thermal and electric fields in many metallic materials are coupled through the thermoelectric effect~\cite{HeSCI17,YanCPL21,QianES19,LuES20,JieCPL23,GaoCPL23}, mainly described by the Seebeck coefficient. The Seebeck coefficient describes the electric potential drop due to an applied temperature difference, which indicates that the carriers transport both heat and electric charge. In this case, applying transformation thermotics to coupled thermoelectric fields requires considering not only heat and electric conduction but also their coupling. If the coupling equation still satisfies the requirement of form invariance under coordinate transformation, the electric, thermal and thermo-electric parameters of the material can be tuned to gain control of coupled thermal and electric fields. The functional manipulation of the electric and thermal fields is inseparable in the presence of coupling terms. Metamaterials thus provide a universal platform to investigate the interplay
of multiple physical fields coupled through the transformation principles
(transformation multiphysics).

\subsubsection{Pseudoconformal mapping}

In principle, the transformation theory reproduces the phenomena in an arbitrarily curved space by using materials with anisotropic and inhomogeneous parameters in the physical space. Despite the unique advantages, anisotropic parameters pose a serious challenge because natural materials are usually isotropic. We refer here to thermotics as an example. Alternative schemes, like scattering-cancellation~\cite{HanPRL14,XuPRL14,LiNM19} and numerical algorithms~\cite{DedeSMO14,ShaNC21,FujiiAPL18}, were proposed to fabricate thermal metamaterials with the desired properties starting from isotropic materials. However, developing a mapping method that eliminates the anisotropy introduced by the transformation theory without relying on scattering-cancellation would be more important. Such a method may handle more complex heat transfer situations than scattering-cancellation technology.

In wave systems, conformal transformation optics~\cite{LeonhardtScience06,XuNP15} provides an illuminating idea as how to eliminate anisotropic parameters. Unlike the general transformation theory, the transformed parameters dictated by conformal mapping are isotropic. Various wave phenomena~\cite{WangPRL17,LiuPRL20,LuNC21,ChenNC22} were designed and realized by isotropic metamaterials with gradient refractive indexes. When thinking of a thermal counterpart, one notices immediately that wave and diffusion systems are fundamentally different. Firstly, the refractive index has no direct analog in thermotics. Secondly, the interface heat-flow matching between the thermal metamaterial designed by conformal transformation and the background material poses serious difficulties. Therefore, establishing a direct thermal analog of conformal transformation optics proved challenging. 

\begin{figure}[!ht]
	\includegraphics[width=1.0\linewidth]{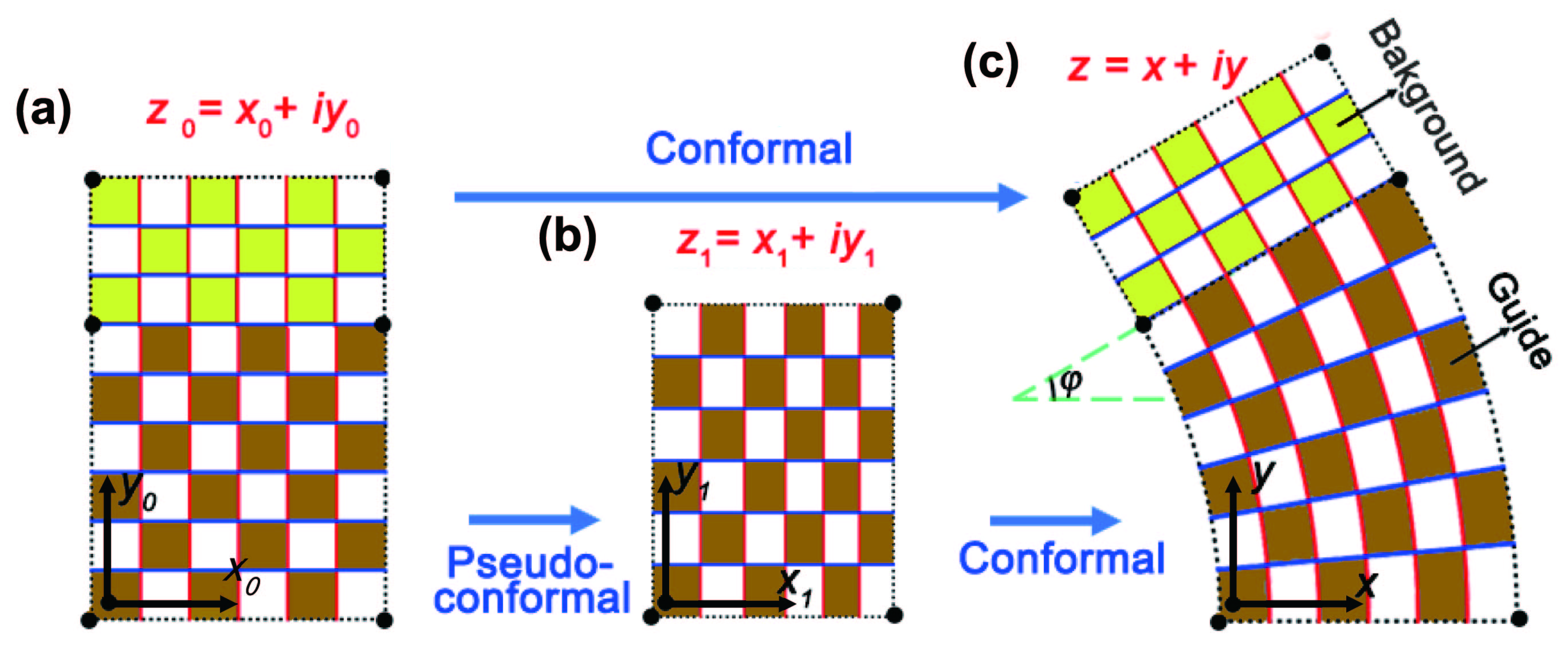}
	\caption{\label{y-f3} pseudoconformal mapping of a heat flux guide. (a)-(c) Schematic diagrams of the virtual, auxiliary, and physical spaces. Adapted from~\citealp{DaiarXiv23}.}
\end{figure}

Recently, the concept of diffusive pseudoconformal mapping~\cite{DaiarXiv23} was proposed to address this issue. This work develops a two-step approach to control a heat flux by means of transformed metamaterials with isotropic thermal conductivity. An example is illustrated in Fig.~\ref{y-f3}, where panels (a)-(c) depict the flux guide in virtual, auxiliary, and physical spaces, respectively. The grid lines are the heat flux streamlines (constant-$x_0$ curves) and the isotherms (constant-$y_0$ curves). The upper rectangle undergoes a rotation and translation transformation to realize a heat flux guide. This process does not change thermal conductivity. The lower rectangle is transformed into an annulus arc to bend the heat flux. A perfect heat flux guide in the physical space should maintain the heat flux continuity at the interface between the two regions. However, this is not usually ensured by conformal mapping, $f:Z_0 \rightarrow Z$, which must satisfy the Cauchy-Riemann equation,$\partial f/\partial Z_0^* = (1/2)(\partial f/\partial x_0 + i\partial f/\partial y_0) = 0$, where $Z_0^*$ is the complex conjugate of $Z_0$, $Z_0 = x_0 + iy_0$, and $Z = f(Z_0) = x + iy$. This equation restricts all curves to be conformal. Suppose a conformal transformation is performed directly on the lower rectangle. In that case, the heat flux in the guide would be unevenly distributed, resulting in a heat flux mismatch at the interface. A two-step pseudoconformal mapping is then implemented to solve this problem. Firstly, a non-conformal mapping is used to transform the virtual space to an auxiliary space ($z_1 = x_1 + iy_1$). The thermal conductivity in the auxiliary space $\bm{\kappa_1}$ should be anisotropic. The second step consists in performing a conformal mapping, to transform the auxiliary space into the physical space. The obtained thermal conductivity in the physical space should also be anisotropic, i.e., $\boldsymbol{\kappa}(Z) = \boldsymbol{\kappa_1}(z_1)= \kappa_0\cdot\text{diag}[(\partial x_1/\partial x_0)/(\partial y_1/\partial y_0),(\partial y_1/\partial y_0)/(\partial x_1/\partial x_0)]$, where $\kappa_0$ is the thermal conductivity in the virtual space. The anisotropy of $\boldsymbol{\kappa}(Z)$ can be eliminated if $\boldsymbol{\kappa}_0$ is diagonally anisotropic. Only the $y_0$-axis component of $\boldsymbol{\kappa_0}$ contributes to heat flux, so the exact value of the $x_0$-axis component is irrelevant. With $\boldsymbol{\kappa}_0 = \text{diag}(0,\kappa_0)$, thermal conductivities in all directions are isotropic, and $\kappa(Z)$ turns out to be
\begin{equation}\label{y15}
 \kappa(Z) = \kappa_0\frac{\phi}{L}\sqrt{x^2+y^2},
\end{equation}
where $\phi$ is the azimuthal angle. The grid lines in all spaces are orthogonal because they are streamlines and isotherms. A non-conformal mapping that maintains the orthogonality of certain curves is called pseudoconformal mapping. The proposed two-step approach includes a pseudoconformal mapping and a conformal mapping, so their combination is still pseudoconformal [Figs.~\ref{y-f3}(a), (c)]. The streamlines are evenly distributed, so the heat flux is matched at the interface when flowing out the guide. The theoretical design of the heat-flux guide was verified experimentally~\cite{XuNSR23}. 

pseudoconformal mapping has been used recently to elucidate the intrinsic geometric relationship between bilayer thermal cloaks~\cite{HanPRL14} and zero-index materials~\cite{LiNM19}. We expect that, due to its general nature, this approach can be easily extended to other diffusion and wave systems.

\subsection{Extended theories}
Transformation theory provides a fundamental scheme for controlling diffusion processes. However, it may require anisotropic, inhomogeneous, and singular parameters that are inconvenient to produce in practice. Therefore, many extended theories have been elaborated to overcome these difficulties. A few are reviewed next.

\subsubsection{Scattering-cancellation theory}

\begin{figure}[!ht]
	\includegraphics[width=1.0\linewidth]{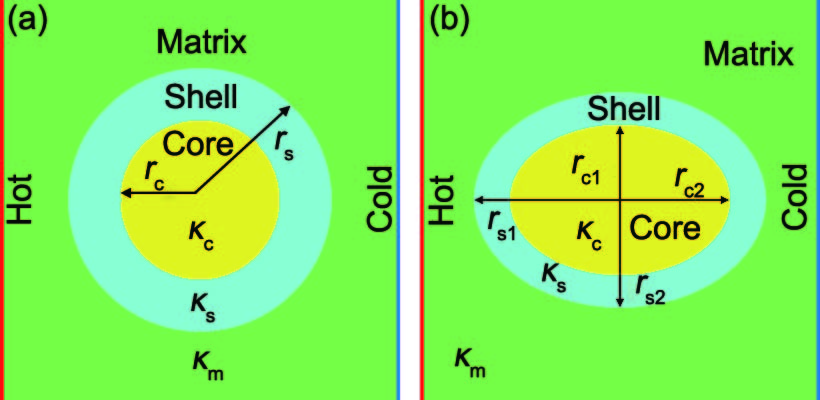}
	\caption{\label{scatter} Two typical core-shell structures: (a) isotropic geometry with anisotropic thermal conductivity; (b) anisotropic geometry with isotropic thermal conductivity. Adapted from~\citealp{XuEPL20}.}
\end{figure}

Since many functions designed by transformation thermotics are based on core-shell structures~\cite{YangPR21,HuangSpringer20,GuoAM22-2,HanIJHMT23,WangJAP18}, calculating the effective thermal conductivity of such structures becomes crucial. The scattering-cancellation theory is a powerful tool to calculate the effective thermal
conductivity of the core-shell structures, especially designed as substitutes for the
inhomogeneous and/or anisotropic materials. Without loss of generality, we discuss two opposite cases: (a) a geometrically isotropic structure with anisotropic material parameters [Fig.~\ref{scatter}(a)], (b)  a geometrically anisotropic structure with isotropic material parameters [Fig.~\ref{scatter}(b)].

We first discuss the case presented in Fig.~\ref{scatter}(a). Since two (2D) and three dimensions (3D) have similar forms (only with slight differences in specific coefficients), we do not discuss them separately but clarify their differences as appropriate. We consider a core-shell structure with inner and outer radii, $r_c$ and $r_s$, and thermal conductivities of $\bm{\kappa}_c={\rm diag}\left(\kappa_{crr},\kappa_{c\theta\theta}\right)$ and $\bm{\kappa}_s={\rm diag}\left(\kappa_{srr},\kappa_{s\theta\theta}\right)$, where $\kappa_{crr}$ and $\kappa_{srr}$ ($\kappa_{c\theta\theta}$ and $\kappa_{s\theta\theta}$) are the radial (tangential) components expressed in cylindrical coordinates, $\left(r,\theta\right)$ in 2D. For 3D $(r,\theta,\phi)$, the thermal conductivity component in the third dimension $\kappa_{\phi\phi}$ is supposed to be identical to $\kappa_{\theta\theta}$ due to the spherical symmetry. In the presence of a horizontal thermal field, the general temperature solution to the heat conduction equation can be expressed as
\begin{equation}\label{3DCGSR}
 T=A_0+\left(A_1r^{u^+}+B_1r^{u^-}\right)\cos\theta,
\end{equation}
where $A_0$ is a constant set to zero for convenience, $A_1$ and $B_1$ are two constants to be determined, and $u^\pm$ are respectively $\pm\sqrt{\kappa_{\theta\theta}/\kappa_{rr}}$ in 2D and $-1/2\pm\sqrt{1/4+2\kappa_{\theta\theta}/\kappa_{rr}}$ in 3D. Then the temperature distributions in the core, $T_c$, shell, $T_s$, and matrix, $T_m$, can be written as~\cite{Milton02}
\begin{subequations}\label{3Tcsb}
	\begin{align}
		T_{c}&=A_{c} r^{u_{c}^+}\cos\theta,\\
		T_{s}&=\left(A_{s}r^{u_{s}^+}+B_{s}r^{u_{s}^-}\right)\cos\theta,\\
		T_{m}&=\left(A_{m} r+B_{m} r^{u_m^-}\right)\cos\theta,
	\end{align}
\end{subequations}
where $A_c$, $A_s$, $B_s$, and $B_m$ are four constants determined by boundary conditions, and $A_m$ is the applied temperature gradient. Because we mainly focus on a linear thermal field, logarithmic terms in Eq.~(\ref{3Tcsb}) can be ignored. In 2D, $u_{c}^+=\sqrt{\kappa_{c\theta\theta}/\kappa_{crr}}$, $u_{s}^\pm=\pm\sqrt{\kappa_{s\theta\theta}/\kappa_{srr}}$, and $u_{m}^-=-1$; in 3D, $u_{c}^+=-1/2+\sqrt{1/4+2\kappa_{c\theta\theta}/\kappa_{crr}}$, $u_{s}^\pm=-1/2\pm\sqrt{1/4+2\kappa_{s\theta\theta}/\kappa_{srr}}$, and $u_{m}^-=-2$.

The boundary conditions require that temperature and normal heat flux are continuous, that is
\begin{subequations}\label{3BC11}
	\begin{align}
		T_{c}\left(r=r_{c}\right)&=T_{s}\left(r=r_{c}\right),\\
		T_{m}\left(r=r_{s}\right)&=T_{s}\left(r=r_{s}\right),\\
		-\kappa_{crr}\dfrac{\partial T_{c}}{\partial r}\left(r=r_{c}\right)&=-\kappa_{srr}\dfrac{\partial T_{s}}{\partial r}\left(r=r_{c}\right),\\
		-\kappa_{m}\dfrac{\partial T_{m}}{\partial r}\left(r=r_{s}\right)&=-\kappa_{srr}\dfrac{\partial T_{s}}{\partial r}\left(r=r_{s}\right),
	\end{align}
\end{subequations}
where $\kappa_m$ is the thermal conductivity of the matrix. Substituting Eq.~(\ref{3Tcsb}) into Eq.~(\ref{3BC11}) yields
\begin{subequations}\label{3BC1}
	\begin{align}
		A_{c}r_{c}^{u_{c}^+}&=A_{s}r_{c}^{u_{s}^+}+B_{s}r_{c}^{u_{s}^-},\\
		A_{m}r_{s}+B_{m}r_{s}^{u_m^-}&=A_{s}r_{s}^{u_{s}^+}+B_{s}r_{s}^{u_{s}^-},\\
		-\kappa_{crr} u_{c}^+A_{c}r_{c}^{u_{c}^+-1}&=-\kappa_{srr}\left(u_{s}^+A_{s}r_{c}^{u_{s}^+-1}+u_{s}^-B_{s}r_{c}^{u_{s}^--1}\right),\\
		 -\kappa_{m}\left(A_{m}+u_m^-B_{m}r_{s}^{u_m^--1}\right)&=-\kappa_{srr}\left(u_{s}^+A_{s}r_{s}^{u_{s}^+-1}+u_{s}^-B_{s}r_{s}^{u_{s}^--1}\right).
	\end{align}
\end{subequations}
Solving these four equations, we can derive the analytical expressions of $A_c$, $A_s$, $B_s$, and $B_m$. scattering-cancellation means that the effect of the core-shell structure vanishes, namely $B_m=0$. Therefore, by imposing $B_m=0$, we can derive the thermal conductivity requirement,
\begin{equation}\label{3km}
	\begin{split}
	&\kappa_{e}=\kappa_{m}\\
	 &=\kappa_{srr}\dfrac{u_{s}^+\left(u_{c}^+\kappa_{crr}-u_{s}^-\kappa_{srr}\right)-u_{s}^-\left(u_{c}^+\kappa_{crr}-u_{s}^+\kappa_{srr}\right) \left(r_{c}/r_{s}\right)^{u_{s}^+-u_{s}^-}}{u_{c}^+\kappa_{crr}-u_{s}^-\kappa_{srr}-\left(u_{c}^+\kappa_{crr}-u_{s}^+\kappa_{srr}\right)\left(r_{c}/r_{s}\right)^{u_{s}^+-u_{s}^-}},
	\end{split}
\end{equation}
where $\kappa_e$ is the effective thermal conductivity of the core-shell structure. This calculation method is not limited to one shell. For $n$ shells, we can derive the effective thermal conductivity with $2n+2$ boundary conditions similar to Eq.~(\ref{3BC11}), where $n+1$ equations describe the temperature continuity, and the other $n+1$ equations indicate the normal heat flux continuity. An alternative method consists in calculating the effective thermal conductivity shell by shell, recursively. For example, we can use Eq.~(\ref{3km}) to calculate the effective thermal conductivity of the core plus the first shell $\kappa_{e1}$. Then the core plus the first shell is treated as a new core with thermal conductivity $\kappa_{e1}$. We then use Eq.~(\ref{3km}) to calculate the effective thermal conductivity of the new core plus the second shell $\kappa_{e2}$, and so on for all $n$ shells.

We address now scattering-cancellation in the structure of Fig.~\ref{scatter}(b). Again, we do not discuss the 2D and 3D cases separately because they lead to similar conclusions. We consider a confocal core-shell structure with isotropic inner and outer thermal conductivities $\kappa_c$ and $\kappa_s$, respectively. The semi-axis of the core (shell) is $r_{ci}$ ($r_{si}$) along the $x_i$ axis, where $i=1,2$ in 2D and  $i=1,2,3$ in 3D. The conversion between the Cartesian coordinates $x_i$ and the elliptic (or ellipsoidal) coordinates $\rho_j$, is given by
\begin{equation}\label{3ellipse}
	\sum_i\dfrac{x_i^2}{\rho_j+r_{ci}^2}=1,
\end{equation}
with $j=1,2$ in 2D and $j=1,2,3$ in 3D. Accordingly,  $\rho_1$, with $\rho_1>-r_{ci}^2$, denotes an elliptic curve (ellipsoidal surface). For example, $\rho_1=\rho_c=0$ and $\rho_1=\rho_s$ represent the inner and outer curves (surfaces) of the shell. We also apply an external thermal field along the $x_i$ axis. Then the temperature fields of the core, $T_{ci}$, shell, $T_{si}$, and matrix, $T_{mi}$, are~\cite{Milton02}
\begin{subequations}\label{3GS1}
	\begin{align}
		T_{ci}&=A_{ci}x_i,\\[0.6em]
		T_{si}&=\left[A_{si}+B_{si}\phi_i\left(\rho_1\right)\right]x_i,\\
		T_{mi}&=\left[A_{mi}+B_{mi}\phi_i\left(\rho_1\right)\right]x_i,
	\end{align}
\end{subequations}
where $A_{ci}$, $A_{si}$, $B_{si}$, and $B_{mi}$ are four constants determined by boundary conditions, $A_{mi}$ is the applied temperature gradient, and $\phi_i\left(\rho_1\right)=\displaystyle\int_{\rho_{\rm c}}^{\rho_1}\left(\left(\rho_1+r_{ci}^2\right)g\left(\rho_1\right)\right)^{-1}{d}\rho_1$ with $g\left(\rho_1\right)=\prod\limits_i\left(\rho_1+r_{ci}^2\right)^{1/2}$.

The boundary conditions expressing the temperature and normal heat flux continuity are
\begin{subequations}\label{3BC}
	\begin{align}
		T_{ci}\left(\rho_1=\rho_{c}\right)&=T_{si}\left(\rho_1=\rho_{c}\right),\\
		T_{mi}\left(\rho_1=\rho_{s}\right)&=T_{si}\left(\rho_1=\rho_{s}\right),\\
		-\kappa_{c}\dfrac{\partial T_{ci}}{\partial \rho_1}\left(\rho_1=\rho_{c}\right)&=-\kappa_{s}\dfrac{\partial T_{si}}{\partial \rho_1}\left(\rho_1=\rho_{c}\right),\\
		-\kappa_{m}\dfrac{\partial T_{mi}}{\partial \rho_1}\left(\rho_1=\rho_{s}\right)&=-\kappa_{s}\dfrac{\partial T_{si}}{\partial \rho_1}\left(\rho_1=\rho_{s}\right),
	\end{align}
\end{subequations}
where $\kappa_m$ is the thermal conductivity of the matrix. Two mathematical identities help handle Eq.~(\ref{3BC}), i.e.,
\begin{subequations}\label{3Skill2}
	\begin{align}
		\dfrac{\partial x_i}{\partial \rho_1}&=\dfrac{x_i}{2\left(\rho_1+r_{ci}^2\right)},\\
		 \dfrac{\partial}{\partial\rho_1}\left[\phi_i\left(\rho_1\right)x_i\right]&=\dfrac{x_i}{2\left(\rho_1+r_{ci}^2\right)}\phi_i\left(\rho_1\right)+\dfrac{x_i}{\left(\rho_1+r_{ci}^2\right)g\left(\rho_1\right)} \notag\\
		 &=\dfrac{x_i}{2\left(\rho_1+r_{ci}^2\right)}\left(\phi_i\left(\rho_1\right)+\dfrac{2}{g\left(\rho_1\right)}\right).
	\end{align}
\end{subequations}
Based on Eq.~(\ref{3Skill2}), Eq.~(\ref{3BC}) can be rewritten as
\begin{subequations}\label{3ebound}
	\begin{align}
		A_{ci}&=A_{si}+B_{si}\phi_i\left(\rho_{c}\right),\\
		 A_{mi}+B_{mi}\phi_i\left(\rho_{s}\right)&=A_{si}+B_{si}\phi_i\left(\rho_{s}\right),\\
		 -\kappa_{c}A_{ci}&=-\kappa_{s}\left(A_{si}+B_{si}\phi_i\left(\rho_{c}\right)+\dfrac{2B_{si}}{g\left(\rho_{c}\right)}\right),\\
		 &-\kappa_{m}\left(A_{mi}+B_{mi}\phi_i\left(\rho_{s}\right)+\dfrac{2B_{mi}}{g\left(\rho_{s}\right)}\right)\nonumber\\
		 &=-\kappa_{s}\left(A_{si}+B_{si}\phi_i\left(\rho_{s}\right)+\dfrac{2B_{si}}{g\left(\rho_{s}\right)}\right).
	\end{align}
\end{subequations}
At this point we introduce the shape factors of the core, $L_{ci}$, and the shell, $L_{si}$, along the $x_i$ axis,
\begin{subequations}\label{3Lcs}
	\begin{align}
		 L_{ci}&=\dfrac{g\left(\rho_{c}\right)}{2}\displaystyle\int_{\rho_{c}}^{\infty}\dfrac{{d}\rho_1}{\left(\rho_1+r_{ci}^2\right) g\left(\rho_1\right)},\label{3Lcs1}\\
		 L_{si}&=\dfrac{g\left(\rho_{s}\right)}{2}\displaystyle\int_{\rho_{s}}^{\infty}\dfrac{{d}\rho_1}{\left(\rho_1+r_{ci}^2\right) g\left(\rho_1\right)},\label{3Lcs2}
	\end{align}
\end{subequations}
with $g\left(\rho_{c}\right)=\prod\limits_ir_{ci}$ and $g\left(\rho_{s}\right)=\prod\limits_ir_{si}$, subject to the conditions $\sum\limits_iL_{ci}=\sum\limits_iL_{si}=1$. In two dimensions, the shape factors can be further reduced to $L_{c1}={r_{c2}}/(r_{c1}+r_{c2})$, $L_{c2}={r_{c1}}/({r_{c1}+r_{c2}})$, $L_{s1}={r_{s2}}/({r_{s1}+r_{s2}})$, $L_{s2}={r_{s1}}/({r_{s1}+r_{s2}})$. According to Eqs.~(\ref{3Lcs}), $\phi_i\left(\rho_{c}\right)$ and $\phi_i\left(\rho_{s}\right)$ can be reformulated as
\begin{subequations}
	\begin{align}
		 \phi_i\left(\rho_{c}\right)&=\int_{\rho_{c}}^{\rho_{c}}\dfrac{{d}\rho_1}{\left(\rho_1+r_{ci}^2\right)g\left(\rho_1\right)}=0,\\
		 \phi_i\left(\rho_{s}\right)&=\left(\int_{\rho_{c}}^{\infty}-\int_{\rho_{s}}^{\infty}\right)\dfrac{{d}\rho_1}{\left(\rho_1+r_{ci}^2\right)g\left(\rho_1\right)}
		 =\dfrac{2L_{ci}}{g\left(\rho_{c}\right)}-\dfrac{2L_{si}}{g\left(\rho_{s}\right)}.
	\end{align}
\end{subequations}
we can now solve Eq.~(\ref{3BC}) to determine $A_{ci}$, $A_{si}$, $B_{si}$, and $B_{mi}$. Since scattering-cancellation requires the core-shell structure not to influence the matrix thermal field, we impose $B_{mi}=0$. Solving for $B_{mi}=0$ yields the thermal conductivity requirement
\begin{equation}\label{3kh1}
 \kappa_{e}=\kappa_{m}=\kappa_{s}\dfrac{L_{ci}\kappa_{c}+\left(1-L_{ci}\right)\kappa_{s}+\left(1-L_{si}\right)\left(\kappa_{c}-\kappa_{s}\right)f}{L_{ci}\kappa_{c}+\left(1-L_{ci}\right)\kappa_{s}-L_{si}\left(\kappa_{c}-\kappa_{s}\right)f},
\end{equation}
where $\kappa_e$ is the effective thermal conductivity of the core-shell structure, and $f=g\left(\rho_{c}\right)/g\left(\rho_{s}\right)=\prod\limits_ir_{ci}/r_{si}$ denotes the area (volume) fraction in two dimensions (three dimensions). This method can easily be extended to calculate the effective thermal conductivity of a core-shell structure with $n$ shells, as we did for the geometrically isotropic case.

\subsubsection{Effective medium theory}

The effective medium theory is another common approach to handle with the complex parameters derived from transformation thermotics. Metamaterials with unconventional thermal conductivities can be designed by combining fundamental units made of isotropic and homogeneous materials. We introduce now typical examples of layered structures of this class, both symmetric and asymmetric.

\begin{figure}[!ht]
	\includegraphics[width=1.0\linewidth]{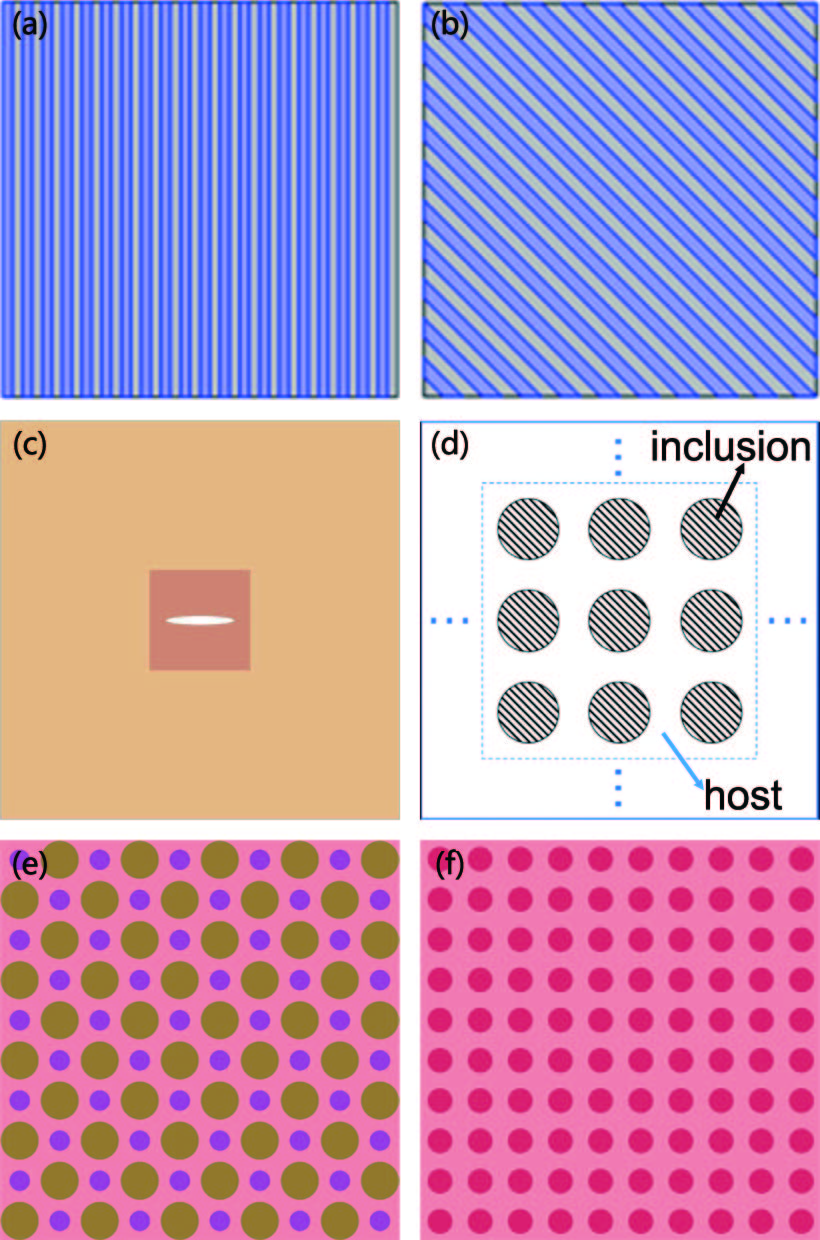}
	\caption{\label{emt} Three typical structures. (a), (b) Layered structures. (c), (d) Asymmetric structures: single and periodic particle arrangement embedded in a matrix. (e) and (f) Symmetric structures featuring particle interaction. Adapted from~\citealp{VemuriAPL13,YangAPL17,DaiIJHMT20,XUPRAP19Th,XuEPJB19-1}.}
\end{figure}

We first introduce layered structures, often employed to study anisotropy effects [Fig.~\ref{emt}(a)]. We consider layers of two homogeneous materials, respectively with thermal conductivities $\kappa_a$ and $\kappa_b$ and widths $w_a$ and $w_b$, stacked alternately. More layers lead to more homogeneous parameters. The effective thermal conductivity perpendicular, $\kappa_s$, and  parallel, $\kappa_p$, to the layers are
\begin{subequations}\label{1ksp}
	\begin{align}
		\kappa_{s}&=\dfrac{w_{a}+w_{b}}{w_{a}/\kappa_{a}+w_{b}/\kappa_{b}},\\
		\kappa_{p}&=\dfrac{w_{a}\kappa_{a}+w_{b}\kappa_{b}}{w_{a}+w_{b}}.
	\end{align}
\end{subequations}
corresponding to the thermal conductivity tensor $\bm{\kappa}$
\begin{equation}\label{1kt}
	\bm{\kappa}=\left[
	\begin{matrix}
		\kappa_{s} & 0\\
		0 & \kappa_{p}
	\end{matrix}
	\right].
\end{equation}

If the layered structure is rotated counterclockwise by an angle $\theta_0$ [$\theta_0=\pi/4$ in Fig.~\ref{emt}(b)], the corresponding Jacobian transformation matrix can be written as
\begin{equation}\label{1Rt}
	\bm{J}=\left[
	\begin{matrix}
		\cos\theta_0 & -\sin\theta_0\\
		\sin\theta_0 & \cos\theta_0
	\end{matrix}
	\right],
\end{equation}
so that the transformed thermal conductivity tensor $\bm{\kappa}'$ reads
\begin{equation}\label{1kRt}
	\bm{\kappa}'=\dfrac{\bm{J}\bm{\kappa}\bm{J}^\dag}{\det\bm{J}}=\left[
	\begin{matrix}
		\kappa_{s}\cos^2\theta_0+\kappa_{p}\sin^2\theta_0 & \left(\kappa_{s}-\kappa_{p}\right)\sin\theta_0\cos\theta_0\\
		\left(\kappa_{s}-\kappa_{p}\right)\sin\theta_0\cos\theta_0 & \kappa_{s}\sin^2\theta_0+\kappa_{p}\cos^2\theta_0
	\end{matrix}
	\right].
\end{equation}
Layered structures can thus be assembled to realize artificial materials with the desired anisotropic thermal conductivity~\cite{VemuriAPL13,VemuriAPL14-1,VemuriAPL14-2,YangAPL14,ZhouEPL23}.

Thermal conductivity in natural materials varies locally, so that its spatial landscape is hard to control. To this end, mixing two materials is a convenient technique to obtain the desired thermal conductivity. The basic structure is presented in Fig.~\ref{emt}(c). An elliptical (or ellipsoidal) particle with thermal conductivity $\kappa_p$ is embedded in a square matrix with thermal conductivity $\kappa_m$, thus forming a single particle structure that is typically asymmetric. Then the whole structure is placed into another bigger matrix with thermal conductivity $\kappa_e$, i.e., the effective thermal conductivity of the particle (with thermal conductivity $\kappa_p$) plus the matrix (with thermal conductivity $\kappa_m$). On applying a thermal field along the $x_i$ axis, the thermal field inside the particle becomes
\begin{equation}\label{3Ga}
	T_{p}=\dfrac{-G_0 \kappa_{m}}{\kappa_{p} L_{p}+\kappa_{m}\left(1-L_{p}\right)}r\cos\theta+T_{0},
\end{equation}
where $-G_0$ is the temperature gradient, $L_p$ the shape factor of the particle along the $x_i$ direction, and $T_0$ a constant. Then the average heat flux in the particle, $\left\langle J_{p} \right\rangle$, can be expressed as
\begin{equation}\label{3Jar}
	\left\langle J_{p} \right\rangle=-\kappa_{p}\left\langle G_{p} \right\rangle=\kappa_{p}\dfrac{G_0\kappa_{m}}{\kappa_{p} L_{p}+\kappa_{m}\left(1-L_{p}\right)},
\end{equation}
where $G_p$ is the temperature gradient in the particle. Here and in the following, $\left\langle \cdots \right\rangle$ denotes a spatial average. Similarly, the average heat flux in the matrix, $\left\langle J_{m} \right\rangle$, can be written as
\begin{equation}\label{3Jm}
	\left\langle J_{m} \right\rangle=-\kappa_{m}\left\langle G_{m} \right\rangle=\kappa_{m} G_0,
\end{equation}
where $G_m$ is the temperature gradient in the matrix. We assume the particle to be small enough to approximate $G_m\approx -G_0$. Then the effective thermal conductivity of the single particle structure can be calculated as
\begin{equation}\label{3E}
	\kappa_{e}=-\dfrac{\left\langle J \right\rangle}{\left\langle G \right\rangle}=-\dfrac{f_{p}\left\langle J_{p}\right\rangle+f_{m}\left\langle J_{m}\right\rangle}{f_{p}\left\langle G_{p} \right\rangle+f_{m}\left\langle G_{m} \right\rangle},
\end{equation}
where $J$ and $G$ are the heat flux and temperature gradient through the resulting composite particle structure, and $f_p=1-f_m$ is the area (volume) fraction in 2D (3D). Based on Eqs.~(\ref{3Jar}) and~(\ref{3Jm}), Eq.~(\ref{3E}) reduces to
\begin{equation}\label{3E2}
	\kappa_{e}=\dfrac{f_{p} \varepsilon_{p} \kappa_{p}+f_{m} \kappa_{m}}{f_{p}\varepsilon_{p}+f_{m}},
\end{equation}
with $\varepsilon_{p}$ can be calculated as
\begin{equation}
	\varepsilon_{p}=\dfrac{\left\langle G_{p} \right\rangle}{\left\langle G_{m} \right\rangle}=\frac{\kappa_{m}}{\kappa_{p} L_{p}+\kappa_{m}\left(1-L_{p}\right)}.
\end{equation}
Since the particle and the matrix are not equivalent, this theory applies to asymmetric structures, and is also known as the Maxwell-Garnett theory~\cite{HuangPR06}. The above analysis, valid for a single particle structure, can also be extended to periodic particle structures [Fig.~\ref{emt}(d)]. Nevertheless, we assume that the area/volume fraction of the particle is small enough for deriving the effective thermal conductivity, so the theory is approximate. A more accurate treatment would be based on the Rayleigh method~\cite{DaiIJHMT20}.

Symmetric structures are also used in practice. For example, in Fig.~\ref{emt}(e) two types of particles are embedded in a uniform matrix to form a symmetric structure. Calculating their effective thermal conductivity is our next problem [Fig.~\ref{emt}(f)].

The total effective thermal conductivity for the pattern of Fig.~\ref{emt}(e) can be calculated by generalizing Eq.~(\ref{3E2}), that is,
\begin{equation}\label{3E2BRUL}
	\kappa_{e}=\dfrac{f_{a} \varepsilon_{a} \kappa_{a}+f_{b} \varepsilon_{b} \kappa_{b}+f_{m} \kappa_{m}}{f_{a}\varepsilon_{a}+f_{b}\varepsilon_{b}+f_{m}},
\end{equation}
where the subscripts $a$ and $b$ denote two types of particles. Similarly, the total effective thermal conductivity for the pattern of Fig.~\ref{emt}(f) is
\begin{equation}\label{3E2BRUR}
	\kappa_{e}=\dfrac{f_{c} \varepsilon_{c} \kappa_{c}+f_{m} \kappa_{m}}{f_{c}\varepsilon_{c}+f_{m}},
\end{equation}
where the subscript $c$ denotes the particles in Fig.~\ref{emt}(f). Since the effective thermal conductivity of the particles in Figs.~\ref{emt}(e) and (f) is the same, for the total effective thermal conductivities of the two patterns to be equal, one must equate the right-hand sides of Eqs.~(\ref{3E2BRUL}) and (\ref{3E2BRUR}),
\begin{equation}\label{3E2BRULR}
	\dfrac{f_{a} \varepsilon_{a} \kappa_{a}+f_{b} \varepsilon_{b} \kappa_{b}+f_{m} \kappa_{m}}{f_{a}\varepsilon_{a}+f_{b}\varepsilon_{b}+f_{m}}=\dfrac{f_{c} \varepsilon_{c} \kappa_{c}+f_{m} \kappa_{m}}{f_{c}\varepsilon_{c}+f_{m}},
\end{equation}
hence
\begin{equation}\label{3E2reducedext} f_{a}\varepsilon_{a}\left(\kappa_{m}-\kappa_{a}\right)+f_{b}\varepsilon_{b}\left(\kappa_{m}-\kappa_{b}\right)=f_{c}\varepsilon_{c}\left(\kappa_{m}-\kappa_{c}\right).
\end{equation}
Each term in Eq.~(\ref{3E2reducedext}) can be effectively treated as representing a thermal dipole associated with each particle species~\cite{XuEPJB19}. Under the condition of Eq.~(\ref{3E2reducedext}), the embedded particles of Figs.~\ref{emt}(e) and (f) produce equivalent dipole effects. Moreover, with reference to the structure of Fig.~\ref{emt}(e), when $\kappa_c=\kappa_m$, the inserted particles do not influence the matrix temperature field, which amounts to thermal transparency~\cite{XUPRAP19Th}. This method for calculating the effective thermal conductivity of symmetric structures, also known as the Bruggeman theory~\cite{HuangPR06}, can also be extended to regular patterns of $n$ particles species.

\subsubsection{Extreme anisotropy theory}

\begin{figure}[!ht]
	\includegraphics[width=1.0\linewidth]{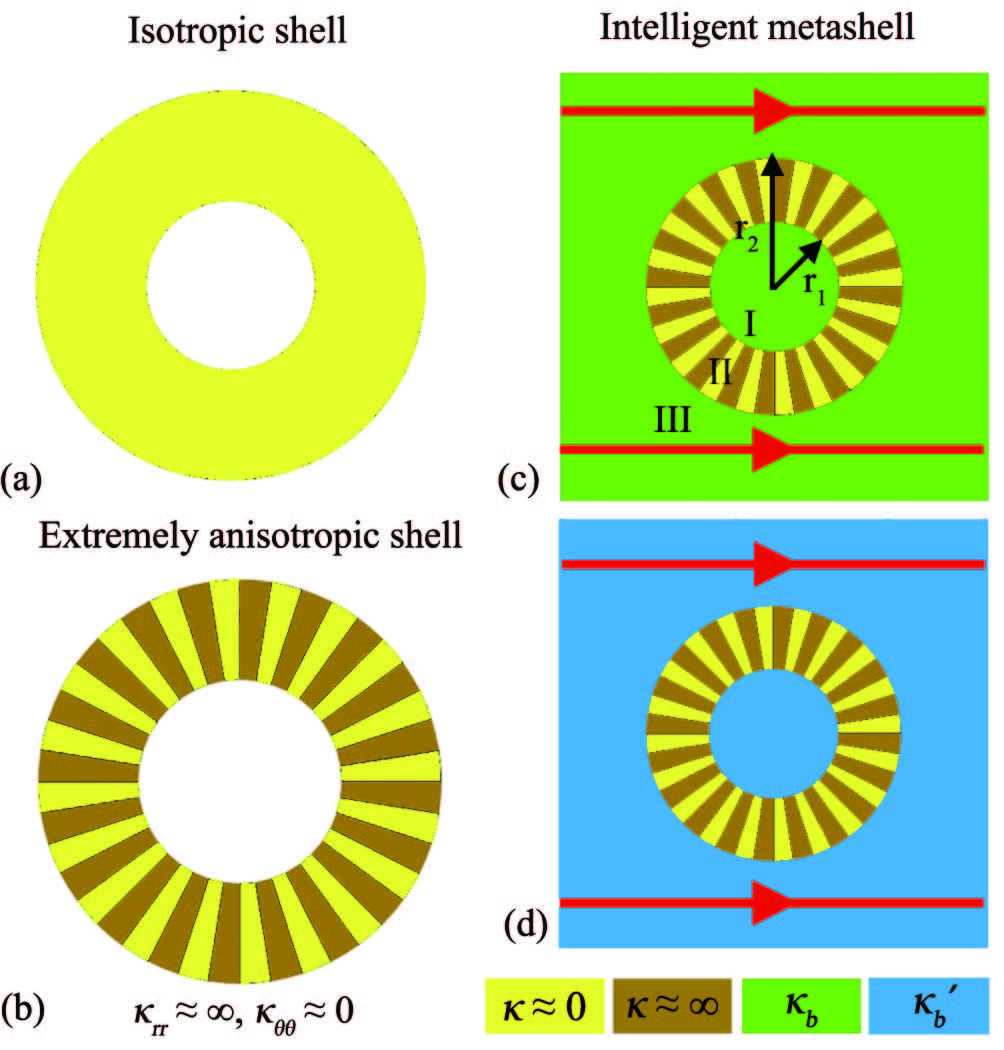}
	\caption{\label{y-f1} Schematic illustration of an extremely anisotropic shell. (a) Isotropic shell with near-zero thermal conductivity. (b) Extremely anisotropic shell with thermal conductivity $\text{diag}(\infty,0)$. (c), (d) Intelligent metashells operating in different backgrounds. Arrows indicate the heat fluxes. The background thermal conductivities of (c) and (d) are $\kappa_b$ and $\kappa_b'$, respectively. Adapted from~\citealp{YangPRAp20}}.
\end{figure}

Transformation thermotics provides a powerful method to manipulate heat flux. However, with the rapid shift toward intelligent metamaterials, the limitations of traditional transformation-thermotics-based metamaterials
become apparent. Such metamaterials can only work in a fixed background with a regular shape. If the background changes or the shape is irregular, transformation-thermotics-based metamaterials are often of little practical use. Recently, a new class of extremely anisotropic metamaterials raised the interest of researchers in different fields of physics, such as electromagnetism~\cite{ZhangJO18,ZhangPRL19-1,HuangJO22}, acoustics~\cite{WuPRAp19,FakheriPRAp20}, thermotics~\cite{SunOE19,BaratiSedehPRAp20,ChenIJTS22,DaiCSTE23,XuPRAp19-1,YangPRAp20}, and mass diffusion~\cite{ZhangPRAp23}. These metamaterials are expected to alleviate the limitations of the earlier metamaterials. In thermotics, for example, extreme anisotropy means almost infinite thermal conductivity in one direction and zero in the other one(s). Most remarkably, this effect can be obtained by using two isotropic materials arranged alternately in a multilayered structure [Figs.~\ref{y-f1}(a), (b)].

By embedding materials with almost infinite thermal conductivity into one with zero thermal conductivity, one can further realize an extremely anisotropic metashell with thermal conductivity $\bm{\kappa} = {\rm{diag}}(\infty,0)$ [Figs.~\ref{y-f1}(b)]. Such a shell is intelligent in 2D since it can adaptively respond to a changing background. In such a setup, stationary heat conduction processes are governed by the Fourier law, which in cylindrical coordinates reads
\begin{equation}\label{y1}
	\frac{1}{r}\dfrac{\partial}{\partial r}\left(r\kappa_{rr}\dfrac{\partial T}{\partial r}\right)+\frac{1}{r}\dfrac{\partial}{\partial\theta}\left(\frac{\kappa_{\theta\theta}}{r}\dfrac{\partial T}{\partial\theta}\right)=0,
\end{equation}
where the thermal conductivity of the metashell is $\bm{\kappa}_s=\text{diag}(\kappa_{rr}, \kappa_{\theta\theta})=\text{diag}(\infty,0)$. The whole system is divided into two parts, including the background materials (regions I and III) and the metashell (region II) [see Fig.~\ref{y-f1}(c)]. The general solution to Eq.~(\ref{y1}) can be expressed as
\begin{equation}\label{y2}
	\begin{split}
	T=A_0+B_0\ln r&+\sum_{i=1}^{\infty}\left[A_i\cos(i\theta)+B_i\sin(i\theta)\right]r^{in}\\
	&+\sum_{j=1}^{\infty}\left[C_j\cos(j\theta)+D_j\sin(j\theta)\right]r^{-jn},
	\end{split}
\end{equation}
where $n=\sqrt{\kappa_{\theta\theta}/\kappa_{rr}}$. The constants $A_0$, $B_0$, $A_j$, $B_j$, $C_k$, and $D_k$ are determined by appropriate boundary conditions. The temperature and heat flux are continuous at the boundaries, and the temperature distribution outside the metashell must be unperturbed, that is
\begin{subequations}\label{y3}
	\begin{align}
		T_1 (r\rightarrow 0)\ &\ \text{is finite},\\
		T_1 (r=r_1) &=T_s (r=r_1),\\
		T_s (r=r_2) &=T_3 (r=r_2),\\
		-\kappa_b\dfrac{\partial T_1}{\partial r} (r=r_1) &=-\kappa_{r}\dfrac{\partial T_s}{\partial r} (r=r_1),\\
		-\kappa_{r}\dfrac{\partial T_s}{\partial r} (r=r_2) &=-\kappa_b\dfrac{\partial T_3}{\partial r} (r=r_2),\\
		T_3 (r\rightarrow\infty) &=-\nabla T r\cos \theta,
	\end{align}
\end{subequations}
where $T_1$, $T_3$, and $T_s$ are the temperature fields of region I, region III, and the shell, respectively; $r_1$ and $r_2$ are the inner and outer radii of the shell; $\kappa_b$ is the thermal conductivity of the background and $-\nabla T$ is a uniform thermal gradient. Substituting Eq.~(\ref{y2}) into
Eqs.~(\ref{y3}) yields the effective thermal conductivity
$\kappa_{e}$ of regions I and II combined,
\begin{eqnarray}\label{y4}
	\kappa_{e} = n\kappa_{r}\frac{\left(1+f^{n}\right)\kappa_{b} + \left(1-f^{n}\right)n\kappa_{r}}{\left(1-f^{n}\right)\kappa_{b} + \left(1+f^{n}\right)n\kappa_{r}},
\end{eqnarray}
with $f = \left(r_{1}/r_{2}\right)^{2}$. Owing to $\kappa_{rr} \approx \infty$ and $\kappa_{\theta\theta} \approx 0$, Eq.~(\ref{y4}) can be summarized as
\begin{equation}\label{y5}
	\kappa_{e} \approx \kappa_{b}.
\end{equation}
This result implies that the extremely anisotropic metashell adapts itself to the thermal landscape without perturbing it [Figs.~\ref{y-f1}(c), (d)]. Although the background thermal conductivity changes, $\kappa_b \rightarrow \kappa_b'$, the heat fluxes denoted by red arrows in region III remain unaltered. In this regard, the metashell is termed intelligent or chameleonlike~\cite{XuPRAp19-1,YangPRAp20}. Due to its thermal conductivity, $\text{diag}(\infty,0)$, the metashell considered here usually works as a thermal concentrator. Materials satisfying this parametric form are also called transformation-invariant materials in analogy with certain wave metamaterials~\cite{ZhangPRL19}. Such metamaterials preserve their materials parameters under arbitrary coordinate transformation, which means a transformed metashell is still intelligent. Let us consider an arbitrary coordinate transformation $r' = F(r,\theta)$ and $\theta' = \Theta(r,\theta)$, where ($r, \theta$) are the coordinates in the virtual space and ($r', \theta'$) the coordinates in the physical space, and the transformed thermal conductivity is $\bm{\kappa}' = \bm{J}\bm{\kappa}_0\bm{J}^{\dagger}/\det\bm{J}$, with $\bm{J} = \partial(r', \theta')/\partial(r, \theta)$ denoting the relevant Jacobian matrix. The detailed expression of the transformed thermal conductivity is
\begin{equation}\label{y6}
	\bm{\kappa}'=\frac{1}{\text{det}\bm{J}}\left[
	\begin{matrix}
		\kappa_{11} & \kappa_{12}\\
		\kappa_{21} & \kappa_{22}
	\end{matrix}
	\right],
\end{equation}
with
\begin{subequations}
	\begin{align} \kappa_{11}=&\kappa_{rr}\left(\frac{\partial{r'}}{\partial{r}}\right)^{2}+\kappa_{\theta\theta}\left(\frac{\partial{r'}}{r\partial{\theta}}\right)^{2},\\
	\kappa_{12}=& \kappa_{21} = \kappa_{rr}\left(\frac{\partial{r'}}{\partial{r}}\right)\left(\frac{r'\partial{\theta'}}{\partial{r}}\right)+\kappa_{\theta\theta}\left(\frac{\partial{r'}}{r\partial{\theta}}\right)\left(\frac{r'\partial{\theta'}}{r\partial{\theta}}\right), \\
	\kappa_{22}=& \kappa_{rr}\left(\frac{r'\partial{\theta'}}{\partial{r}}\right)^{2}+\kappa_{\theta\theta}\left(\frac{r'\partial{\theta'}}{r\partial{\theta}}\right)^{2},
	\end{align}
\end{subequations}
where $\bm{\kappa}_0$ is the thermal conductivity tensor of the transformation-invariant metamaterial. Imposing $\bm{\kappa}_0 = \text{diag}(\kappa_{rr}, \kappa_{\theta\theta}) = \text{diag}(\infty,0)$, we immediately obtain the eigenvalues of Eq.~(\ref{y6}),
\begin{subequations}\label{y7}
	\begin{align}
		\lambda_1 &= \frac{\kappa_{rr}}{\text{det}\bm{J}}\left[\left(\frac{\partial{r'}}{\partial{r}}\right)^{2} + \left(\frac{r'\partial{\theta'}}{\partial{r}}\right)^{2}\right]\approx\infty,\\
		\lambda_2 &= \frac{\kappa_{\theta\theta}}{\text{det}\bm{J}}\approx0.
	\end{align}
\end{subequations}
Metamaterials with extreme anisotropy do not change their parameters under an arbitrary coordinate transformation. Accordingly, intelligent metadevices with other thermal functions can be realized with transformation-invariant metamaterials, not limited to thermal concentrators, such as thermal rotators~\cite{YangPRAp20}.

\begin{figure}[!ht]
	\includegraphics[width=1.0\linewidth]{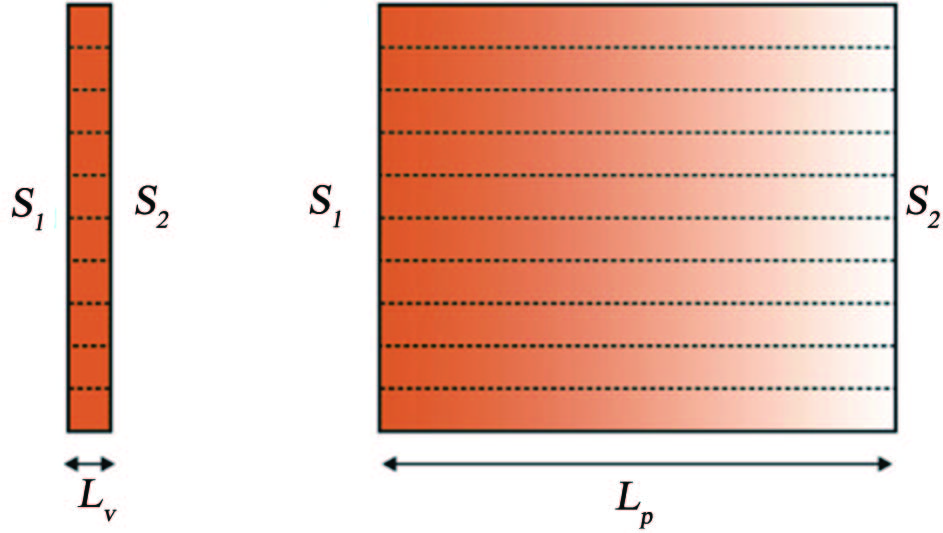}
	\caption{\label{y-f2} Schematic diagram of a thermal null medium. Adapted from~\citealp{BaratiSedehPRAp20}.}
\end{figure}

Extremely anisotropic parameters can also be obtained by coordinate transformation [Fig.~\ref{y-f2}]. A 2D slab of near-zero width $L_v$ in the virtual space is mapped to a slab of finite width $L_p$ in the physical space. The corresponding coordinate transformation is
\begin{subequations}\label{y8}
	\begin{align}
	x' =& x,~(x<0)\\
	x' =& \frac{L_p}{L_v}x,~(0<x<L_v)\\
	x' =& x - L_v +L_p.~(x>L_v)
	\end{align}
\end{subequations}
According to the transformation rules $\bm{\kappa}' = \bm{J}\kappa\bm{J}^{\dag}/\text{det}\bm{J}$, where $\bm{J} = \partial(x', y')/\partial(x, y)$, the transformed thermal conductivity is $\bm{\kappa}' = \kappa\cdot \text{diag}(L_p/L_v, L_v/L_p)$. Since $L_v \approx 0$, $\bm{\kappa}'$ is changed to $\kappa\cdot\text{diag}(\infty, 0)$. An extremely anisotropic metamaterials obtained by this method is also called a null medium~\cite{BaratiSedehPRAp20,SunOE19,ChenIJTS22,FakheriPRAp20}.

Although null media have the same parameters as the transformation-invariant metamaterials, this transforming method helps design arbitrarily shaped metamaterials. We take thermotics as an example and consider the concentrating transformation in the cylindrical coordinate system. Three arbitrarily shaped contours $R_2(\theta)$, $R_c(\theta)$, and $R_1(\theta)$ divide the concentrator into three regions. A larger region $r<R_c(\theta)$ is squeezed into a smaller one, $r'<R_1(\theta)$, to enhance the heat flux, and the region $R_c(\theta)<r<R_2(\theta)$ is stretched into the region $R_1(\theta)<r'<R_2(\theta)$. The heat fluxes previously located in $r<R_c(\theta)$ are now compressed in $r'<R_1(\theta)$. The transformation is written as
\begin{subequations}\label{y9}
	\begin{align}
		r' &=\frac{R_1(\theta)}{R_c(\theta)}r,~[r'<R_1(\theta)]\\ r'&=\frac{\left[R_2(\theta)-R_1(\theta)\right]r}{R_2(\theta)-R_c(\theta)}\nonumber\\
		 &+\frac{\left[R_1(\theta)-R_c(\theta)\right]R_2(\theta)}{R_2(\theta)-R_c(\theta)},[R_1(\theta)<r'<R_2(\theta)]\\
		\theta' &=\theta.
	\end{align}
\end{subequations}
Three contours are supposed to have the same shape factor $R(\theta)$, that is, $R_2(\theta) = r_2R(\theta)$, $R_c(\theta) = r_cR(\theta)$, and $R_1(\theta) = r_1R(\theta)$, where $R(\theta)$ is an arbitrary continuous function with a period of $2\pi$. $r_2$, $r_c$, and $r_1$ are three constants which satisfy the condition of $r_1<r_c<r_2$. By combining Eqs.~(\ref{y9}) and the transformation rules, the transformed thermal conductivity in the region $R_1(\theta)<r'<R_2(\theta)$ can be expressed as
\begin{equation}\label{y10}
	\bm{\kappa}'=\kappa_b\left[
	\begin{matrix}
		\kappa_{rr} & \kappa_{r\theta}\\
		\kappa_{\theta r} & \kappa_{\theta\theta}
	\end{matrix}
	\right],
\end{equation}
with components
\begin{subequations}\label{y11}
	\begin{align}
		\kappa_{rr}&=\frac{(r_2-r_c)r'+r_2(r_c-r_1)R(\theta)}{(r_2-r_c)r'}\nonumber\\
		 &+\frac{r_2^2(r_c-r_1)^2\left[dR(\theta)/d\theta\right]^2}{\left[(r_2-r_c)^2r'+r_2(r_c-r_1)(r_2-r_c)R(\theta)\right]r'},\\
		\kappa_{r\theta}&=\kappa_{\theta r}=\frac{r_2(r_1-r_c)dR(\theta)/d\theta}{(r_2-r_c)r'+r_2(r_c-r_1)R(\theta)},\\
		\kappa_{\theta\theta}&=\frac{(r_2-r_c)r'}{(r_2-r_c)r'+r_2(r_c-r_1)R(\theta)}.
	\end{align}
\end{subequations}
A null medium requires $r_c \approx r_2$, so Eqs.~(\ref{y11}) can be simplified as
\begin{subequations}\label{y12}
	\begin{align}
		\kappa_{rr}&=\frac{1}{\Delta},\\
		\kappa_{r\theta}&=\kappa_{\theta r}=\frac{dR(\theta)/d\theta}{R(\theta)},\\
		\kappa_{\theta\theta}&=\Delta,
	\end{align}
\end{subequations}
where $\Delta \approx 0$ and $(dR(\theta)/d\theta)/R(\theta)$ is a finite quantity. The passive and stable heat conduction process is governed by the Fourier law $\nabla\cdot(-\kappa\nabla T) = 0$. By substituting Eqs.~(\ref{y10}) and (\ref{y12}) into the Fourier law, the resulting Laplace equation, expressed in cylindrical coordinates, reads
\begin{equation}\label{y13}
	\begin{split}
		&\frac{1}{r}\frac{\partial T}{\partial r} + \frac{\partial^{2}T}{\partial r^{2}}+\Delta\frac{2}{r}\kappa_{r\theta}\frac{\partial^{2}T}{\partial r\partial\theta}+\Delta\frac{1}{r}\frac{\partial \kappa_{\theta r}}{\partial\theta}\frac{\partial T}{\partial r} \\
		&+\Delta^{2}\frac{1}{r^{2}}\frac{\partial^{2}T}{\partial\theta^{2}}=0.
	\end{split}
\end{equation}
$\kappa_{r\theta}$ has little effect on heat transfer due to the near-zero value of $\Delta$. The exact value of the off-diagonal components is not essential. Therefore, the extremely anisotropic parameter ${\rm diag}(\infty,0)$ is obtained by setting the off-diagonal components to zero. With such material parameters one can realize an arbitrarily shaped thermal concentrator~\cite{BaratiSedehPRAp20}. In conclusion, thermal null media share the same parameters as transformation-invariant metamaterials. Metashells with such extremely anisotropic parameters do help overcome the limitations of traditional transformation-thermotics metamaterials, in that they work well also in a changing background with an irregular shape. Moreover, due to the robustness of their transformation invariance, metashells with extreme anisotropy can be designed to perform a wide range of thermal functions.

\subsubsection{Numerical optimization theory}

The transformation theory~\cite{LeonhardtScience06,PendryScience06,FanAPL08,ChenAPL08} links the physical space with a parametric space. The physical field can be controlled on demand by suitably tuning the material properties. However, as mentioned above, the material parameter patterns obtained via transformation theory usually have nonuniformity and anisotropy characteristics~\cite{NarayanaPRL12,YuPNAS14} difficult to find in natural bulk materials. With the effective medium theory, inhomogeneity and anisotropy are obtained by taking advantage of the anisotropy associated with the material geometries. However, when the governing equation of a system fails to satisfy form invariance under coordinate transformation, the transformation theory becomes inapplicable. In this case, one may have recourse to the scattering-cancellation theory~\cite{HePRE13,XuPRL14,MaPRL14,HanPRL14}. With that technique natural bulk materials can be utilized to realize, for instance, a bilayer cloak that maintains its core temperature uniform, without perturbing the temperature field of the background. The rationale of this method consists in solving the stationary heat conduction equation by implementing the continuity of normal heat flow and temperature at the junction of different materials. Due to the geometric symmetry of the boundary conditions~\cite{YangAM15}, the solution to the heat conduction equation can be significantly simplified. On the other hand, if the geometry of the material is not symmetric, the boundary conditions are also not symmetric and an analytical solution to the governing equation may no longer exist. Recently, the emergence of optimization methods~\cite{PopaPRA09,XuSCIS13,DedeSMO14,DedeIEEE15,PeraltaSR17,GuoAM22,LuoATE22,WangJAP22} has made up for the limitations of the transformation theory (inhomogeneous and anisotropic) and the scattering-cancellation theory (single component, simple shape). The optimization method helps select the size and shape of the natural bulk materials best suited to reproduce the thermal properties determined via transformation theory. At the same time, when the scattering-cancellation theory does not admit of analytical solution, the optimization method can be used to define an optimal device target function and
numerically determine the relevant material parameters. Many distinct optimization methods are being employed to design diffusion metamaterials, including particle swarm optimization~\cite{KennedyIEEE95,PoliSI07,AlekseevDP17,AlekseevDP17-2,AlekseevCMMP18,AlekseevIJHMT19,JinIJHMT21}, topology optimization~\cite{BendsoeCMAME88,CakoniIP18,FujiiAPL18,FujiiAPL19,FujiiIJHMT19,FujiiIJHMT20,ShaJAP20,ShaNC21,ZhuIJHMT21,FujiiPRE22,HirasawaIJHMT22,ShaNPJ22,ShaMTP22,SvanbergIJNME87,LuoATE22}, and machine learning~\cite{HuPRX20,HuNE20,LuoResearch20,QianNP20,KudyshevAPR20,LuIJHMT22,JiIJHMT22,JiIJHMT22-2,JinAM23,LinScience18,ShaltoutScience19,PeurifoySA18,OuyangCPL20,OuyangFoP21,BoeRMP22}. To comprehend the importance of optimization techniques for creating diffusion metamaterials, we present a range of illustrative studies.

\begin{figure}[!ht]
	\includegraphics[width=1.0\linewidth]{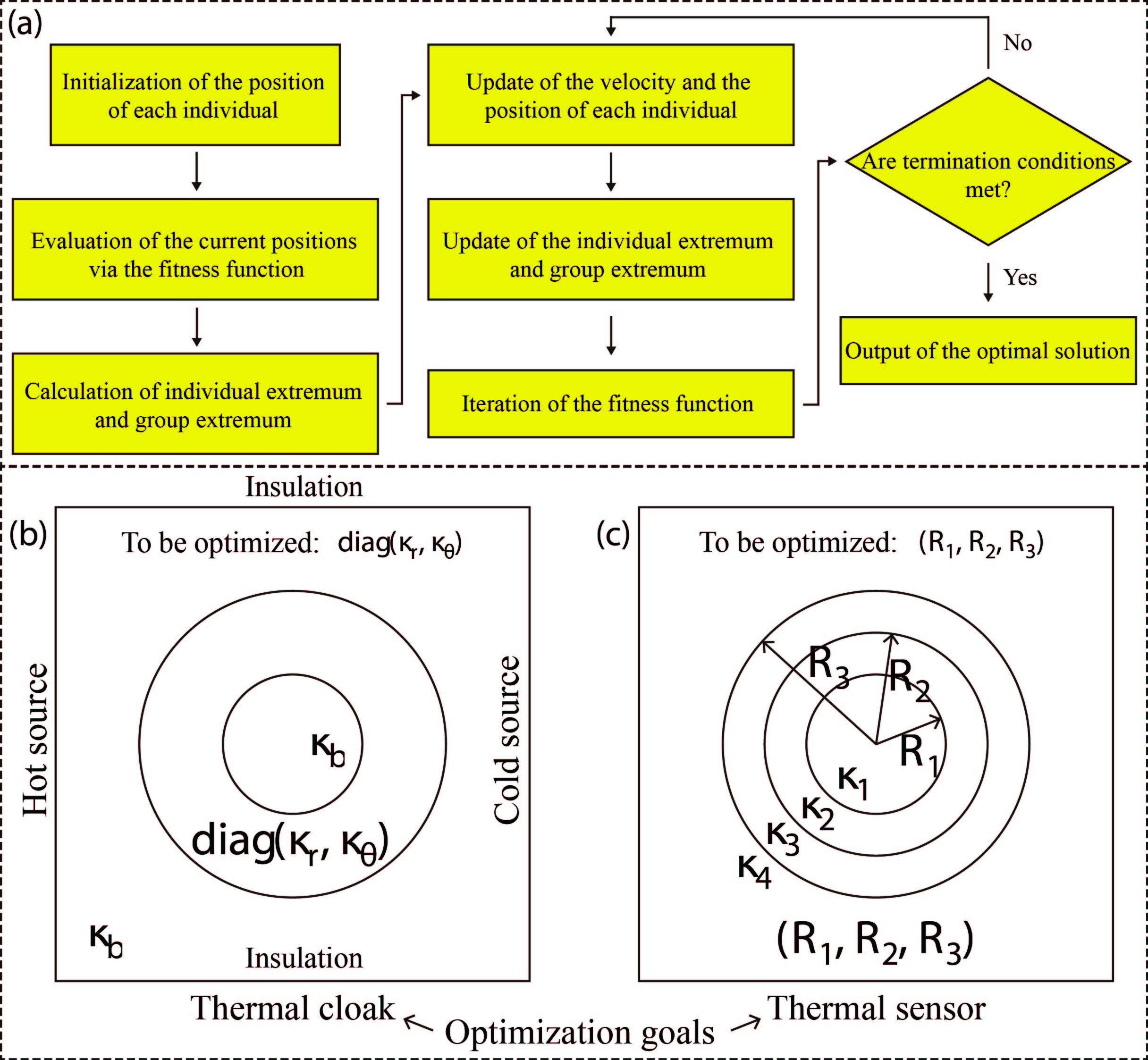}
	\caption{Particle-swarm optimization (PSO) thermal metamaterials. (a) Algorithm flow chart. (b) Schematic diagram of a PSO-based thermal cloak. (c) Schematic diagram of a PSO-based thermal sensor. Boundary conditions in (b) and (c) are the same.}
	\label{1}
\end{figure}

{\em (i)} Particle swarm optimization (PSO) methods originated from the work of Kennedy (a social psychologist) and Eberhart (an electrical engineer), who thought of computational intelligence in analogy with social interactions~\cite{KennedyIEEE95}. As illustrated in Fig.~\ref{1}(a), the procedure begins with the initialization of the positions of each individual in accessible solution space, where each position represents a potential solution to the optimization problem with a corresponding fitness function quantifying the optimization effect. The fitness functions of the current position of each individual are then calculated, and the individual- and group-extremum points are identified. Here, individual- and group-extremum points are the positions with the minimum fitness functions that individuals and the group respectively have experienced up to the current time. Then, the positions of each individual are updated based on the shared information of the swarm (individual- and group-extremum points). The minimum fitness functions of each individual are iteratively evaluated. Finally, if the selected fitness function meets the predefined termination condition, the optimal solution is outputted. If not, the loop will continue. The PSO can be used as an inverse solver and is suitable to address problems where other classical methods yield untenable solutions. Recently, it has been used as an effective tool for the inverse design of diffusion metamaterials~\cite{AlekseevDP17,AlekseevDP17-2,AlekseevCMMP18,AlekseevIJHMT19,JinIJHMT21}. The geometric or material parameters can be inversely designed when the demanded metamaterial properties are known. For instance, Alekseev \emph{et al}.~\cite{AlekseevDP17,AlekseevDP17-2,AlekseevCMMP18,AlekseevIJHMT19} optimized this way the material properties of different shells to be used in a thermal cloaking device [see Fig.~\ref{1}(b)]. However, by forward solving the stationary heat conduction equation, one can also find the analytical solution for the shell material properties. In this regard, the advantages of the PSO method have not been fully exploited, yet. In 2021, Jin \emph{et al}.~\cite{JinIJHMT21} applied the PSO method to design a bilayer thermal sensor from natural bulk materials [Fig.~\ref{1}(c)], a problem hard to tackle by means of analytical tools. In a previous attempt~\cite{JinIJHMT20}, a thermal sensor based on an anisotropic monolayer structure had required extreme parameters almost impossible to engineer. In this more recent work, the authors chose natural bulk materials as components of a bilayer thermal sensor and optimized the three radii, $R_1$, $R_2$, and $R_3$, defining the geometry of Fig.~\ref{1}(c). Starting with a forward approach, they considered the Laplace equation for the stationary heat conduction. However, they ended up with multiple unknowns nonlinearly coupled with each other: the forward analytical method was of no help to calculate any radius. They noticed that the radii, $R_1$, $R_2$, and $R_3$, are expectedly the main factors affecting their performance. To set up an inverse problem in the PSO framework, these authors defined two optimization functions for the accuracy and non-disturbance of their device,
\begin{equation}\label{111}
	\bm{\Psi_{s}}=\frac{1}{N_s}\sum_{i=1}^{N_s} \left|T\left(i\right)-T_{ref}\left(i\right)\right|,
\end{equation}
\begin{equation}\label{22}
	\bm{\Psi_{b}}=\frac{1}{N_b}\sum_{i=1}^{N_b} \left|T\left(i\right)-T_{ref}\left(i\right)\right|,
\end{equation}
where $i$, $T$, $T_{ref}$, $N_s$, and $N_b$ are, respectively, the node index, temperature data in bilayer thermal sensor, temperature data in the pure background, number of nodes in the detected and background regions after discretization. The fitness function is the sum of Eq.~(\ref{111}) and Eq.~(\ref{22}),
\begin{equation}\label{33}
	\bm{\Psi}=\bm{\Psi_{s}}+\bm{\Psi_{b}}.
\end{equation}
As a starting step of the algorithm flow chart of Fig.~\ref{1}(a), Jin \emph{et al}. initialized $N$ individual $\bm{R}_j^0$ with $j=1,2,...,N$ in the solution domain $\bm{K}$,
\begin{equation}\label{44}
	\begin{split}
		&\bm{K}\!=\\
		&\left\{\bm{R}\!=\!\left(R_1,R_2,R_3\right)\!:\!R_{min}\leq R_i<R_j\leq R_{max},i<j;i,j\!\in\!\left\{1,2,3\right\}\right\}.
	\end{split}
\end{equation}
The process of iterating particles' velocity and position over time is characterized as
\begin{equation}\label{55}	 \bm{V}_j^{i+1}=w\bm{V}_j^i+c_1d_1\left(\bm{P}_j^i-\bm{R}_j^i\right)+c_2d_2\left(\bm{P}_g-\bm{R}_j^i\right),
\end{equation}
\begin{equation}\label{66}
	\bm{R}_j^{i+1}=\bm{R}_j^i+\bm{V}_j^{i+1},
\end{equation}
where $i$ is the iteration number, $j$ the particle label, $\bm{P}_j^i$ the individual extremum point of the $j$-th particle at the $i$-th iteration, and $\bm{P}_g$ the group extremum point. Furthermore,  $w$ is the inertia weight, commonly chosen as a decreasing function; $c_1$ and $c_2$ represent empirical constant; $d_1$ and $d_2$ are random numbers between 0 and 1. After an appropriate number of iterations, the minimum fitness function is finally met; hence, the optimal values of three radii are designed. The resulting bilayer thermal sensors have excellent performance. This work demonstrates the extraordinary advantage of the PSO method in the design of thermal metamaterials, when analytical approaches are not viable~\cite{ZhangPRD22}.

\begin{figure}[!ht]
	\includegraphics[width=1.0\linewidth]{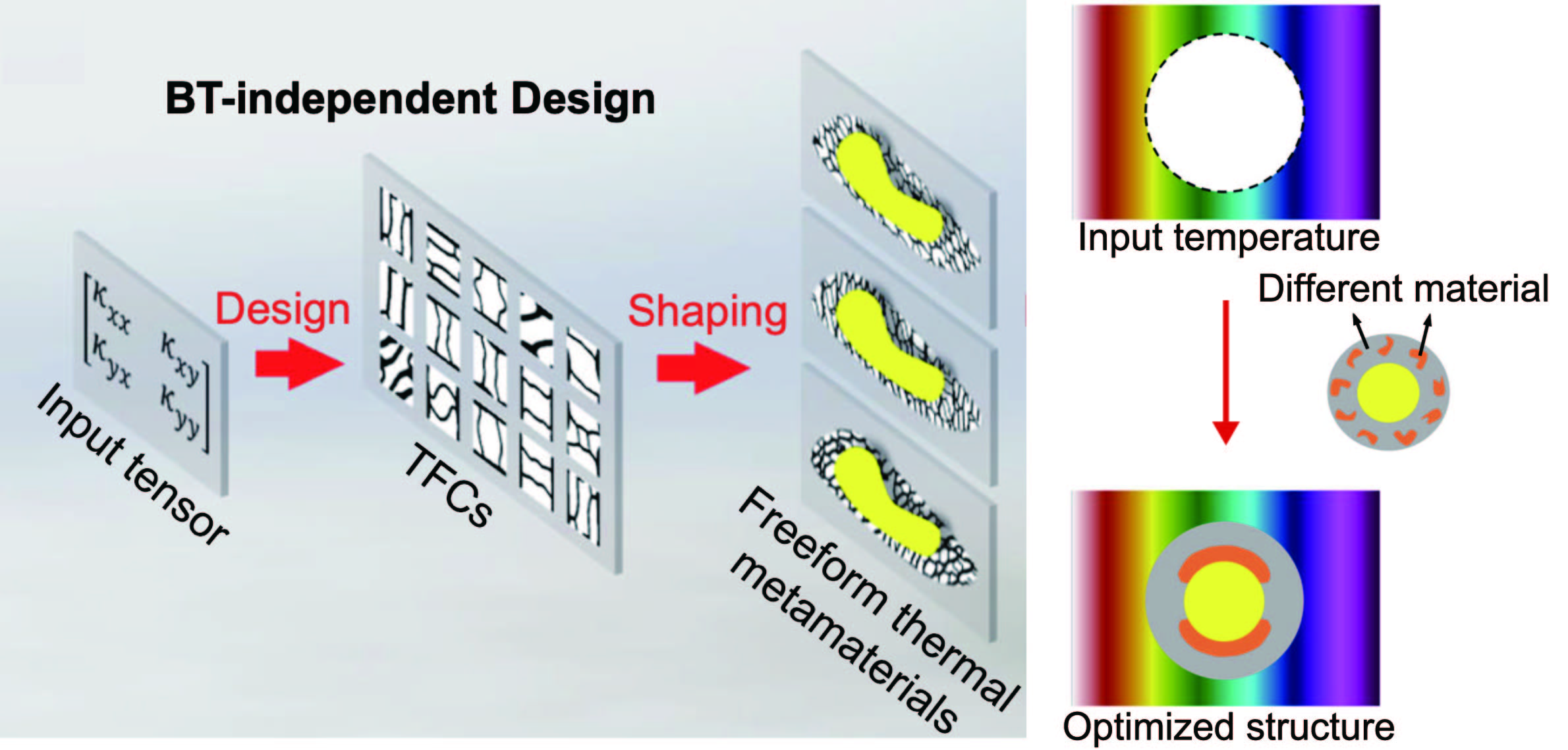}
	\caption{Topology-optimized thermal metamaterials. Adapted from~\citealp{ShaNC21}.}
	\label{2}
\end{figure}

{\em (ii)} Recently, researchers have used a more flexible algorithm, named topology optimization~\cite{SvanbergIJNME87,BendsoeCMAME88,CakoniIP18} to design thermal metamaterials. Thermal metamaterials, or metadevices, can achieve high thermal performance only through an optimized distribution of components made of natural bulk materials. This new approach led to many excellent works. For instance, Fujii \emph{et al}.~\cite{FujiiAPL18,FujiiAPL19,FujiiIJHMT19,FujiiIJHMT20,FujiiPRE22,HirasawaIJHMT22} optimized the layout of two bulk materials [copper and polydimethylsiloxane (PDMS)] in the shell-like metamaterial region and designed a series of powerful thermal metadevices, including classical thermal cloaks~\cite{FujiiIJHMT19}, multifunctional thermal metadevices~\cite{FujiiIJHMT20}, and multiphysical thermal concentrators~\cite{FujiiPRE22}. Multifunctional thermal metadevices designed by topology optimization demonstrate that these optimization problems have two objective functions. For example, the thermal cloak-concentrator works as a thermal concentrator in the core region (objective function 1) without perturbing the background temperature distributions (objective function 2). Sometimes devices need to work in a complex multiphysical environment. This motivated Fujii \emph{et al}. to improve their cloak-concentrator device to simultaneously operate in the presence of both thermal and electric fields~\cite{FujiiPRE22}. Two objective functions are defined for cloaking temperature and voltage in the background, and two more for the heat flux and directed electric current in the core region. At the end of the optimization process, an optimized structure made of copper and PDMS is obtained. However, none of the above metadevices have ever been realized experimentally. In 2022, Hirasawa \emph{et al}.~\cite{HirasawaIJHMT22} fabricated a topology optimized metamaterial made of copper, steel, and PDMS resorting to a precision cutting machine. Topology optimization is also an effective tool for designing transformation-theory based thermal metamaterials. As mentioned above, such metamaterials usually require inhomogeneous and anisotropic material parameters, nowhere to be found in nature. Topology optimization may help overcome this difficulty. Sha \emph{et al}.~\cite{ShaNC21} proposed a new concept of topological functional cells, thus making the realization of inhomogeneous and anisotropic materials possible. Their approach consists of three steps. First, the thermal conductivity distribution in the metamaterial region is calculated based on the transformation theory. Second, the functional cells are topologically optimized to achieve the desired thermal conductivity tensor (topological functional cells). Third, such cells are assembled together to engineer a new thermal metamaterial with the desired inhomogeneous and anisotropic conductivity [Fig.~\ref{2}]. The optimization algorithm for a single topological functional cell is formulated as,
\begin{equation}
	\begin{split}
		&{\rm min}~ C = \frac{1}{|V|}\sum_{e=1}^N \rho_e, \\
		&{\rm such~that}: \mathbf{\rm K}(\rho_e) \mathbf{\rm T} = \mathbf{\rm Q}, \\
		&G=f((\kappa_{lm}^{\rm Output} - \kappa_{lm}^{\rm Input})^2) = 0, \\
		&0 \leq \rho_e \leq 1,~e = 1,~2...N,\\
		&\kappa_{lm}^{\rm Input} = \begin{pmatrix} \kappa_{11}^{\rm Input} & \kappa_{12}^{\rm Input} \\ \kappa_{21}^{\rm Input} & \kappa_{22}^{\rm Input} \end{pmatrix}(l,m=1,2),\\
		&\kappa_{lm}^{\rm Output} = \begin{pmatrix} \kappa_{11}^{\rm Output} & \kappa_{12}^{\rm Output} \\ \kappa_{21}^{\rm Output} & \kappa_{22}^{\rm Output} \end{pmatrix}(l,m=1,2),
	\end{split}
\end{equation}
where $\rho_e$ represents a continuous design variable, that ranges from $\rho_e=0$ (material 1) to $\rho_e=1$ (material 2). $N$ is the amount of $\rho_e$, and $|V|$ is the volume of the entire topological functional cell. $\mathbf{\rm K}(\rho_e)$, $\mathbf{\rm T}$, and $\mathbf{\rm Q}$ are globally heat conduction, temperature, and heat load matrices, respectively. $\kappa_{lm}^{\rm Input}$ and $\kappa_{lm}^{\rm Output}$ are target and optimized thermal conductivity tensors, respectively. $f$ is a function to weigh the difference between $\kappa_{lm}^{\rm Output}$ and $\kappa_{lm}^{\rm Input}$. The distribution of $\rho_e$ is optimized by the classical method of moving asymptotes. Recently, the attainable range of anisotropic thermal conductivity has been significantly expanded by topology-optimized functional cell design~\cite{ShaNPJ22}. Topology optimization makes diffusion metamaterial design more affordable.

\begin{figure}[!ht]
	\includegraphics[width=1.0\linewidth]{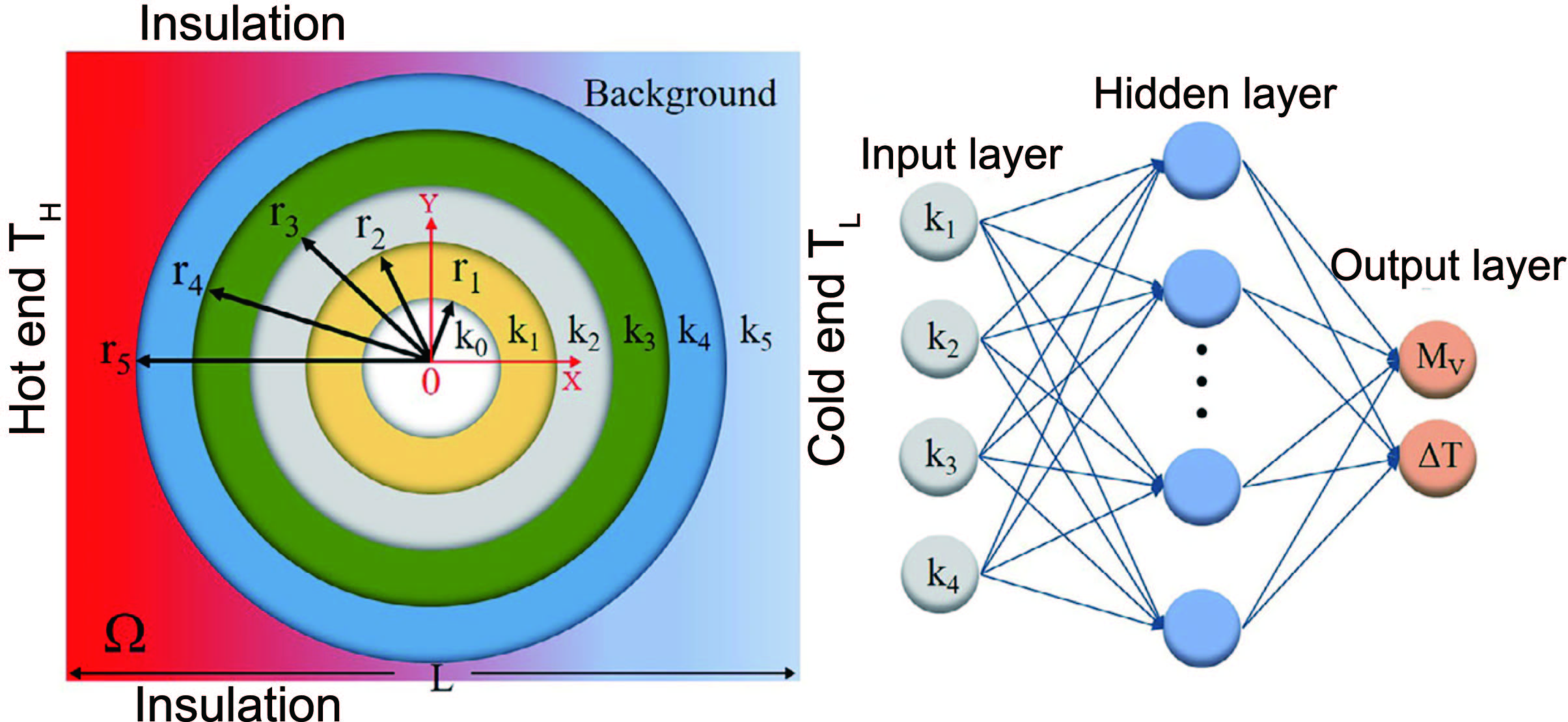}
	\caption{Framework of a machine learning-based thermal cloak. Adapted from~\citealp{JiIJHMT22}.}
	\label{3}
\end{figure}

{\em (iii)} Thermal metamaterials are rapidly developing distinctive features like multiphysics, multi-functions, and intelligence~\cite{HuPRX20,HuNE20,QianNP20,JinAM23}. Therefore more sophisticated design tools are needed than offered by the traditional analytical theories, including transformation theory. In particular, when the design variables are too many or not directly related to the metamaterial performance, getting an analytical form for the material parameter solution becomes increasingly difficult. Machine learning algorithms may offer an alternative approach. Recently, a new class of intelligent materials combining artificial intelligence algorithms with material design has received great attention from researchers in optics~\cite{LinScience18,ShaltoutScience19,QianNP20}, nanotechnology~\cite{PeurifoySA18}, and acoustics~\cite{LuoResearch20}. However, compared with wave systems, diffusion physics lacks controllable degrees of freedom, which apparently limits the applications of this new technology to the diffusion metamaterials. We take again the thermal counterpart as an example. Existing work is confined so far to the inverse design of geometry or material parameters. Ji \emph{et al}.~\cite{JiIJHMT22} optimized a four-layer thermal cloak by inverse design through a machine learning algorithm [Fig.~\ref{3}]. These authors considered the thermal conductivities, $k_1,k_2,k_3$, and $k_4$ of four isotropic materials (four middle layers) as the input of an artificial neural network, and defined the outputs as two objective functions, meant to quantify the performance of a thermal cloak. One objective function evaluates the uniformity of the temperature in the cloaked region,
\begin{equation}
	\Delta T = |T_{x=r_1}-T_{x=-r_1}|,
\end{equation}
the other one is to characterize the thermal neutrality in the background region,
\begin{equation}
	M_V = \frac{\int_{\Omega}|T(x,y,z)-T_r(x,y,z)|d\Omega}{\int_{\Omega}d\Omega},
\end{equation}
where $\Omega$ is the domain $r > r_5$, and $T_r$ is the temperature distribution of a uniform medium. After training an artificial neural network via 10000 design samples, an inverse-mapping between output \{$\Delta T$, $M_V$\} and input ($k_1,k_2,k_3,k_4$) was established. Finally, the thermal conductivities of four isotropic materials were calculated for optimal cloaking performance, and a sample composed of copper and PDMS was designed based on the effective medium theory. The resulting device guaranteed uniform temperature distributions inside the cloaked region without perturbing the background thermal field. However, such a machine-learning based thermal cloak is still a static device, lacking dynamic adaptability. Subsequently, again with the aid of artificial intelligence and more advanced hardware, a new thermal metamaterial was designed with parameters autonomously adapting to a variable environment~\cite{JinAM23}. A micro infrared camera is used to monitor the temperature of a bilayer structure. Via a computing system with a well-trained artificial neural network, the monitored temperature information is associated with the spinning angular velocity of the core region (through a stepper motor), dynamically adjusting the thermal conductivity of the core region~\cite{XuNC20}. In particular, the input and output are connected with four hidden layers as follows
\begin{equation}
	\begin{cases}
		\bm{H}^{(i+1)}={\rm ReLU} \left(\bm{W}^{(i)}\bm{T}^{(i)}+\bm{b}^{(i+1)}\right),~i~=~0\\
		\bm{H}^{(i+1)}={\rm ReLU} \left(\bm{W}^{(i)}\bm{H}^{(i)}+\bm{b}^{(i+1)}\right),~0~<~i~<~4\\
		\omega_1={\rm ReLU} \left(\bm{W}^{(i)}\bm{H}^{(i)}+\bm{b}^{(i+1)}\right),~i~=~4
	\end{cases}
\end{equation}
where $\bm{H}^{(i)}$ are the activations of the current layer, ${\rm ReLU}\left(\rm a\right)={\rm max}\left(0,\rm a\right)$ represents the rectified linear unit function, $\bm{W}^{(i)}$ and $\bm{b}^{(i)}$ are the weights and biases for neurons in the $i$-th layer. Such design links the ambient temperature information with its thermal functions through a machine learning intelligent device with sensing-feedback performance. Self-adaptive metamaterials represent one of the most promising advances in the field. In view of existing intelligent metamaterials in wave systems, machine-learning algorithm has injected new inspiration and vitality into metamaterial design, that is, enabling metamaterials to have the ability to think and make decisions like humans. Taking the development of optical cloak as an example, a traditional optical cloak can only have stealth effects within a specific frequency range of the transverse wave band. Once the background light field undergoes dynamic changes, the originally designed cloak will no longer work. Only very recently, Qian \emph{et al}.~\cite{QianNP20} proposed the concept of a self-adaptive optical cloak in microwave bands, enabled by deep learning. Experimental results show that such a cloak works with a robust invisible performance against the ever-changing ambient light with no human intervention. Significantly, the inclusion of the machine-learning algorithm in metamaterials enhanced the ability and flexibility of controlling light fields. Looking back on recent advances in machine learning-based metamaterials in longitudinal wave systems, like in acoustic systems, Luo \emph{et al}.~\cite{LuoResearch20} developed a probability-density-based deep-learning algorithm for capturing all plausible metastructures for the desired transmission performance. Here, the machine-learning algorithm also helps automatic solutions for inverse structural or parametric design of functional metamaterials in wave systems. By contrast, we do not have to consider spectral characteristics when designing machine learning-assisted metamaterials in diffusion systems. In particular, it is difficult to design a full-band intelligent optical stealth cloak, while a thermal cloak can work well under different ambient temperatures. Only one aspect has to be considered when designing intelligent thermal materials, namely their self-adaptation to the different environmental components, an aspect recently explored for deep learning-based self-enhanced thermal materials~\cite{JinAM23}.

\subsubsection{Topological states and non-Hermitian physics}

Another intriguing topic is the topological properties of thermal diffusion. The classification of different phases of matter has been a long-standing issue in condensed matter physics. Before the 1980s, it was widely believed that the Landau paradigm based on spontaneous symmetry breaking could explain all phase transitions. However, the discovery of the quantum Hall effect~\cite{KlitzingPRL80, Klitzing05} challenged this perspective because the quantum Hall effect does not break any symmetries. Later, the quantum Hall transition was reinterpreted as a topological phase transition and the Hall conductance as a topological invariant~\cite{ThoulessPRL82}. The last twenty years witnessed the flourishing of topological physics, and plenty of topological phases of matter have been theoretically predicted and experimentally discovered~\cite{HasanRMP10, QiRMP11, ChiuRMP16, BansilRMP16}, first for gapped systems. A quantum spin Hall insulator with time-reversal symmetry exhibits robust helical edge state~\cite{KanePRL05-1,KanePRL05-2}, as observed in HgTe/CdTe quantum well structures~\cite{BernevigSci06, KonigSci07}. A quantum anomalous Hall insulator requiring no external magnetic fields was initially proposed in a honeycomb model~\cite{HaldanePRL88} and then experimentally realized in thin films of chromium-doped ${\rm{(Bi,Sb)_{2}Te_{3}}}$~\cite{YuSci10, ChangSci13}. Other phases occur in gapless systems. Weyl semimetal hosts quasiparticles satisfying the Weyl equation in high energy physics~\cite{WanPRB11, ArmitageRMP18}. The surface states of Weyl semimetals consist of topologically protected Fermi arcs, which contributes to many exotic transport properties. 

Outside condensed matter physics, topological effects have been quickly incorporated into the phenomenology of classical wave systems, which originated new areas of research. The most typical and significant example is topological photonics~\cite{LuNP14, OzawaRMP19, PriceJPP22}. Based on photonic crystals, Haldane and Raghu proposed that electromagnetic waves in spatially periodic structures can exhibit nontrivial topological properties~\cite{HaldanePRL08, RaghuPRA08}. The basic principle is the mapping between the Maxwell equations in electrodynamics and the Schr${\rm{\ddot{o}}}$dinger equation in quantum mechanics. Topological phases that have not been produced in condensed matter physics, could be produced in photonic system due to their great tunability. For example, a 3D Chern insulator characterized by the Chern vector was demonstrated in a gyromagnetic photonic crystal without condensed matter counterpart~\cite{LiuNat22}.

Another example of the combination between topological physics and artificial structures is topological acoustics~\cite{YangPRL15, XueNRM22}. Early studies on topological acoustics mostly replicated fundamental topological phases from condensed matter physics. However, in recent years, the rapid development of topological acoustics has surpassed the advances of condensed matter physics. This field investigates now novel exotic topological phases of matter, well  beyond those obtained in solids. For example, almost simultaneously with the theoretical proposal, fragile topology has been experimentally characterized in an acoustic metamaterial~\cite{PeriSci19}. All in all, the encounter of topological physics with classical metamaterials turned out to be beneficial for both fields. 

Another physical framework we will discuss in the context of metamaterials is the so-called non-Hermitian physics. In quantum mechanics, a basic postulate is that the Hamiltonian describing a system must be Hermitian to ensure observability, so its eigenvalue is a real number. However, people find that most physical systems in nature are not isolated and will inevitably couple with the surrounding environment. Although people develop many approaches to characterize these systems in Hermitian physics, the effective non-Hermitian Hamiltonian provides a simpler and more intuitive method to understand the dynamics of open systems~\cite{ChingRMP98,AshidaAP20,BergholtzRMP21,FruNat21,SebNat22}. Therefore, the study of non-Hermitian physics has been a frontier in recent decades.
\subsection{Experimental methods}

Metamaterial fabrication means achieving inhomogeneous, anisotropic, and other extreme material parameters rarely found in natural materials. One way to overcome this challenge is suggested by the effective medium theories introduced above. The first experimental demonstration of thermal
metamaterials used layered structures to realize anisotropic material
parameters~\cite{NarayanaPRL12}. The experimental research was later
extended to transient regimes by designing the effective parameters of mass
density and heat capacity~\cite{SchittnyPRL13}. As an example, we use the isotropic materials B and C to construct an alternate layered structure A. Its parameters in the perpendicular, $A_\perp$, and parallel direction, $A_\parallel$, are different, i.e., $A_\perp=1/ (f_B / A_B+f_C / A_C )$ and $A_\parallel=f_B A_B+f_C A_C$. $f_B$ and $f_C=1-f_B$ are their area/volume fractions. One can choose air as material C to achieve extreme anisotropy with only one natural material. Here, interfacial resistance may slightly affect these parameters and reduce the effective anisotropy~\cite{LiJAP10,ZhengPRAp20}, but this effect can be easily compensated by fine tuning the fractions $f_B$ and $f_C$.

The Maxwell-Garnett and Bruggeman theories are commonly used to model artificial isotropic but inhomogeneous parameters. The Maxwell-Garnett theory is often invoked when dealing with systems where unconnected particles are embedded in an isotropic matrix. For example, by punching unconnected holes (with holes as particles made of air) periodically into a plate, a continuous parameter can be achieved by grading the area fraction of the holes. The maximum particle fraction is less than 1 in the Maxwell-Garnett theory. As for the Bruggeman theory, we cannot distinguish the overlapping particles and the matrix. The particle fraction can vary from 0 to 1, and the material parameters are symmetric functions of the component fractions. Moreover, the embedded particle can be elliptical and ordered-oriented to achieve anisotropy, as confirmed by a thermal conduction experiment~\cite{TianIJHMT21}. The effective medium theory has been applied to many distinct problems, such as heat transfer and mass diffusion, with remarkable cross-fertilization advantages.

The most common and effective way to detect the temperature field in thermal systems is through infrared thermal imaging. In most experiments conducted near room temperature, the heat loss due to thermal radiation from the surface of a sample is much less than thermal conduction and convection, making infrared thermal imaging a commonly interference-free method. However, infrared thermal imaging can be cheated to get effective higher readings~\cite{WangPRAp20,ZhangCPL23}. Indeed, cavity unit has an effective high emissivity similar to that of an ideal blackbody and allows resolving infrared patterns at a uniform temperature. The properties of thermal metamaterials are often determined experimentally by infrared imaging.

Thermal cloaking has been extensively studied in conduction systems~\cite{NarayanaPRL12,LanOE16-1,ZhouES19,JiES20,GuoES20,HouCPL21,GaoCPL21,FengJAP22}, mainly having recourse to the effective medium theory. Li \emph{et al.}~\cite{LiPRL15} used a shape-memory alloy to switch the connection of layered structure at a specific temperature, thus realizing a macroscopic thermal diode. Its temperature dependent parameters made such a device adaptive to different environments.

Thermal convection is a fundamental mechanism of heat transport in moving fluids or solids. However, controlling the matter convection can be very difficult due to the nonlinearities in the relevant Stokes equation. Jin \emph{et al.}~\cite{JinPNAS23} produced a tunable metamaterial in a idealized conduction-convection system governed by Darcy's law. According to the transformation rules~\cite{DaiJAP18} and the effective medium theory, they achieved the desired thermal conductivity by choosing appropriate fractions of different materials, and simultaneously controlled the diffusion coefficients by means of layered structures. By increasing the pressure, convection was continuously tuned, and the heat flux amplification factor increased, which was validated by infrared imaging and ink streamlines. As for moving structures, speed can be accurately determined by motors, making the effective medium theory solvable for rotating structures \cite{LiOE20}. Upon increasing the motor speed, the effective thermal conductivity of a metadevice was experimentally varied over a wide range~\cite{LiNM19,LiAM20}. Note that, contrary to the case of fluid layers, thermal interfacial resistance between moving layers of solid materials greatly impacts the resulting thermal properties.

Some external gain or loss systems also brought new insight. Guo \emph{et al.} integrated thermoelectric thermal sensors with active heaters and coolers to adapt to changing environments~\cite{GuoAM22}. Each semiconductor unit reads its instantaneous temperature and adjusts its current to maintain a constant heat or cold source. Thus, the programmable metasurface marginally depended on the operational conditions. In another experiment, the adjustable heat source was introduced utilizing a well-contacted copper cylinder held at a fixed temperature~\cite{XuPRE19}. The fixed temperature boundaries provided by the copper cylinder could be treated as a thermal dipole that compensates for background distortions caused by material mismatches. Since the distance and temperature difference between two copper cylinders could be easily altered, this thermal dipole is tunable for different situations.

Experiments of particle diffusion are much more challenging than heat transfer because particle concentrations cannot be imaged directly. Despite observed novel functions, such as concentrating and cloaking particle diffusion, and transforming chemical waves~\cite{ZhangATS22}, most studies still focus on theoretical design and simulations rather than on experiments. Zeng \emph{et al.}~\cite{ZengSP13} fabricated an actual diffusion cloak for chloride ions. Based on the effective medium theory, the cloak was made of five layers of concrete with different compositions and diffusion coefficients. An ion-selective electrode measured the concentration of chloride ions in the vertical test lines, and a slight cloak effect was demonstrated. Li \emph{et al.}~\cite{LiAS22} proposed a diffusion metamaterial with ``Plug and Switch'' modules for function switching. The background diffusion coefficient was set according to the Bruggeman theory. The kernel module could be a cloak, concentrator, or selector. The cloak module was realized as a bilayer structure based on scattering-cancellation, whereas the concentrator and selectormodules were also layered structures but designed by means of the transformation theory. The concentration of Fe$^{2+}$ in the experiment was detected by measuring the catalytic rate of the Fenton reagent on an organic dye. The protein test of inside and outside areas validated the ion selection of Cu$^{2+}$ and K$^+$.

The transformation theory for plasmas devices is still in its inception; more preparation work is needed to motivate experimental projects. Since plasma transport involves controlling particle diffusion and electromagnetic effects, further advances in multiphysics metamaterials are soon to be expected.

\section{Thermal conduction}

\begin{figure}[!ht]
	\includegraphics[width=1.0\linewidth]{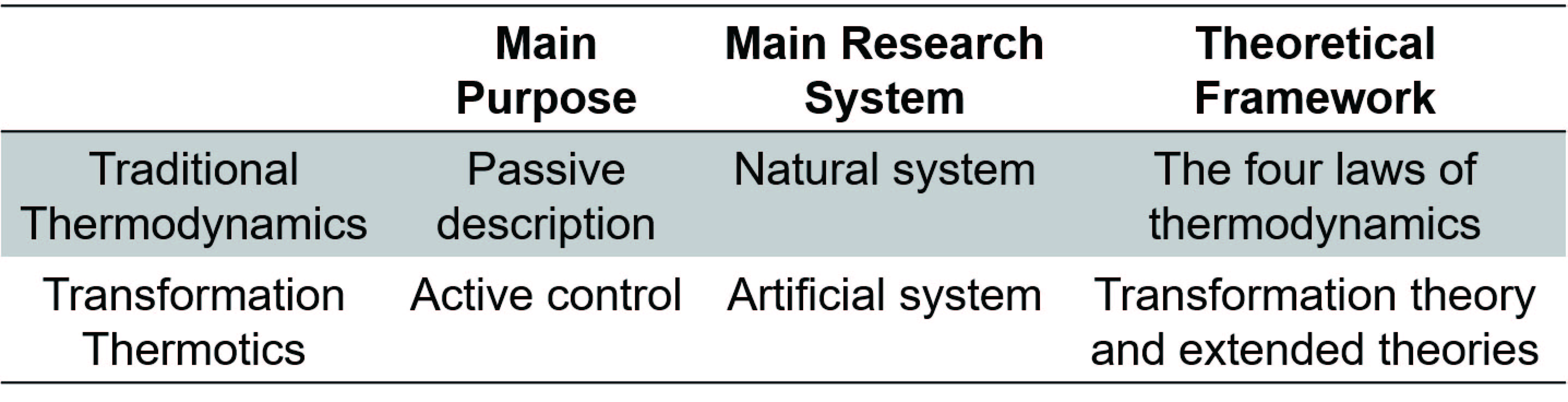}
	\caption{\label{table} Comparison between traditional thermodynamics and transformation thermotics. ``Passive description'' refers to the inability to modify the heat behavior of natural systems, but rather to understand them according to the four laws of thermodynamics. In contrast, ``active control'' implies the ability to adjust the heat behavior by designing artificial systems based on the transformation theory and its extended theories, which paves the way to the nascent technology of thermal metamaterials and metamaterial-based devices. Adapted from~\citealp{YangPR21}.}
\end{figure}

Macroscopic thermodynamics provides the most fundamental theoretical
description of mass and heat transport. Therefore, it is worth comparing its
framework with that of transformation thermotics, which in turn has distinct objectives, systems, and formalisms [Fig.~\ref{table}]. In this section, we comprehensively review the research on transformation thermotics based on thermal conduction from three perspectives, namely, the fundamental theory related to the transformation principle, current applications based on metamaterials or metadevices, and the emergence of related novel physics such as nonlinear, nonreciprocal, and topological effects.

\subsection{Theory and transformation principles}

The transformation theory has elicited considerable research interest since its proposal~\cite{LeonhardtScience06,PendryScience06}. By combining metamaterials with transformation theory, researchers have successfully designed cloaking devices in the zero-frequency range of electromagnetic waves, i.e., working for electrostatic and magnetostatic fields~\cite{WoodJPCS07,GomorySci12}, and demonstrated them in experiments~\cite{ZhangAIPAdv11,NarayanaAM2012,YangPRL12}. Since the governing equation (Laplace) of heat transfer in the steady state is analogous to electrostatics, transformation theory should also be applicable to thermal conduction.

Inspired by transformation optics, Fan \emph{et al}.~\cite{FanAPL08} and Chen \emph{et al}.~\cite{ChenAPL08} proposed thermal cloaking in the steady state, marking the beginning of transformation thermotics. Subsequently, Guenneau \emph{et al}. extended the formalism to include
transient heat conduction~\cite{GuenneauOE12}. The general governing equation of heat conduction with a heat source is given by
\begin{equation}\label{heatcond}
 \rho C \frac{\partial T}{\partial t} + \nabla\cdot(-\kappa\nabla T) = Q,
\end{equation}
where $\rho$, $C$, $T$, $\kappa$, and $Q$ denote the density, heat capacity, temperature, thermal conductivity, and external heat source, respectively. According to the transformation theory mentioned in Sec. II, the equation under a 2D coordinate transformation from ($x, y$) to ($x', y'$) becomes
\begin{equation}\label{heatcond'}
 \rho' C' \frac{\partial T}{\partial t} + \nabla'\cdot(-\kappa'\nabla' T) = Q',
\end{equation}
where the transformation rules are $\rho' C' = \rho C/\text{det}\bm{J}$, $\bm{\kappa}' = \bm{J}\kappa \bm{J}^{\dag}/\text{det}\bm{J}$, and $Q' = Q/\text{det}\bm{J}$.

\begin{figure}[!ht]
	\includegraphics[width=1.0\linewidth]{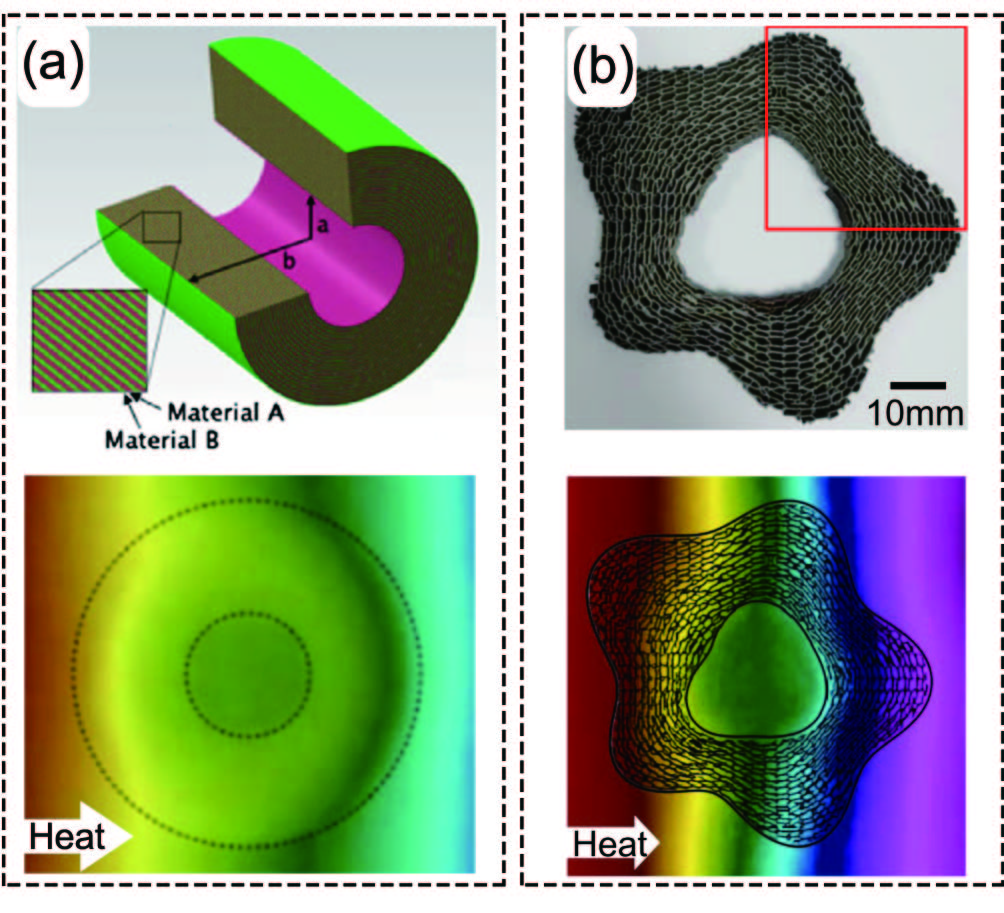}
	\caption{\label{fig-Thermetacond} Thermal metamaterials for thermal conduction. (a) Schematic diagram of a thermal cloak with a layered structure (upper). Experimental results of a thermal cloak (lower). Adapted from~\citealp{NarayanaPRL12}. (b) 3D-printed thermal cloak obtained by numerical optimization (upper) and its experimental results (lower). Adapted from~\citealp{ShaNC21}.}
\end{figure}

Transformation theory inevitably leads to inhomogeneous, anisotropic, and
even singular parameters, which seem to defy experimental realization.
However, the emergence of thermal metamaterials made it possible to remove
such roadblocks. As anticipated above, the effective medium theory, a powerful tool for predicting the physical properties of artificial composite materials, applies to thermal metamaterials. Narayana \emph{et al}. first used thermal metamaterials and the effective medium theory to fabricate a layered structure~\cite{NarayanaPRL12} that could achieve the desired anisotropic thermal conductivity [Fig.~\ref{fig-Thermetacond}(a)]. Their experiments confirmed that this structure could effectively realize the functions predicted by the transformation theory. Later on, transient thermal metamaterials were also experimentally demonstrated~\cite{SchittnyPRL13}.

\subsection{Applications: Metamaterials and metadevices}

Transformation thermotics has been enriched with a wide range of applications, such as intelligent and multifunctional thermal metamaterials. Based on its
framework, advanced complementary tools have been developed, including
effective medium theory and numerical algorithms, to predict and demonstrate
unprecedented thermal phenomena.

Intelligent thermal metamaterials were conceived to adaptively respond to changing environments, such as chameleon-like concentrators~\cite{XuPRAp19-1} and rotators~\cite{YangPRAp20}. Their intelligence is owed to the highly anisotropic thermal conductivity of their constituents. Besides, thermal null media enable the design of arbitrarily shaped metamaterials~\cite{BaratiSedehPRAp20,SunOE19,LiuATS18}, also composed of units with highly anisotropic parameters. Even if designed materials with extreme anisotropy are nearly impossible to obtain, recent studies suggest that multilayer graphene and Van der Waals thin films may offer viable alternatives~\cite{KimNat21,SunNM19,QianNM21,YinFoP21}.

Scattering-cancellation technology provides another popular approach to the
design of thermal metamaterials~\cite{XuPRL14}. This method is based on directly solving the Laplace equation, primarily in cylindrical or spherical coordinates, and requires isotropic layers with different parameters \cite{CummerPRL08}. This technique is commonly used to analyze bilayer structures, where the environment, shell, and core can be made of the same perforated material
with different hole densities. According to the effective medium theory, the resulting field can be uniform as if there were no shells or
cores~\cite{HanPRL14,HanAM14}. Intuitively, the inner layer of the bilayer structure serves as a thermal insulation wall protecting the core region, while the outer layer acts as a compensating substrate to ensure undisturbed background thermal fields. Initially, the bilayer structure had unsatisfactory performance under transients. Later, Farhat \emph{et al}.~\cite{FarhatSR15} solved this problem and broadened the application scope of the scattering-cancellation technique. This method is advantageous for realizing various thermal metamaterials and helps reproduce the effects predicted via transformation theory.

Numerical optimization algorithms constitute another area of research. With the concept of topology optimization, Dede \emph{et al}. achieved a variety of controls on heat flux through thermal composite structures~\cite{DedeSMO14} made of elliptical inclusions embedded in the matrix. Subsequently, Fujii \emph{et al}. simplified the structures by utilizing the covariance matrix adaption evolution strategy and designed a thermal cloak with natural materials~\cite{FujiiAPL18}. Furthermore, printable freeform thermal metamaterials have been proposed~\cite{ShaNC21}[Fig.~\ref{fig-Thermetacond}(b)], whereby, with two-step topology optimization, a thermal cloak can be fabricated by assembling functional cells consisting of Die steel and PDMS. This leads to the
conclusion that optimization algorithms bypass the restrictions of fixed shapes of metamaterials. With the rapid advancement of machine-learning techniques~\cite{MaACSNano18,LiNC19}, optimization algorithms have gradually become an essential approach to thermal metamaterial design.

\begin{figure}[!ht]
	\includegraphics[width=1.0\linewidth]{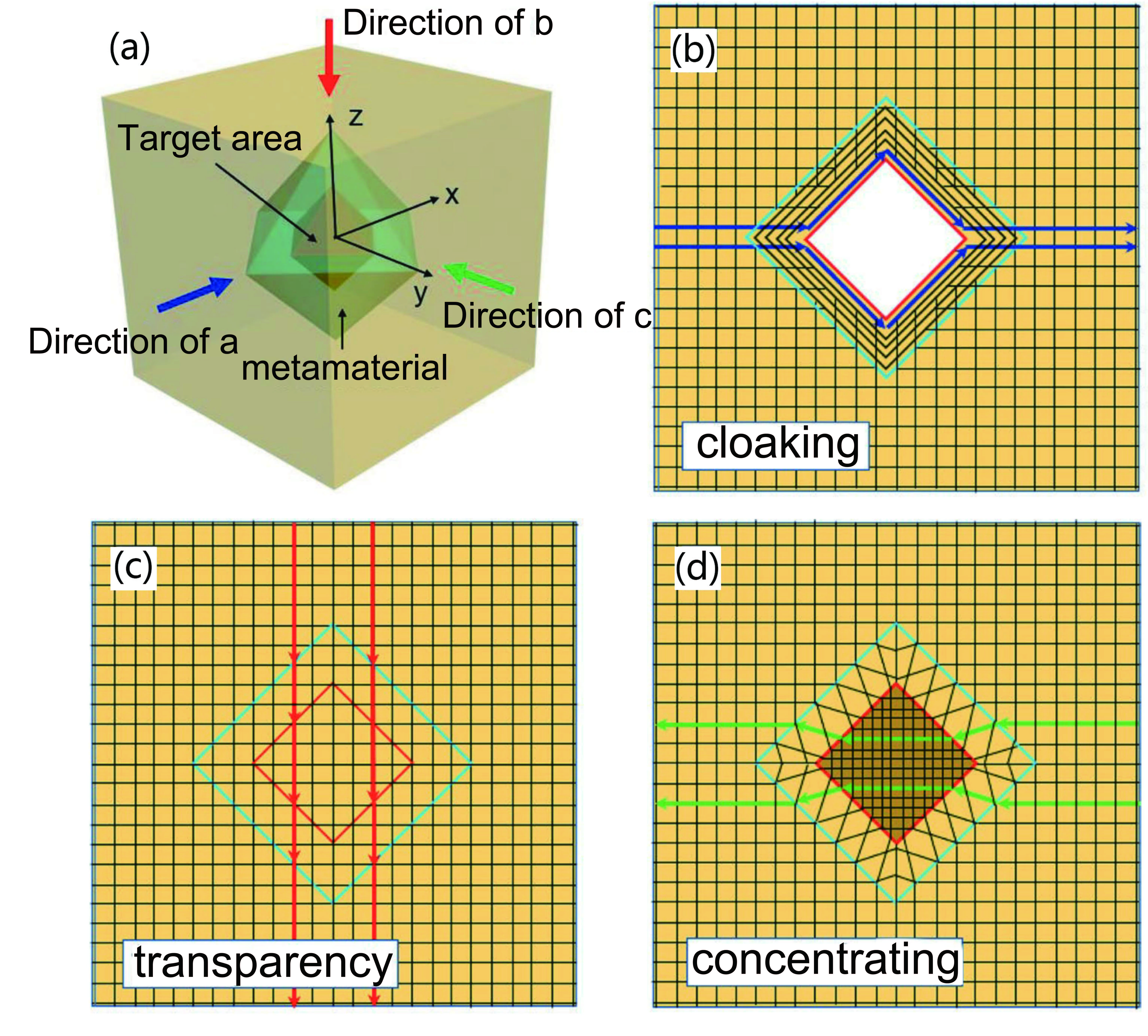}
	\caption{Schematic diagram of path-dependent thermal metadevices. (b) Cloaking, (c) transparency, and (d) concentration profile when the heat flow crosses the 3D metadevice along orthogonal directions as shown in (a); Adapted from~\citealp{SuAM21}.}
	\label{SuAM21}
\end{figure}

Various thermal metamaterials with different functionalities, such as thermal cloaking, concentration, and rotation, have already been demonstrated due to the development of transformation thermotics. However, most existing work focuses on geometrically isotropic (circular) structures, limiting the potential diversity of functions. However, structure anisotropy seems to provide additional degrees of freedom in regulating directional heat transfer [Fig.~\ref{SuAM21}]. Su \emph{et al}.~\cite{SuAM21} proposed a path-dependent thermal metadevice beyond Janus characteristics, which can exhibit three distinct thermal behaviors in different heat flow directions. Specifically, temperature gradient flows along the $x$, $y$, and $z$ axes [blue, green, and red arrows in Fig.~\ref{SuAM21}(a)] correspond respectively to thermal cloaking [Fig.~\ref{SuAM21}(b)], transparency [Fig.~\ref{SuAM21}(c)], and concentration [Fig.~\ref{SuAM21}(d)]. This unique versatility is due to an ingenious design that compresses the material along the $x$ axis and stretches it along the $y$ axis. This three-function thermal metadevice has been successfully validated by a proof-of-concept experiment of anisotropic in-plane conduction. We remind that the transformation theory can be extended also to arbitrary shapes with extremely anisotropic parameters mentioned before~\cite{BaratiSedehPRAp20}.

\begin{figure}[!ht]
	\includegraphics[width=1.0\linewidth]{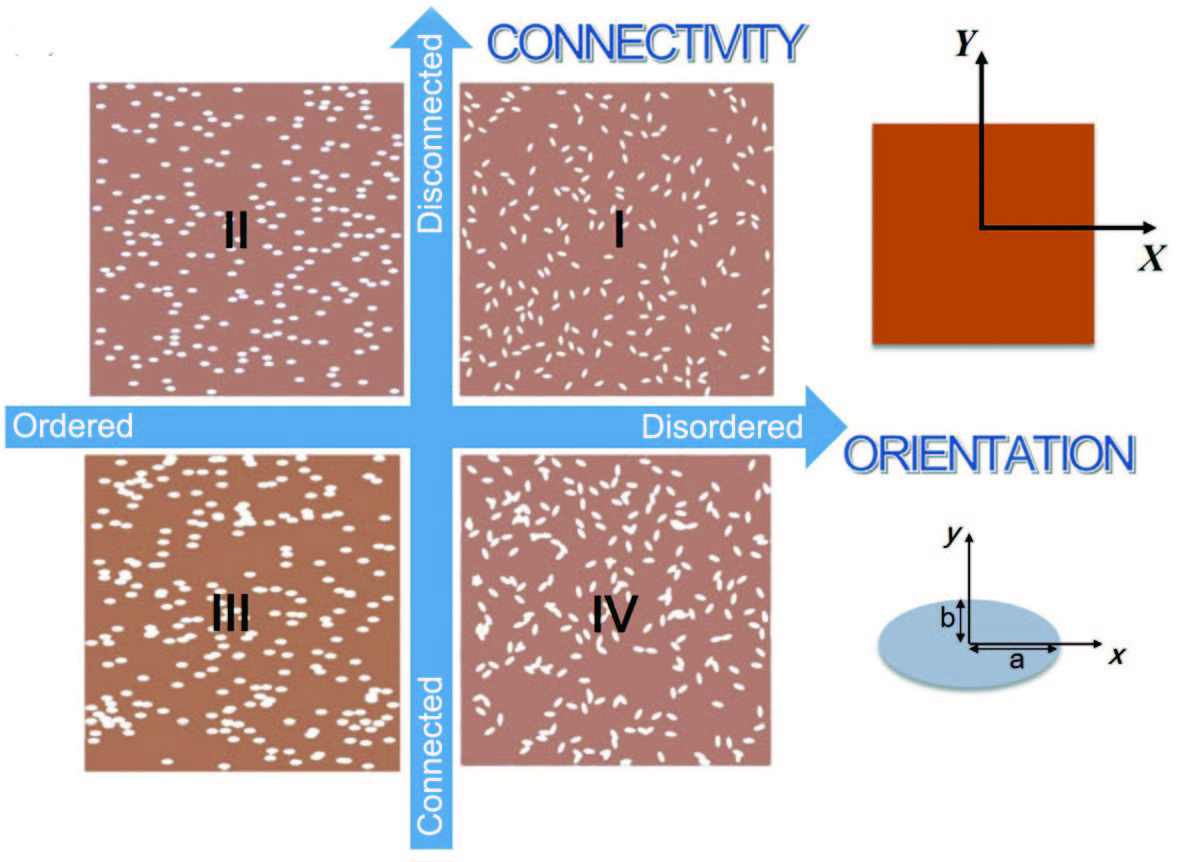}
	\caption{Heterogeneous composites are categorized in the four-quadrant system according to particle orientation and connection. Adapted from~\citealp{TianIJHMT21}.}\label{YangAPL17}
\end{figure}

Recent advances in transformation thermotics offer new means for thermal manipulation in artificial structures. Unfortunately, its strict requirements on the
parameters of the constituents (i.e., anisotropy, inhomogeneity, or even singularity) limit practical application. To make it feasible, effective medium theories were developed to solve geometric anisotropy problems, like the calculation of effective thermal conductivity of geometrically anisotropic structures. Yang \emph{et al}.~\cite{YangAPL17} derived the effective thermal conductivity of a single particle structure [Fig.~\ref{emt}(c)] by solving the Laplace equation in elliptical coordinates. These results were verified by experiments and finite element simulations. For multi-particle structures, the computational complexity of their approach would grow untenable. Fortunately, Tian \emph{et al}.~\cite{TianIJHMT21} developed a generalized effective medium theory to predict the effective thermal conductivity of heterogeneous systems containing elliptical (or ellipsoidal) inclusions. Four models [Fig.~\ref{YangAPL17}] were discussed, which revealed the underlying mechanisms of the relevant effective medium theories and provided a different approach to achieve nonlinear modulation and enhancement of thermal conductivity. Their simulation results agree well with the theoretical predictions in the range of low area fraction, which is limited by the inherent assumptions and the availability of mathematical forms of effective medium theory. Lu \emph{et al}.~\cite{LuIJHMT22} explored the power of deep learning to bypass the mathematical constraints of a particular formulation of the effective medium theory. By using a transfer learning algorithm with a small number of samples (less than 100), the accuracy of thermal conductivity calculations could be greatly improved. This approach saves time and resources and is especially beneficial for hard-to-access data sets. This robust optimization algorithm can also be applied to other non-elliptic geometrically anisotropic structures~\cite{JiIJHMT21}. Recently, the concept of traversing full-parameter anisotropic space~\cite{ShaNPJ22} has been proposed to describe the set of effective thermal conductivity tensors for any mixed material structure. A spatial traversal method for an anisotropic thermal conductivity tensor with full parameters based on topology optimization was proposed. The micro-structures with the desired effective thermal conductivity tensor were obtained through topological optimization, namely, topological functional cells. A series of topological functional cells were designed for copper and PDMS, whose effective thermal conductivity tensor can traverse their entire parameter space. Schemes based on topology optimization will significantly reduce the difficulty of designing functional metadevices.

\begin{figure}[!ht]
	\centering
	\includegraphics[width=1.0\linewidth]{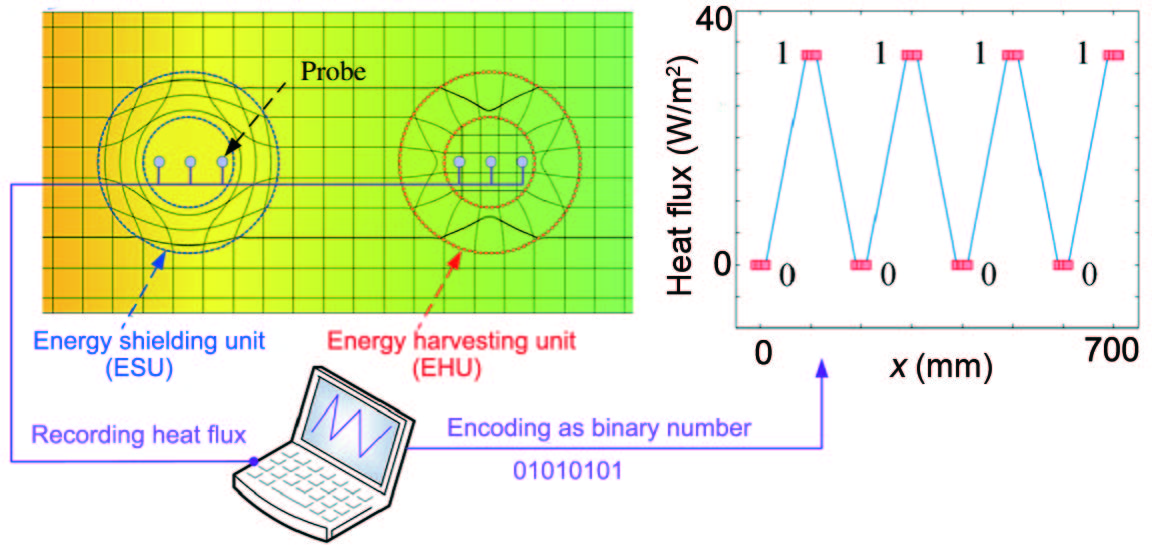}
	\caption{Schematic diagram of thermal coding. Adapted from~\citealp{HuPRAp18}.}
	\label{FigL7}
\end{figure}

Moreover, the development and popularization of advanced coding technology has been a long-standing and prominent issue in materials and information science, providing the basis for modern communication and computing. Coding techniques have been well studied and implemented in wave fields ranging from electromagnetism to acoustics. The concept of thermal coding has been introduced based on the notion of thermal cloak and thermal concentrator, as pioneered by Hu \emph{et al}.~\cite{HuPRAp18}. Because of the heat flow difference at their centers, the thermal cloak and the thermal concentrator can be defined respectively as the energy shielding unit and the energy harvesting unit [Fig.~\ref{FigL7}]. Thus, they can be regarded as equivalent to the binary data bit 0 and 1. A temperature gradient is applied to the coding unit. If the heat flow in the central region is higher than the reference value, the output is the 1 state; if it is lower than the reference value, the output is the 0 state.

The advances in thermal metamaterial technology offer new avenues to thermal information processing. Different configurations of the shape-memory alloy insets and traditional thermal media can be used to fabricate coding units working in the appropriate operating temperature range~\cite{ZhouIJHMT21}. Other external fields can also be used to control the thermal unit function~\cite{HeCPL23} so as to modulate thermal coding. For example, a reconfigurable two-phase thermal metamaterial consisting of a translucent liquid and controllable micro-magnetic particles can be freely reconfigured by a magnet. Thus, a complex unit array composed of multiple reconfigurable thermal response units can realize accurate and reconfigurable thermal coding/decoding information transmission under the control of external magnetic field~\cite{XuAMT21}. Additionally, a light field can be used to control the transmission of heat information. Thermal metamaterials designed by developing regionalized transformation methods have been proposed to realize encrypted thermal printing, whereby the encoded (static or dynamic) information is accessible from infrared images, but hidden in natural light~\cite{HuAM19}. Guo \emph{et al}. proposed that programmable thermal functions can be realized through the automatic response of thermoelectric heat sources and the real-time control of a driving voltage. An implementation of this technique is the adaptive metasurface platform~\cite{GuoAM22}.

Moreover, to break the limitations due to the fixed functions of each unit in a traditional thermal coding devices, a programmable all-thermal coding strategy based on temperature-dependent transformation thermotics has recently been proposed and experimentally implemented~\cite{Lei23-2}. A switchable cloak-concentrator component acts as the coding unit, and the binary signal can be distinguished and modulated by the heat flux divergence characteristics of each unit. Temperature-responsive phase-changing materials implement iterative coding operations (programmability) with adaptive external stimuli and internal response mechanisms. The thermal field realizes the whole coding process without having recourse to external fields, thus making it an all-thermal coding process. Furthermore, Yang \emph{et al} proposed a spatiotemporal thermal binary coding based on dynamic metamaterials~\cite{YangPRAp23}. By spatiotemporally modulating the shell of a code unit, they achieved thermal coding tunability in the time dimension, so as to improve the information storage ability. The design of thermal information metamaterials opens up new horizons toward advanced thermal management.


\subsection{Nonlinear thermal conductivity, non-Hermitian thermal topology, and asymmetric thermal conduction}

\begin{figure}[!ht]
	\includegraphics[width=1.0\linewidth]{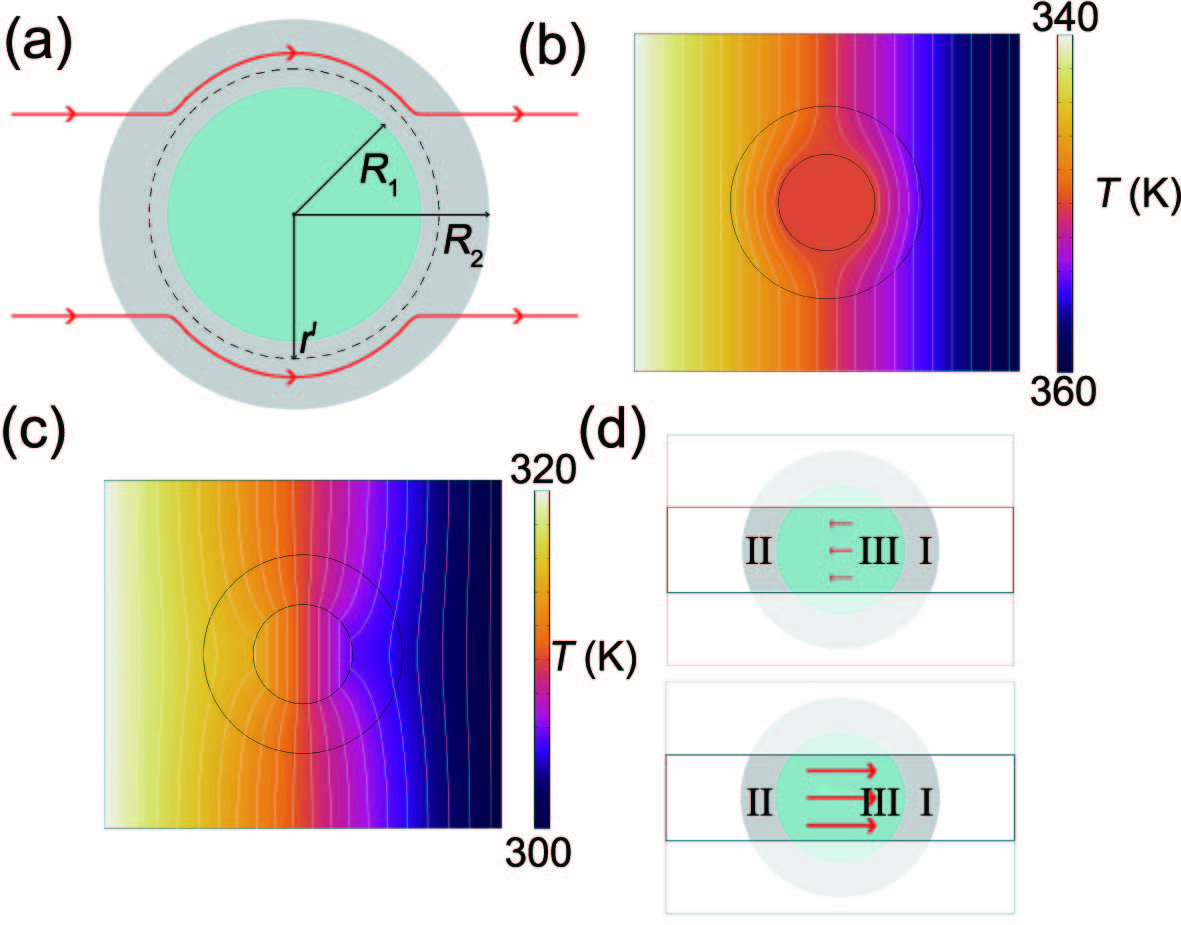}
	\caption{(a) Schematic view of a thermal cloak with radii $R_1$ and $R_2$. (b)-(c) Switchable thermal cloaks obtained by 2D finite-element simulations: (b) switch on at temperatures above 340~K and (c) switch off at temperatures below 320~K. (d) Sketch of a thermal diode delimited by the rectangular area enclosed between solid black lines. Adapted from~\citealp{LiPRL15}.}\label{LiPRL15}
\end{figure}

The above discussion mostly focused on linear materials, i.e., thermal conductivity was assumed to be temperature independent. Nevertheless, nonlinear phenomena are ubiquitous, and the underlying mechanisms are essential for understanding and designing complex systems. Exploiting nonlinearity has been quite successful in many fundamental areas of physics, such as optics, mechanics, and acoustics. In analogy with the dependence of polarization on the intensity of the electric field in nonlinear optics, the thermal conductivity in nonlinear thermotics should respond to the temperature gradient. For example, the thermal conductivity of a metal increases with temperature because the lattice structure of the metal expands as it heats up, thus increasing the heat transport capacity of hot electrons. Based on this fact, Li \emph{et al}.~\cite{LiPRL15} extended the transformation theory to the steady-state nonlinear heat conduction and proved that the governing equation keeps being formally invariance under any coordinate transformation. These authors considered the annulus sketched in Fig.~\ref{LiPRL15}(a), with radius, $r$,
ranging between $\tilde{R}_1(T)$ and $R_2$, and the coordinate transformation
\begin{equation}
	r^{\prime}=r \frac{R_2-\tilde{R}_1(T)}{R_2}+\tilde{R}_1(T).
\end{equation}
Here, $\tilde{R}_1(T)=R_1\left[1-\left(1+e^{\beta\left(T-T_c\right)}\right)^{-1}\right]$ for type-A cloaks and $\tilde{R}_1(T)=R_1/\left(1+e^{\beta\left(T-T_c\right)}\right)$ for type-B cloaks, which are a kind of typical step functions. Near the critical temperature $T_c$, the value of the function jumps quickly between 0 and 1, thus acting as a switch, with $\beta$ a scaling factor that controls the change speed of the working state near the critical temperature. The finite element numerical simulation [Figs.~\ref{LiPRL15}(b) and~\ref{LiPRL15}(c)] shows that type-A invisibility cloaks switch on (or off) at high (or low) temperatures. In addition, the effective medium theory predicts, and numerical simulations confirm, that two materials with isotropic and temperature-dependent thermal conductivity
can be alternately arranged in layers to approximate any desired thermal conductivity distribution. This conclusion is further extended to the case of transient nonlinear heat conduction~\cite{LiPLA16}. Inspired by switchable thermal cloaks, they further designed a macroscopic thermal diode.  As shown in Fig.~\ref{LiPRL15}(d), the diode device consists of a circular structure divided up into three regions: region I (II) is a segment of the type-A (type-B) cloak, and region III is a thermal conductor. This structure is asymmetric and highly nonlinear, which explains the rectification effect revealed by numerical simulation.

\begin{figure}[!ht]
	\includegraphics[width=0.8\columnwidth]{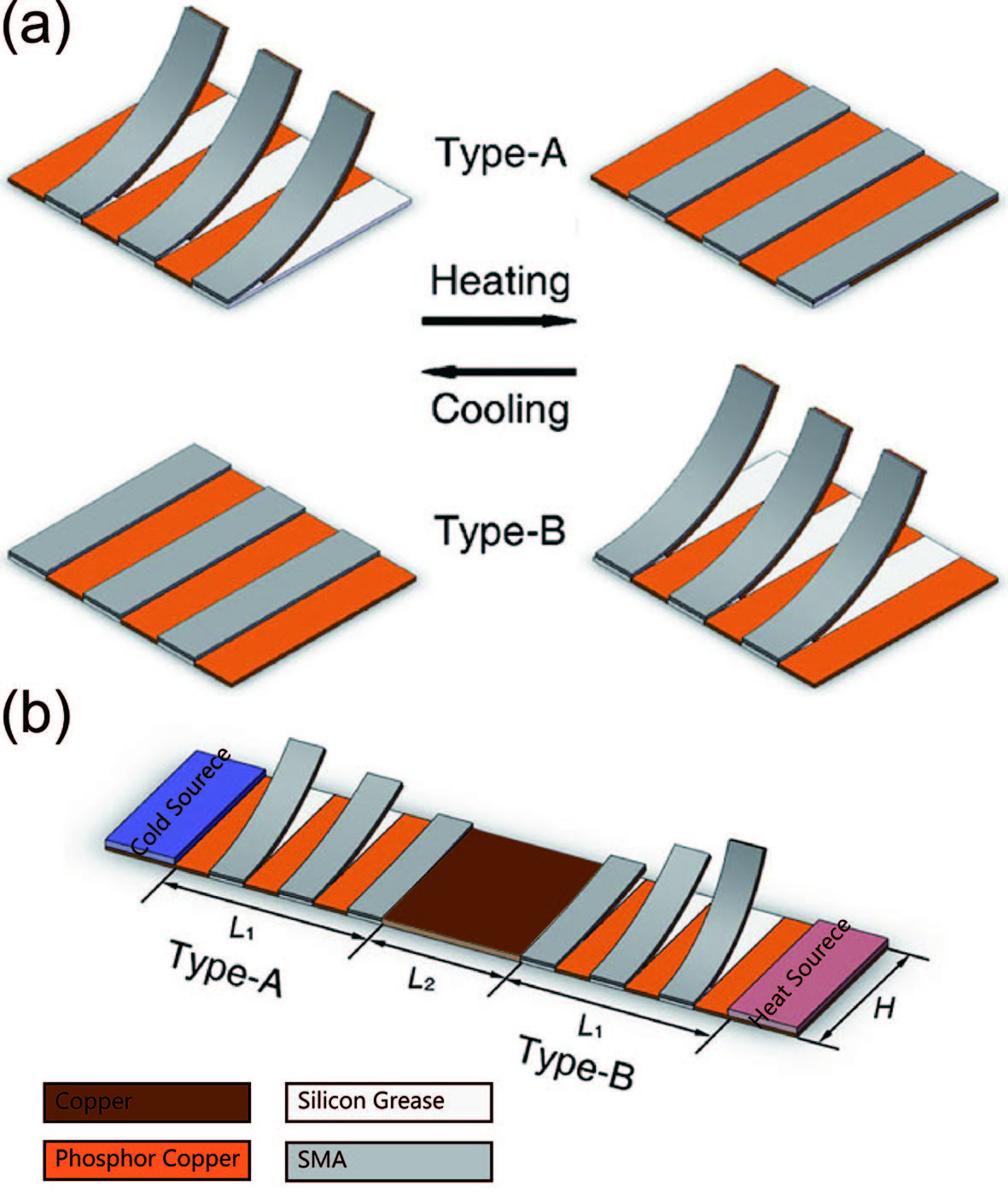}
	\caption{Sketch of zero-energy-consumption thermostat: (a) design and (b) experimental setup. Adapted from~\citealp{ShenPRL16}.}\label{ShenPRL16}
\end{figure}

Innovative experimental research on thermal stability from Olsson \emph{et al}.~\cite{OlssonMST99,OlssonSR19,OlssonU85,OlssonAEM14,OlssonACBE13,OlssonAPL14,OlssonU17} inspired the exploration of thermal stability in thermal metamaterials. A thermal cloak enables its central region to maintain a constant temperature under fixed boundary conditions. However, if the boundary conditions change, the value of this constant temperature may also change. To address this issue, Shen \emph{et al}.~\cite{ShenPRL16} designed a zero-energy-consumption thermostat that keeps the temperature inside the cloak constant, without requiring additional energy input. Two phase-changing materials [Fig.~\ref{ShenPRL16}(a)] are applied to the left (type-A) and right (type-B) sides of the zero-energy-consumption thermostat [Figs.~\ref{ShenPRL16}(b)]. Both shape memory alloys are capable of changing their shape as the temperature varies [Fig.~\ref{ShenPRL16}(a)]: strips of type-A material tilt up an angle below 278.2~K and level off above 297.2~K; for the type-B material, the transition temperature is the same, but the deformation occurs in the opposite direction. The middle area is a good conductor of heat and is also expected to be thermally insulated. By solving the steady-state nonlinear heat transfer equation, Shen and coworkers demonstrated that the temperature in this region depends only on the phase transition temperature of materials A and B, and is no longer affected by the ambient temperature. Their work provides guidance for controlling heat flow without consuming energy and for designing novel metamaterials with temperature- or field-responsive parameters. Along this line, they further designed an intelligent thermal metamaterial~\cite{ShenAPL16} that automatically switches from a cloak (or concentrator) to a concentrator (or cloak) according to the ambient temperature, by combining a homogeneous isotropic material with shape memory alloys.

The classical effective medium approximation plays a critical role in the design of linear thermal metamaterials. Similarly, when materials with nonlinear thermal conductivity are involved, the validity of that approximation should be reassessed. For instance, Su \emph{et al}.~\cite{SuEPL20} formulated a rigorous theory to calculate the effective thermal conductivity of core-shell structures made of materials with nonlinear thermal conductivity, which appears to work well under certain conditions. As an example, they considered thermal radiation under Rosseland's diffusion approximation, a typical scheme implying
nonlinear thermal conductivity, and implemented the concentrator-cloak switch
function. Dai \emph{et al}.~\cite{DaiEPJB18} considered cylindrical particles randomly distributed in a matrix, with temperature-dependent thermal conductivities. Through the Maxwell-Garnett and Bruggeman theories, they calculated the effective nonlinear conductivity of this system. Furthermore, they derived the effective nonlinear conductivity of periodic structures~\cite{DaiIJHMT20}  as a temperature perturbation expansion. In most fields of study, the nonlinear case leads to many novel phenomena that directly differ from the linear case.    

\begin{figure}[!ht]
	\includegraphics[width=1.0\linewidth]{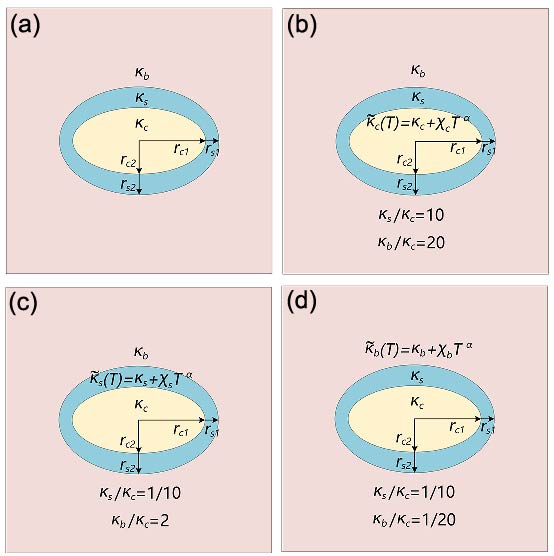}
	\caption{(a) 2D sketch of a composite structure with thermal nonlinear (b) core, (c) shell, and (d) background. Adapted from~\citealp{ZhuangPRE22}.}\label{ZhuangPRE22}
\end{figure}

A convenient method for solving the nonlinear conduction equation was developed by Yang \emph{et al}.~\cite{YangPRE19}. They investigated the nonlinear response of circular core-shell structures embedded in a finite matrix, where the core is made of a weakly nonlinear material and the shell and
matrix of linear materials. Based on a general analytical approach, they predicted nonlinear enhancement effects in two and three dimensions, and proposed an intelligent thermal metadevice, which can automatically switch subject to the external temperature. However, a unified theory for treating nonlinear effects in geometrically anisotropic structures is still lacking. Zhuang \emph{et al}.~\cite{ZhuangPRE22} systematically investigated a category of classical anisotropic composites, namely confocal ellipses or ellipsoids with temperature-dependent parameters. Specifically, they considered a confocal core-shell elliptic structure embedded in a finite background [Fig.~\ref{ZhuangPRE22}(a)]. The semi-major and -minor axis lengths of the core (or shell) were denoted as $r_{c1}$ and $r_{c2}$ (or $r_{s1}$ and $r_{s2}$), respectively. For materials with linear thermal conductivity, we know the effective thermal conductivity of the core-shell structure reads,
\begin{equation}\label{kcs2}
	\kappa_{csi} =\kappa_s\frac{L_{ci}\kappa_{c}+\left(1-L_{ci}\right)\kappa_s+f_1\left(1-L_{si}\right)(\kappa_{c}-\kappa_s)}{L_{ci}\kappa_{c}+\left(1-L_{ci}\right)\kappa_s-f_1L_{si}(\kappa_{c}-\kappa_s)},
\end{equation}
and that of the core-shell structure plus background~\cite{YangAPL17},
\begin{equation}\label{ke2}
	\kappa_{ei} =k_{b} \frac{L_{s i}\kappa_{csi} +\left(1-L_{s i}\right) \kappa_{b}+f_{2}\left(1-L_{s i}\right)(\kappa_{csi} -\kappa_{b})}{L_{s i} \kappa_{csi} +\left(1-L_{s i}\right) \kappa_{b}-f_{2} L_{s i}(\kappa_{csi} -\kappa_{b})}.
\end{equation}
Here, the inner and outer area fractions can be expressed as $f_1=r_{c1}r_{c2}/(r_{s1}r_{s2})$ and  $f_2=\pi r_{s1}r_{s2}/S_0$, where $S_0$ is the total area covered by background, core, and shell. The shape factors $L_{w1}$ and $L_{w2}$ along the direction of the semi-major and -minor axes are respectively $L_{w1}=r_{w2}/\left(r_{w1}+r_{w2}\right)$ and $L_{w2}=r_{w1}/\left(r_{w1}+r_{w2}\right)$, where the subscript $w$ reads $c$ or $s$ for core or shell. The shape factors satisfy the identity $L_{w1}+L_{w2}=1$, and the degree of their deviation from 0.5 indicates the flattening of the ellipses. These parameters describe how the thermal conductivity of the actual material varies with the temperature. For convenience and without sacrificing generality, they considered the three cases shown in Figs.~\ref{ZhuangPRE22}(b)-\ref{ZhuangPRE22}(d), where one part of the ternary structure has a stronger nonlinear thermal conductivity than the other two parts. Consequently, they assumed that the weak nonlinear thermal conductivities are independent of temperature, while the strong nonlinear thermal conductivity is formulated as
\begin{subequations}\label{c-s-b2}
	\begin{align}
		\tilde{\kappa}_{c}\left( T\right)  =\kappa_{c}+\chi_{c} T^{\alpha},\\
		\tilde{\kappa}_{s}\left( T\right)  =\kappa_{s}+\chi_{s} T^{\alpha},\\
		\tilde{\kappa}_{b}\left( T\right)  =\kappa_{b}+\chi_{b} T^{\alpha},
	\end{align}
\end{subequations}
where $\chi_{c}$, $\chi_s$, and $\chi_b$ are the nonlinear coefficients, and $\alpha$ can be any real number. Indeed, the thermal conductivities of natural materials usually vary with temperature, which can be described by Eq.~(\ref{c-s-b2}).

For weakly temperature-dependent thermal conductivity, i.e., $\chi_{c} T^{\alpha}\ll \kappa_c$, $\chi_{s} T^{\alpha}\ll \kappa_s$, or $\chi_{b} T^{\alpha}\ll \kappa_b$, Eqs.~(\ref{c-s-b2}a)-(\ref{c-s-b2}c) can be
substituted into Eqs.~(\ref{kcs2}) and~(\ref{ke2}) separately. Then, they can expand the temperature-dependent effective thermal conductivities of these three schemes up to the $T^{\alpha}$ term by using the Taylor formula as
\begin{subequations}\label{taylor}
	\begin{align}
		\tilde{\kappa}^{(c)}_{e i}\left( T\right)  =\kappa_{e i}+\lambda_{i1} \chi_{c} T^{\alpha}+O\left(T^{ \alpha}\right),\\
		\tilde{\kappa}^{(s)}_{e i}\left( T\right)  =\kappa_{e i}+\lambda_{i2} \chi_{s} T^{\alpha}+O\left(T^{ \alpha}\right),\\
		\tilde{\kappa}^{(b)}_{e i}\left( T\right)  =\kappa_{e i}+\lambda_{i3} \chi_{b} T^{\alpha}+O\left(T^{ \alpha}\right)
	\end{align}
\end{subequations}
and define the nonlinear modulation coefficients as
\begin{subequations}\label{etac-s-b2}
	\begin{align}
		\eta^{(c)}_i&=\left[\tilde{\kappa}^{(c)}_{e i}\left( T\right)-\kappa_{e i}\right]/T^{\alpha}=\lambda_{i1} ,\\
		\eta^{(s)}_i&=\left[\tilde{\kappa}^{(s)}_{e i}\left( T\right)-\kappa_{e i}\right]/T^{\alpha}=\lambda_{i2} ,\\
		\eta^{(b)}_i&=\left[\tilde{\kappa}^{(b)}_{e i}\left( T\right)-\kappa_{e i}\right]/T^{\alpha}=\lambda_{i3} ,
	\end{align}
\end{subequations}
where the parameters $\lambda_{i1}$, $\lambda_{i1}$, and $\lambda_{i1}$ depend on the geometric parameters ($f_1$, $f_2$, $L_{ci}$, and $L_{si}$) and the linear thermal conductivities ($\kappa_c$, $\kappa_s$, and $\kappa_b$).

To further reduce the number of independent variables, Zhuang and coworkers reparameterized the system as,
\begin{subequations}\label{RP}
	\begin{align}
		f_{1}&=\frac{\left(1-2 L_{s 1}\right)\left(1-L_{c 1}\right) L_{c 1}}{\left(1-2 L_{c 1}\right)\left(1-L_{s 1}\right) L_{s 1}},\\
		f_{2}&=\frac{\pi c^{2}\left(1-L_{s 1}\right) L_{s 1}}{\left(1-2 L_{s 1}\right) S_{0}},
	\end{align}
\end{subequations}
where $c$ represents the focal length. Their theory was verified through finite-element simulations.

Topological phenomena in thermal metamaterials are also an intriguing topic. There are two paradigms for revealing topological effects in heat conduction. The first paradigm is based on the discretization of the diffusion equation. It originated from a theoretical work by Yoshida and Hatsugai~\cite{YoshidaSR21}. By combining the Fick law and the equation of continuity, one obtains the 1D diffusion equation of a continuous scalar field $\phi(t,x)$,
\begin{equation}
	\partial_{t}\phi(t,x)=D\partial^{2}_{x}\phi(t,x),
	\label{diffusion_equ}
\end{equation}
where $\partial_{t(x)}$ denotes the derivative with respect to time $t$ (and the spatial coordinate x). $\phi(t,x)$ can be the temperature field, the number density of particles, \emph{etc.} Then, the diffusion equation is spatially discretized to bridge diffusion and electronic transport in quantum physics.

\begin{figure}[!ht]
	\includegraphics[width=\linewidth]{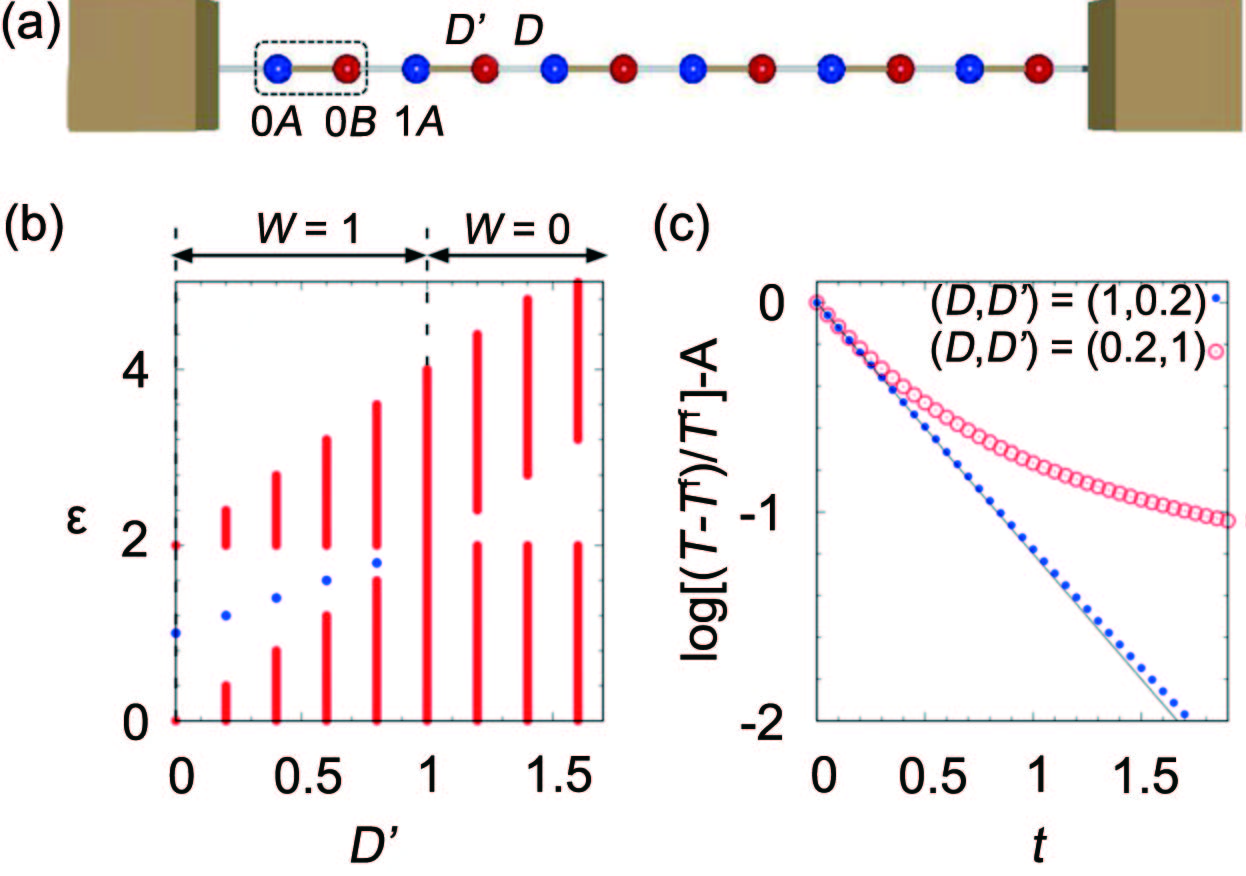}
	\caption{Topological physics in 1D diffusion systems. (a) Diffusive SSH model under fixed boundary condition with $N=6$, where $N$ is the number of unit cells. Boxes denote constant-temperature heat reservoirs. The sites are labeled by ($i,\alpha$), where $i$ refers to the unit cell index and $\alpha$ to the sublattice index. The coupling strengths are denoted by the diffusion coefficients $D'$ or $D$. (b) Fixed-boundary-condition spectrum of the diffusive SSH Hamiltonian with $D=1$ and $N=240$. The dots denote edge states. (c) Time evolution of $\vec{T}_{0A}(t)$ in the topologically nontrivial [$(D, D')=(1,0.2)$] and trivial [$(D, D')=(0.2,1)$] phases, marked respectively by solid and hollow dots. A black line for the function $-(D+D')t$ is plotted for reference. $A={\rm{log}}\{[T_{0A}(0)-T^f]/T^f\}$ is subtracted for a comparison. The initial condition is set as $\vec{T}_{i\alpha}=\delta_{i0}\delta_{{\alpha}A}$. The final equilibrium temperature is $T^f=T_{0A}(t=50)$. Adapted from~\citealp{YoshidaSR21}.}
	\label{YoshidaSR21}
\end{figure}

As an example, Yoshida and Hatsugai used the 1D Su-Schrieffer-Heeger (SSH) model~\cite{SuPRL79} to demonstrate topological phenomena in diffusion systems [Fig.~\ref{YoshidaSR21}(a)]. $\phi(t,x)$ is set as the temperature field. The temperature at each site is described by the vector $\vec{T}=(T_{0A},T_{0B},T_{1A},\cdots,T_{(N-1)B})^\dag$. The diffusion equation can be spatially discretized as
\begin{equation}
	\partial_{t}\vec{T}(t)=-\hat{H}_{\rm{SSH}}\vec{T}(t),
	\label{SSH_eq}
\end{equation}
with diffusive SSH Hamiltonian
\begin{equation}
	\hat{H}_{\rm{SSH}}=\left(\begin{matrix}
		D+D' & -D & 0 & \cdots & -D'
		\\
		-D & D+D' & -D' & \cdots & 0
		\\
		0 & -D' & D+D' & \cdots & 0
		\\
		\vdots & \vdots & \vdots & \ddots & \vdots
		\\
		-D' & 0 & 0 & \cdots & D+D'
		\\
	\end{matrix}\right).
	\label{SSH_Ham}
\end{equation}
One notices that $\hat{H}_{\rm{SSH}}$ are the onsite energy shifts of the prototypical SSH Hamiltonian and, therefore, preserves the topological properties of the 1D SSH model. Here, fixed instead of open boundary
conditions were imposed, because the edge sites of the
chain are connected to constant-temperature heat reservoirs

We discuss now the topological properties of the diffusive 1D SSH model. In the momentum space, the Bloch Hamiltonian reads
\begin{equation}
	\hat{h}_{\rm{SSH}}(k)=\left(\begin{matrix}
		D+D' & -D'-De^{ik}
		\\
		-D'-De^{-ik} & D+D'
		\\
	\end{matrix}\right),
	\label{SSH_Bloch}
\end{equation}
where $k$ is the Bloch vector. This Hamiltonian preserves chiral symmetry and its topological properties are characterized by the winding number
\begin{equation}	 W=-\int_{-\pi}^{\pi}\frac{dk}{4{\pi}i}{\rm{tr}}[\sigma_{3}\hat{h}_{\rm{SSH}}^{'-1}(k)\partial_{k}\hat{h}'_{\rm{SSH}}(k)],
	\label{SSH_winding}
\end{equation}
where $\sigma_{3}$ is the Pauli matrix, $\hat{h}'_{\rm{SSH}}=\hat{h}_{\rm{SSH}}-(D+D')\sigma_{0}$, and $\sigma_{0}$ is the identity matrix. The winding number is 1 (nontrivial topological phase) for $0\enspace {\leq}\enspace D' <1$ and 0 (trivial topological phase) for $D'\enspace {\geq}\enspace 1$. The fixed-boundary-condition spectrum of the diffusive 1D SSH Hamiltonian is depicted in Fig.~\ref{YoshidaSR21}(b). For $W=1$, there exists
an edge state with decay rate $\varepsilon=D+D'$, which points to a bulk-edge
correspondence in diffusion systems. Here, the bulk-edge correspondence implies that the topological property in the bulk determines the character of the edge modes.

Here comes a question: how can we experimentally detect the edge states? One method is to observe the temperature evolution of the edge site. Figure~\ref{YoshidaSR21}(c) shows the time evolution of the temperature at
the edge ($0, A$). When in the nontrivial topological phase [$(D, D')=(1,0.2)$], the temperature $T_{0A}$ follows closely the reference
exponential decay $T_{0A}{\sim}e^{-(D+D')t}$. This peculiar time dependence is a signature of nontrivial topology. In contrast, the temperature evolution deviates from the reference curve in the topologically trivial phase [$(D, D')=(0.2,1)$].

\begin{figure}[!ht]
	\includegraphics[width=1.0\linewidth]{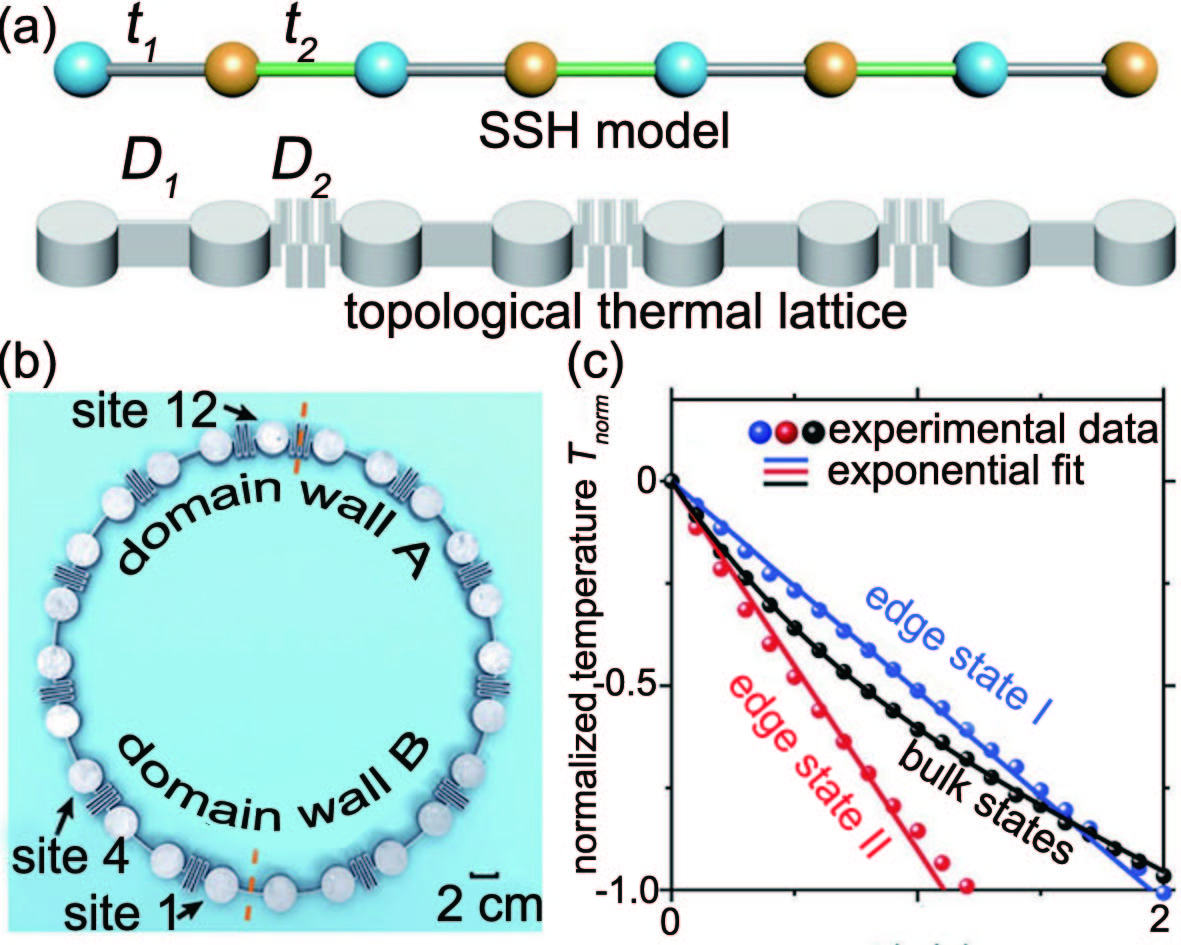}
	\caption{Experimental realization of topological edge states in thermal diffusion systems. (a) Topological thermal lattice as an analogue of the 1D SSH model. The disk in the thermal lattice is the analogous of the atom, while the effective diffusivity of the straight $D_1$, (meandering, $D_2$) channels correspond to the hopping amplitudes $t_1$ ($t_2$) in the SSH model. (b) Photograph of the thermal lattice with domain walls. Two dashed lines indicate two types of domain walls, the center of two edge states. (c) Measured temperature evolution of two edge states and one bulk state. The temperature is normalized as $T_{\rm{norm}}={\rm{ln}}[(T-T_{f})/(T_{i}-T_{f})]$, where $T_i$ is the initial temperature of the heated site and $T_f$ is the room temperature. Adapted from~\citealp{HuAM22}.}
	\label{HuAM22}
\end{figure}

The above conclusions have been validated experimentally by Hu {\it et al.}~\cite{HuAM22}. They constructed a 1D thermal lattice composed of aluminum disks and channels to mimic the 1D SSH model [Fig.~\ref{HuAM22}(a)]. The channels were alternately bent to tailor the effective thermal diffusivity between adjacent disks: $D_1$ for the straight channels, $D_2$ for the bent channels. Then they design two types of domain walls to observe the topological edge states between two topologically distinct thermal lattices, as illustrated in Fig.~\ref{HuAM22}(b). At the domain wall A (B), the topological nontrivial and trivial domains are connected with a meandering (straight) channel, which corresponds to the edge state I (II). To verify the existence of topological edge states, they heated site 1 (12) in the vicinity of domain wall A (B) and then measured the temperature evolution of the heated site [Fig.~\ref{HuAM22}(c)]. One can find that the temperature evolution of the two edge states follows an exponential decay (with the decay rate $0.52\;{\rm{min}}^{-1}$ for edge state I and $0.91\;{\rm{min}}^{-1}$ for edge state II). In contrast, two exponential functions are required to fit the temperature evolution of the bulk state by heating sites 3 and 4, which results from the contribution of several bulk modes.

The above theoretical and experimental works have been limited to 1D systems, but more interesting topological physics lies in the modeling of higher dimensional systems. Liu \emph{et al.} successfully realized a higher-order topological insulator (HOTI) in pure diffusion systems and characterized the thermal behavior of corner states~\cite{LiuarXiv22-1}. HOTIs exhibiting lower-dimensional topological states have attracted much attention from the condensed matter community~\cite{BenalcazarSci17,BenalcazarPRB17,SchindlerSA18,AggarwalNC21}. For example, an $m$D topological insulator with ($m-n$)D topological boundary states is referred to as an $n$th-order topological insulator. Unlike the Benalcazar-Bernevig-Hughes model, characterized by a bulk quadrupole moment, the 2D SSH model works as a HOTI without a multipole moment~\cite{BenalcazarPRB19} and has been realized in classical wave systems~\cite{XieNRP21,XiePRL19,ChenPRL19,ZhangPRL19}.

\begin{figure}[!ht]
	\includegraphics[width=1.0\linewidth]{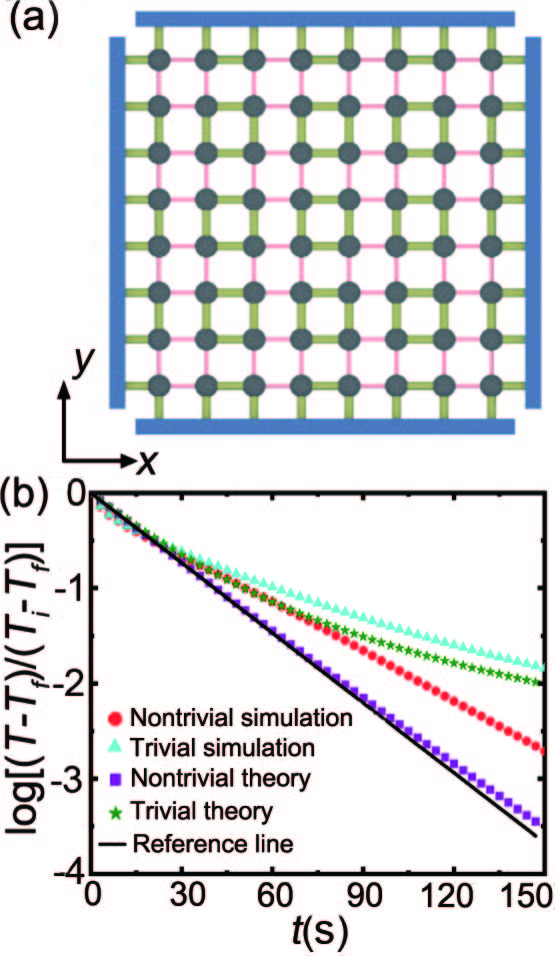}
	\caption{Higher-order topological insulator (HOTI) in thermal diffusion systems. (a) Sketch of the sphere-rod structure with four constant-temperature boundaries set at 293.2~K. Thermal diffusivities of thin and thick rods are denoted as $D_1$ and $D_2$. Black bars denote constant-temperature heat reservoirs. The central black dashed square shows a unit cell, where numbers label the four sublattices. ($n_x,n_y$) indicates the unit cell label. (b) Temperature evolution of a thermal corner state vs. time $t$. The temperature was rescaled as ${\rm{log}}[(T-T_{f})/(T_{i}-T_{f})]$, with $T_{i}=333.2$ K and $T_{f}=293.2$ K. Adapted from~\citealp{LiuarXiv22-1}.}
	\label{LiuarXiv22_1}
\end{figure}

To implement the diffusive counterpart of the 2D SSH model~\cite{LiuarXiv22-1}, Liu \emph{et al.} created a 2D sphere-rod structure with the same density $\rho$, heat capacity $c_{p}$, thermal conductivity $\kappa$, and rod lengths $L$ [Fig.~\ref{LiuarXiv22_1}(a)]. The radii of the intra-cell and inter-cell rods are $R_{0,1}$ and $R_{0,2}$, respectively, while $R$ is the radius of the spheres. The fixed boundary condition is achieved by connecting the boundary rods to constant-temperature heat reservoirs. To investigate the thermal behavior of the corner states, the four corner spheres were initially heated up to 333.2~K. The temperature evolution of one corner sphere is shown in Fig.~\ref{LiuarXiv22_1}(b). The theoretical results were obtained by solving a series of partial differential equations, and COMSOL Multiphysics was employed to simulate the results. Theoretical and simulated temperature evolutions agree well with each other. Additionally, in the nontrivial topological phase the theoretical time dependence of the temperature at the corners follows the expected exponential decay. However, the simulated temperature appears slightly higher than the theoretical prediction. Because heat travels through the rod for a while before reaching the next sphere in simulations, whereas heat instantly travels to the next sphere in theoretical calculations.

However, the tight-binding theory is not precise enough to uncover topological physics of diffusion systems~\cite{QiAM22}. This is because the temperature field is nonlocal and may extend over a wide area. Consequently, they developed a new topological theory for thermal diffusion inspired by classical optics.

\begin{figure}[!ht]
	\includegraphics[width=1.0\linewidth]{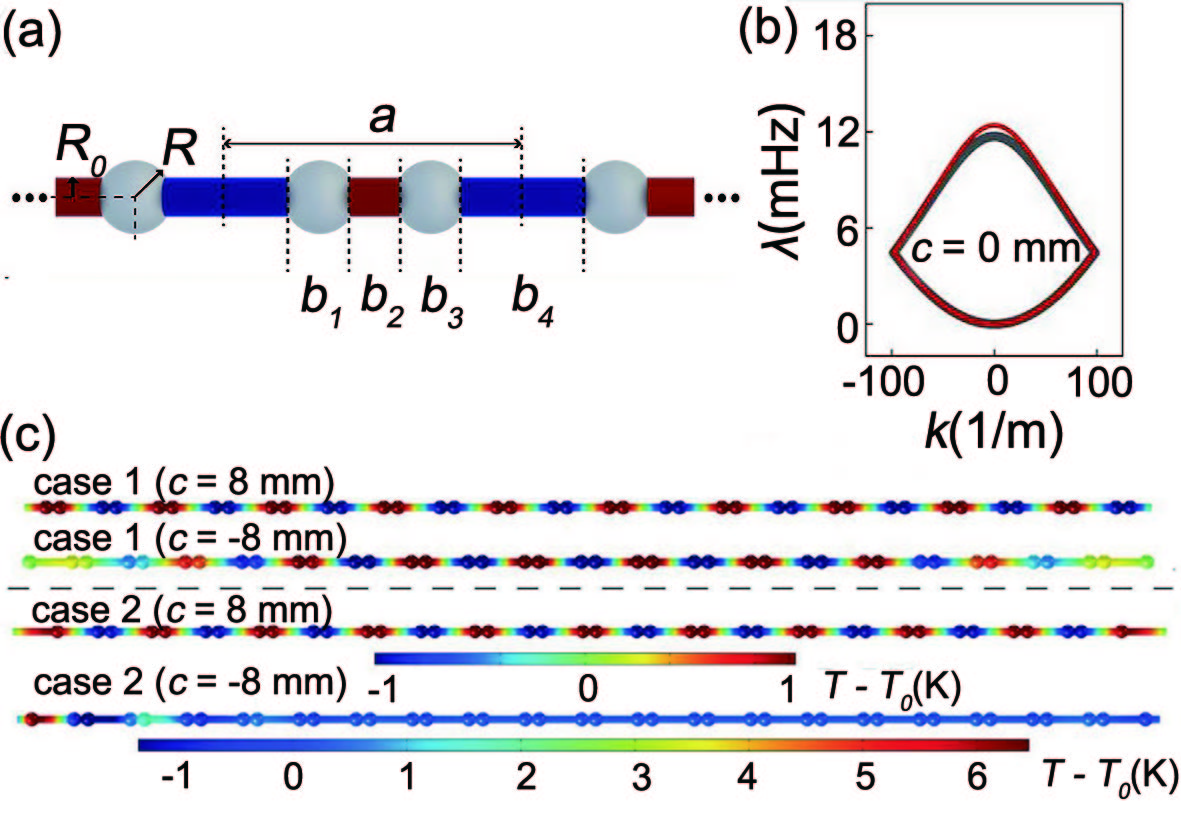}
	\caption{Thermal diffusive topology beyond the tight-binding model. (a) A 3D sphere-rod model. The lengths occupied by two spheres with radius $R$ are $b_1$ and $b_3$ ($=b_1$). The lengths of the (shorter) intracell and (longer) intercell rods of radius $R_0$ are $b_2$ and $b_4$, respectively. The length of a unit is $a$ and the length difference between the two rods is $c=(b_{4}-b_{2})/2$. (b) Band structures of the 1D equivalent model and 3D sphere-rod model. (c) Temperature distributions of states within or at the edges of the bandgap. Adapted from~\citealp{QiAM22}.}
	\label{QiAM22}
\end{figure}

They considered a 3D sphere-rod model with constituents of the same mass density $\rho_{3\rm{D}}$, heat capacity $c_{p3\rm{D}}$, and thermal conductivity $\kappa_{3\rm{D}}$ [Fig.~\ref{QiAM22}(a)]. Band
structure simulations deviate from the
band structure of the 1D SSH model.  To address this issue, they proposed a more refined 1D model
equivalent to their 3D sphere-rod model. The band structures for the 1D equivalent model, depicted in Fig.~\ref{QiAM22}(b) by red dots, closely agree with the 3D sphere-rod model simulation.

The Zak phase, which serves as the cornerstone of 1D topological physics, should be redefined in order to establish the bulk-edge correspondence in diffusion systems. It takes the form of
\begin{equation}
	{\rm{exp}(i\mathcal{Z})}={\rm{sgn}}(\b{U}_{1}),
	\label{Zak_phase}
\end{equation}
where $\b{U}_{1}$ is the value of $U_1={\rm{Im}}[M_{12}{\rm{exp}}(-ik_{2}a)]$ at the upper edge of the lower band, $M$ is the transfer matrix, and $k_{m}=({\lambda}{\rho}_{m}c_{pm}/{\kappa}_{m})^{1/2}$, with $m=1,2$, denoting respectively spheres and rods. The Zak phase can resolve two topologically distinct phases: $\mathcal{Z}=\pi$ for $c<0$ and $\mathcal{Z}=0$ for $c>0$.

However, the existence of edge states may be
affected by the choice of the boundary conditions.  Qi and
coworkers~\cite{QiAM22} analyzed two different boundaries for their model. In case 1, no edge state was found in the eigenvalue simulation, no matter the sign of $c$. The relevant temperature field simulations reported in Fig.~\ref{QiAM22}(c) validate the results. On the other hand, in case 2, the existence of an edge state requires that the equation $[{\rm{Re}}(M_{11})^2-1]^{1/2}=-U_{1}{\rm{sin}}\;{k_{2}b_{5}}$ admits of a solution. Therefore, $\eta=U_{1}{\rm{sin}}\;{k_{2}b_{5}}$ defines the edge state, with ${\eta}<0$ indicating a topologically nontrivial phase. The topological index is thus the sign of $\eta$, which, in view of Eq.~(\ref{Zak_phase}), can be rewritten as
\begin{equation}
	{\rm{sgn}}(\eta)={\rm{exp}(i\mathcal{Z})}{\rm{sgn}}({\rm{sin}}\;{k_{2}b_{5}}).
	\label{eta}
\end{equation}
According to Eq.~(\ref{eta}), temperature field simulations [Fig.~\ref{QiAM22}(c)] confirm the existence of an edge state for $c<0$. The bulk-edge correspondence in diffusion systems was thus rigorously established beyond the scope of the tight-binding model. This work clearly points to a richer topological physics in diffusion systems without
condensed matter counterparts.

The other paradigm to detect topological effects in the heat conduction
consists in realizing the analogue of the lattice model in condensed matter
by means of coupled ring chain structures. By this technique, a unique
non-Hermitian topology, the well-known non-Hermitian skin effect, has been
first demonstrated by Cao and coworkers~\cite{CaoCP21} also in diffusion systems. Topological properties under periodic and open boundary conditions do not conform well in nonreciprocal systems. Thus, the conventional bulk-edge correspondence of Hermitian physics is broken, and non-Bloch band theory has to be established to solve this problem~\cite{TorresPRB18,YaoPRL18, YokomizoPRL19,TorresJPM19,QiangCPL23}. Additionally, it was found that the bulk state is localized at the edge, called non-Hermitian skin effect~\cite{ZhangAPX22}.

\begin{figure}[!ht]
	\includegraphics[width=1.0\linewidth]{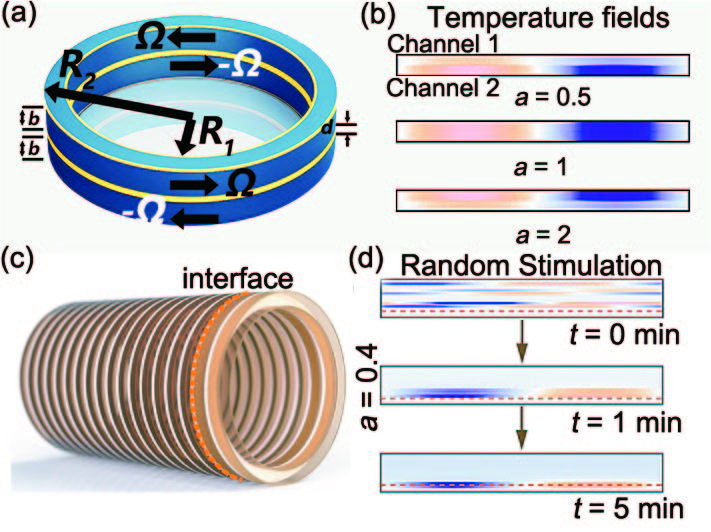}
	\caption{Diffusive skin effect in coupled ring chain structure. (a) Double ring model. $R_1$, $R_2$, and $b$ are the ring inner radius, outer radius and thicknes; $d$ is the interlayer thickness. Adapted from~\citealp{LiSci19}. (b) Temperature field distributions in double ring structure different asymmetric factors $a=\sqrt{h_{12}/h_{21}}$. Blue and orange colors indicate the minimum and maximum temperatures. (c) Sketch of a coupled ring chain structure with interfaces (red dashed curves). (d) Temperature field evolutions for the random initial condition with asymmetric coupling ($a=0.4$). Blue and orange colors indicate the minimum and maximum temperatures. Adapted from~\citealp{CaoCP21}.}
	\label{CaoCP21}
\end{figure}

We begin with the simple double-ring model proposed by Li et
al.~\cite{LiSci19}. As illustrated in Fig.~\ref{CaoCP21}(a), two rings are connected by an interlayer. According to the Fourier law for heat conduction, the coupling equations can be formulated as
\begin{align}	 \frac{{\partial}T_{1}}{{\partial}t}=D_{1}\frac{{\partial}^{2}T_{1}}{{\partial}x^2}+h_{21}(T_2-T_1), \\
	 \frac{{\partial}T_{2}}{{\partial}t}=D_{2}\frac{{\partial}^{2}T_{2}}{{\partial}x^2}+h_{12}(T_1-T_2),
	\label{double_ring_eq}
\end{align}
where $T_{1}$ ($T_{2}$) is the temperature field of the upper (lower) ring and $D_{1}=\kappa_{1}/\rho_{1}C_{1}$ [$D_{2}=\kappa_{2}/(\rho_{2}C_{2})$] the thermal diffusivity of the upper (lower) ring with $\kappa_{1}$ ($\kappa_{2}$), $\rho_{1}$ ($\rho_{2}$), and $C_{1}$ ($C_{2}$) being the respective thermal conductivity, mass density, and heat capacity. The heat exchange rate of the upper (lower) ring is $h_{21}={\kappa_{i}}/(\rho_{1}C_{1}bd)$ [$h_{12}={\kappa_{i}}/(\rho_{2}C_{2}bd)$], where $\kappa_{i}$ denotes the thermal conductivity of the interlayer. Thanks to the periodic structure of the channel, one can assume that the temperature field has a plane wave component $T=Ae^{i({\beta}x-{\omega}t)}$, where $A$ is the amplitude, $\beta$ the wave number, and $\omega$ the complex eigenfrequency. According to the standing wave condition, the wave number has to be discrete with $\beta=n/R$ (where $n$ is the mode order and $R{\approx}R_{1}{\approx}R_{2}$ the radius of the rings). Here the fundamental mode is $n=1$ because only the slowest decaying mode can be observed in the temperature field simulation. Substituting the plane wave solution into Eq.~(\ref{double_ring_eq}), one can obtain an effective Hamiltonian for the double ring model,
\begin{equation}
	{\hat{H}}=\left(\begin{matrix}
		-i({\beta}^2D_{1}+h_{21}) & ih_{21}
		\\
		ih_{12} & -i({\beta}^2D_{2}+h_{12})
	\end{matrix}\right).
	\label{double_ring_Ham}
\end{equation}
The diagonal elements in Eq.~(\ref{double_ring_Ham}) can be unified as a constant by adjusting the diffusivities, $D_1$ and $D_2$. The resulting
Hamiltonian is asymmetric for $h_{21}\neq h_{12}$, that is for $\rho_{1}C_{1}\neq \rho_{2}C_{2}$, which yields an asymmetric distribution
of eigenstates. Consequently, the temperature field distribution for the double ring model will become asymmetric [Fig.~\ref{CaoCP21}(b)].

Next, a chain structure of coupled rings can be designed to realize a 1D nonreciprocal SSH model~\cite{CaoCP21}. A heat funnel model with an interface between two mirrored SSH chains was devised to gain flexible control of the skin mode. The schematic diagram of the heat funnel model is drawn in Fig.~\ref{CaoCP21}(c). Thermal conductivity, mass density, and heat capacity
of rings and interlayers are denoted respectively by ($\kappa_{n}, \rho_{n}, C_{0}$) and ($\kappa_{in}, \rho_{0}, C_{0}$), $n$ and $in$ being the ring and interlayer suffices. Accordingly, the heat exchange rates between ring $n$ and $n+1$ are $h_{n,n+1}=\kappa_{in}/(\rho_{n+1}C_{0}bd)$ and $h_{n+1,n}=\kappa_{in}/(\rho_{n}C_{0}bd)$. To meet the requirements of the diffusive nonreciprocal SSH model, Cao and coworkers tuned the parameters of rings and interlayers as
\begin{subequations}
	\begin{align}
		&\rho_{n}=
		\begin{cases}
			a^{n-1}\rho_{0},{\enspace}n=1,3,5,\cdots   \\
			a^{n}\rho_{0},{\enspace}n=2,4,6,\cdots
		\end{cases},\\
		&\kappa_{in}=a^{n}\kappa_{i0},{\enspace}n=1,2,3,4,\cdots.
		\label{SSH_parameter}
    \end{align}
\end{subequations}
the constant $a$ playing the role of an asymmetry factor. Then, the
temperature field was simulated to illustrate the heat funneling effect [Fig.~\ref{CaoCP21}(d)]. For $a=0.4$, the temperature field appeared to concentrate on the interface between rings 10 and 11. 

\begin{figure}[!ht]
	\includegraphics[width=0.8\linewidth]{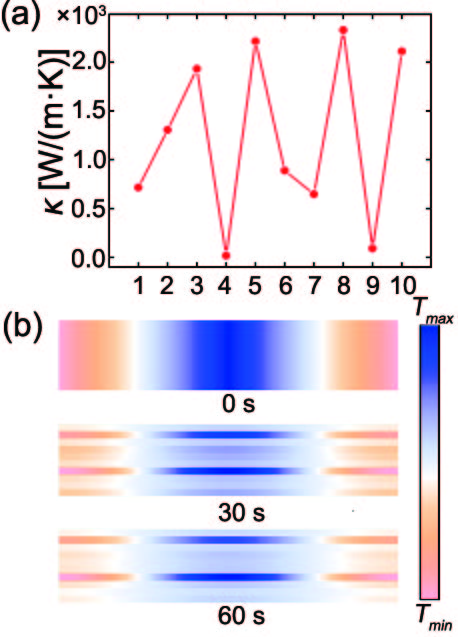}
	\caption{Diffusive Anderson localized state in a coupled ring chain. (a) Adjusted thermal conductivity of each ring. (b) Temperature field vs. time with the initial temperature decreasing linearly from its maximum in the middle of the channel, $T_h$, down to its minimum, $T_l$, at the channel endpoints. Adapted from~\citealp{LiuarXiv22-2}.}
	\label{LiuarXiv22_2}
\end{figure}

The 1D Aubry-Andr$\acute{\rm e}$-Harper (AAH) model~\cite{Aubry80, Harper55} is a case study of quasicrystal (an ordered but non-periodic phase) in condensed matter physics. It exhibits Anderson localized states above a finite critical value of an incommensurate onsite potential. Recently, it has been implemented also in diffusion systems
consisting of coupled ring chain structures~\cite{LiuarXiv22-2}, by tuning
the thermal conductivity of each ring so as to map the Hamiltonian of the
coupled ring chain into an AAH Hamiltonian [Fig.~\ref{LiuarXiv22_2}(a)].

In the temperature field simulation of the localized state [Fig.~\ref{LiuarXiv22_2}(b)], the field tends to concentrate around the fourth and ninth ring, referred to as double localization centers. This novel phenomenon arises from the proximity of decay rates of the slowest and second slowest branch.

\begin{figure}[!ht]
	\includegraphics[width=1.0\linewidth]{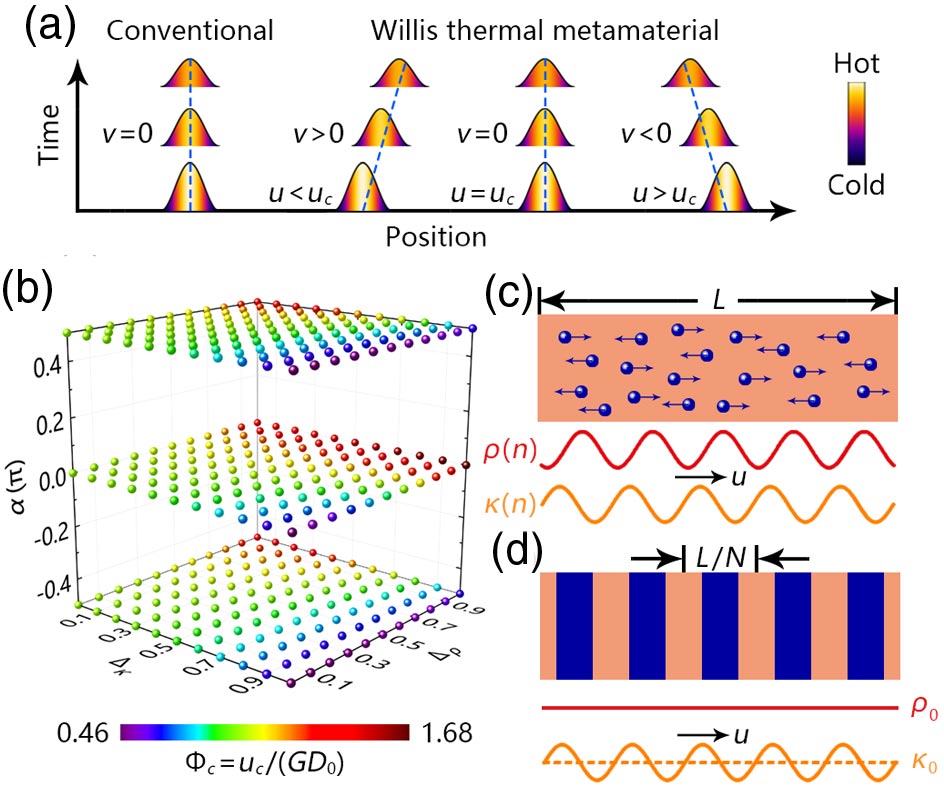}
	\caption{Thermal Willis coupling. (a) Comparison between a conventional material and a Willis thermal metamaterial. (b) Critical point determining the propagation direction of a wavelike temperature field. Schematic diagrams of (c) a two-parameter modulation model and (d) a single-parameter modulation model. Adapted from~\citealp{XuPRL22-2}.}
	\label{williscoupling}
\end{figure}

Asymmetric heat conduction is of great interest due to its potential applications to energy collection, waste heat recovery, and heat rectification. However, space-reversal symmetry prevents heat from diffusing asymmetrically, making asymmetric transport in heat conduction a difficult
problem. Xu \emph{et al}. proposed a novel mechanism based on thermal Willis coupling to induce asymmetric heat conduction~\cite{XuPRL22-2}. Willis thermal metamaterials with spatiotemporal modulation were designed to control the propagation of a heat pulse in a specific direction [Fig.~\ref{williscoupling}(a)]. The propagation direction is highly tunable, determined by a critical point of spatiotemporal modulation [Fig.~\ref{williscoupling}(b)].

Heat conduction in a uniform medium is governed by the Fourier law, $\rho C\partial T/\partial t+\nabla\cdot\left(-\kappa\cdot\nabla T\right)=S$, where $\rho$, $C$, $\kappa$, $S$, and $T$ denote the mass density, heat capacity, thermal conductivity, heat power density, and temperature, respectively. In the common case of a passive system, $\left(S=0\right)$,
with homogeneous parameters, the heat pulse does not propagate due to
space-reversal symmetry [Fig.~\ref{williscoupling}(a)], and the temperature
distribution remains unchanged over time.

Spatiotemporal modulation is therefore considered to break symmetric diffusion. Since heat capacity is difficult to control, the mass density, $\rho(n)$, and thermal conductivity, $\kappa(n)$, were modulated in both space and time.
\begin{equation}\label{TWCrho}
	\rho(n)=\rho_0(1+\Delta_\rho\cos(Gn+\alpha)),
\end{equation}
\begin{equation}\label{TWCkappa}
	\kappa(n)=\kappa_0(1+\Delta_\kappa\cos(Gn)),
\end{equation}
where $\rho_0$ and $\kappa_0$ are two constants, $\Delta_\rho$ and $\Delta_\kappa$ are the dimensionless variation amplitudes, $\alpha$ is the phase difference between two spatiotemporal parameters, $G$ is the modulation wave number, and $n=x-ut$ is the generalized coordinate with $u$ being the modulation speed. The heat transport along the chain obeys
the 1D equation
\begin{equation}\label{TWCG}
	\rho(n)C_0\dfrac{\partial T}{\partial t}+u(\rho(n)-\rho_0)C_0\dfrac{\partial T}{\partial x}+\dfrac{\partial}{\partial x}\left(-\kappa(n)\dfrac{\partial T}{\partial x}\right)=0.
\end{equation}
Mass conservation causes a local advection heat flux, $u(\rho(n)-\rho_0)C_0\partial T/\partial x$, but the time-averaged effect is still zero, so no directional advection appears. Xu and coworker~\cite{XuPRL22-2} then further introduced the wavelike temperature field, $T={e}^{{i}(\beta x-\omega t)}+T_0$, where $\beta$ and $\omega$ are the wave number and frequency, respectively, and $T_0$ is the balanced temperature set as zero for brevity. Due to the spatiotemporal periodicity of the Willis thermal metamaterial, the Bloch-Floquet theorem can be applied to describe the temperature distribution,
\begin{equation}\label{TWCT}
	T=\epsilon(n){e}^{{i}(\beta x-\omega t)}+T_0=\left(\sum_s\epsilon_s{e}^{{i}sGn}\right){e}^{{i}(\beta x-\omega t)}+T_0,
\end{equation}
where $\epsilon(n)$ is the Bloch-Floquet modulation coefficient. Substituting Eq.~(\ref{TWCT}) into Eq.~(\ref{TWCG}) yields $2s+1$ component equations. As long as $s$ is large enough, the final numerical results are nearly accurate. For the sake of simplicity in analytical calculation, $s$ was limited to $0,\pm1$. Then, Eq.~(\ref{TWCG}) could be homogenized as
\begin{equation}\label{TWCH}
	\rho_0C_0\dfrac{\partial T_0}{\partial t}-K\dfrac{\partial^2T_0}{\partial x^2}-R\dfrac{\partial^2\ T_0}{\partial t^2}-W\dfrac{\partial^2T_0}{\partial t\partial x}=0,
\end{equation}
where $T_0$ describes the outline of the wave-like temperature field. The homogenized parameters $K$, $R$, and $W$ were expressed as
\begin{equation}\label{TWCK} K=\kappa_0\left(1+\dfrac{-\Delta_\kappa^2+\Delta_\rho^2\Phi^2+2\Delta_\kappa\Delta_\rho\Phi^2\cos\alpha}{2\left(1+\Phi^2\right)}\right),
\end{equation}
\begin{equation}\label{TWCR}
	R=\dfrac{\kappa_0}{u^2}\dfrac{\Delta_\rho^2\Phi^2}{2\left(1+\Phi^2\right)},
\end{equation}
\begin{equation}\label{TWCW} W=\dfrac{\kappa_0}{u}\dfrac{\Delta_\rho^2\Phi^2+\Delta_\kappa\Delta_\rho\Phi^2\cos\alpha}{1+\Phi^2},
\end{equation}
and $\Phi=u/\left(GD_0\right)=uL/\left(2\pi N D_0\right)$ is a dimensionless parameter. Energy conservation requires the universal 1D heat conduction equation, $\rho_0C_0\partial_tT_0+\partial_xJ=0$, with $J$ being the heat flux, to be satisfied. By comparing this equation with Eq.~(\ref{TWCH}), the Fourier law should be modified as
\begin{equation}\label{TWCMF}
	-\frac{R}{\rho_0C_0}\frac{\partial J}{\partial t}+J=-K\frac{\partial T_0}{\partial x}-W\frac{\partial T_0}{\partial t}.	
\end{equation}
The term $\partial_tJ$ is mathematically related to thermal relaxation, and the term $\partial_tT_0$ illustrates the source of the thermal Willis coupling. In other words, the heat flux is associated with the temperature gradient $\partial_xT_0$ and the rate of temperature change $\partial_tT_0$. This is analogous to the acoustic Willis coupling: the classical elastic wave equation cannot account for acoustic propagation in inhomogeneous materials. Equation~(\ref{TWCMF}) was derived based on two assumptions, i.e., $s=0,\pm1$ and $\beta\ll G$. The errors caused by these two assumptions can be addressed by examining numerical results and do not affect the conclusion.

\begin{figure}[!ht]
	\includegraphics[width=1.0\linewidth]{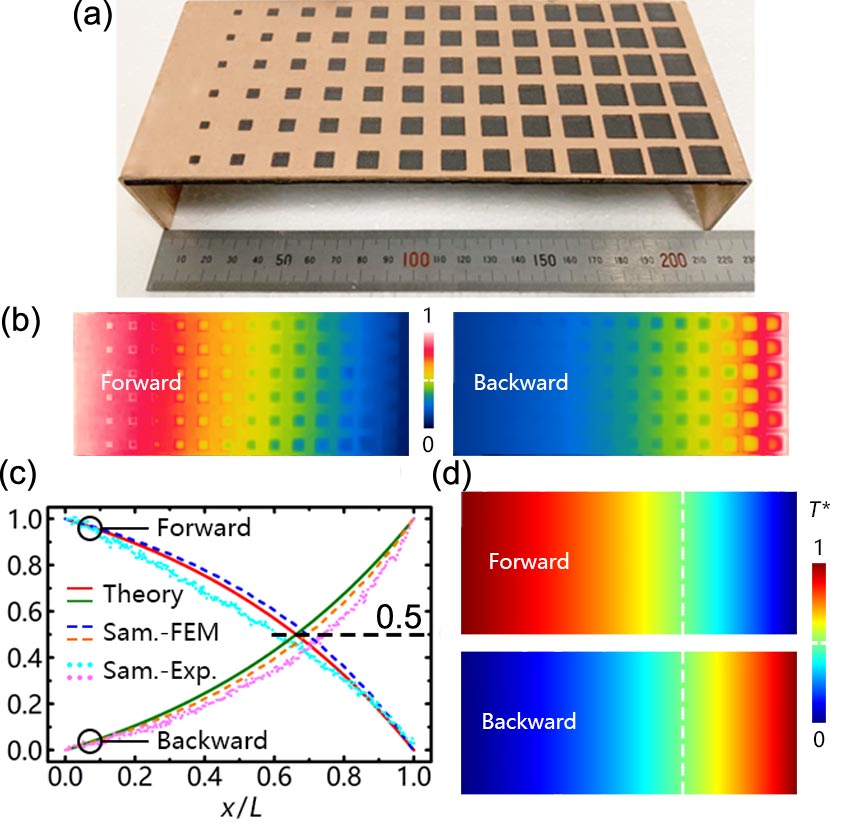}
	\caption{Imitated advection in a graded heat-conduction metadevice. (a) A practical sample. (b) Measured forward and backward temperature profiles. (c) Temperature distributions along the $x$ axis. (d) Temperature profiles in the presence of actual advection. Adapted from~\citealp{XuNSR23}.}
	\label{graded}
\end{figure}

Spatiotemporal modulation offers a promising approach to achieve asymmetry, but dynamic control is challenging to implement. To avoid this difficulty, Xu \emph{et al}. proposed an alternative mechanism based on graded parameters~\cite{XuNSR23}.

In electromagnetic wave systems, a graded refractive index can generate a phase gradient, which acts on the propagating wave as an additional effective momentum applied. Thus, even when an electromagnetic wave is incident normally, the bending effect can still be observed. Although heat conduction has a distinct mechanism from wave propagation, a graded thermal conductivity can also produce a similar bias, which behaves like an imitated advection.

The governing equation in a graded heat-conduction metadevice is analyzed to reveal the underlying mechanism,
\begin{equation}\label{GHCMG}
	\rho\left(x\right)C\left(x\right)\dfrac{\partial T}{\partial t}+\dfrac{\partial}{\partial x}\left(-\kappa\left(x\right)\dfrac{\partial T}{\partial x}\right)=0,
\end{equation}
where $\rho\left(x\right)$, $C\left(x\right)$, and $\kappa\left(x\right)$ are the graded mass density, heat capacity, and thermal conductivity, respectively.  Eq.~(\ref{GHCMG}) can be further simplified as
\begin{equation}\label{GHCMGR}
	\frac{\partial T}{\partial t}-\frac{1}{\rho\left(x\right)C\left(x\right)}\frac{\partial\kappa\left(x\right)}{\partial x}\frac{\partial T}{\partial x}-\frac{\kappa\left(x\right)}{\rho\left(x\right)C\left(x\right)}\frac{\partial^2T}{\partial x^2}=0.
\end{equation}
A conventional conduction-advection equation has a form of
\begin{equation}\label{GHCMA}
	\dfrac{\partial T}{\partial t}+v_0\dfrac{\partial T}{\partial x}-D_0\dfrac{\partial^2T}{\partial x^2}=0,
\end{equation}
where $v_0$ is the advection velocity and $D_0$ is the thermal diffusivity. Comparing Eqs.~(\ref{GHCMGR}) and~(\ref{GHCMA}), the term related to $\partial_xT$ in Eq.~(\ref{GHCMGR}) can be attributed to an imitated advection with the imitated advection velocity of $v_i$,
\begin{equation}\label{GHCMI} v_i=\frac{-1}{\rho\left(x\right)C\left(x\right)}\frac{\partial\kappa\left(x\right)}{\partial x}.
\end{equation}
Equation~(\ref{GHCMI}) demonstrates that a graded thermal conductivity with $\partial_x\kappa\left(x\right)\neq0$ is essential for the simulated advection. To obtain a constant effective advection velocity, Xu and coworkers~\cite{XuNSR23}
further assumed the following exponential form for the material parameters,
\begin{equation}\label{GHCMK}
	\kappa\left(x\right)=\kappa_0{e}^{\alpha x},
\end{equation}
\begin{equation}\label{GHCMR}
	\rho\left(x\right)C\left(x\right)=\rho_0C_0{e}^{\alpha x},
\end{equation}
where $\kappa_0$ and $\rho_0$ are constants representing thermal conductivity and mass density, respectively, and $\alpha$ is a constant parameter gradient. In this way, the imitated advection velocity [i.e., Eq.~(\ref{GHCMI})] becomes
\begin{equation}\label{GHCMIC}
	v_{i0}=-\alpha D_0.
\end{equation}

A graded heat-conduction metadevice was fabricated [Fig.~\ref{graded}(a)] to demonstrate the imitated advection experimentally. The measured temperature profiles are presented in Fig.~\ref{graded}(b). The temperature distribution [Fig.~\ref{graded}(c)] indicates that the average temperature $T^*=0.5$ is not at the center of the sample $x/L=0.5$, which is a direct evidence of imitated advection. For comparison, the temperature profiles with real advection are also shown in Fig.~\ref{graded}(d). Inspired by black holes that can trap light, Xu \emph{et al}. further applied the imitated advection to realize thermal trapping. To this end, two 2D graded heat-conduction metadevices were fabricated, with the imitated advection pointing towards the center. 

Finally, we mention that spatiotemporal modulation can also induce
advection-like effects~\cite{TorrentPRL18,XuPRE21,Ordonez-MirandaPRAp21}, thus contributing to asymmetric heat transfer. Traditional nonlinear schemes considering temperature-dependent thermal conductivity have also been applied to realize asymmetric control of wavelike temperature fields~\cite{liPRB21,ShimokusuIJHMT22,Ordonez-MirandaPRB22,LiuPRAp22}. These results significantly advance ability the control of nonreciprocal heat transfer.

\section{Thermal conduction and convection}

Convection is one of the primary modes of heat transfer and is always accompanied with heat conduction~\cite{JuAM23}. However, applying the transformation theory in the presence of both is challenging due to the complexity of hydrodynamic equations. In 2011, the first attempt to control fluid flow in transformation media was proposed~\cite{UrzhumovPRL11}, using the Darcy law to describe the motion of the fluid flow. Later on, Guenneau \emph{et al.} proposed to manipulate heat and mass diffusion by
transforming the Fourier and Fick equations~\cite{GuenneauAIPAdv15}. They did
prove the form invariance of the thermal conduction-convection equation under
coordinate transformations, but were unable to implement it into actual thermal metamaterials. Inspired by these works, Dai \emph{et al.} proposed a regime for transforming thermal conduction-convection~\cite{DaiJAP18}, realizing thermal cloaking, concentrating, and rotating by controlling creeping flow and heat flux simultaneously, under certain conditions. We review these important results in this section.

\subsection{Theory and transformation principles}

In view of Darcy's law for an incompressible pore-filling fluid, the temperature
field is governed by the conduction-convection law
\begin{subequations}\label{heatconv}
	\begin{align}
		(\rho C)_e\frac{\partial T}{\partial t} + \nabla\cdot [-&\kappa_e\nabla T + (\rho C)_f\bm{v}T] = 0,\\
		&\nabla\cdot\bm{v} =0,\\
		\bm{v}& = -\frac{\eta}{\mu}\nabla P,
	\end{align}
\end{subequations}
where $(\rho C)_e$, $\kappa_e$, and $\bm{v}$ are the effective mass density and heat capacity product, effective thermal conductivity, and velocity, respectively. $\eta$, $\mu$, and $P$ are the permeability, dynamic viscosity, and pressure. The average volume method defines the above effective parameters as $(\rho C)_e = \phi(\rho C)_f + (1 - \phi)(\rho C)_s$ and $\kappa_e = \phi\kappa_f + (1 - \phi)\kappa_s$, where $\phi$ is the porosity, and the subscripts $s$ and $f$ denote solid and fluid components, respectively. Equation~(\ref{heatconv}) is form-invariant under coordinate transformations, with the transformation rules
\begin{subequations}\label{transru-poro}
	\begin{align}
		(\rho C)'_s &= \frac{(\rho C)'_e - \phi (\rho C)_f}{1 - \phi},\\
		\kappa_s'&= \frac{\kappa_e' - \phi\kappa_f}{1 - \phi},\\
		\bm{\eta}' &= \frac{\bm{J}\eta \bm{J}^{\dag}}{\text{det}\bm{J}},
	\end{align}
\end{subequations}
where $(\rho C)'_e = (\rho C)_e/\text{det}\bm{J}$ and $\bm{\kappa}_e' = \bm{J}\kappa_e\bm{J}^{\dag}/\text{det}\bm{J}$. The Reynolds number of the fluid creeping flow is assumed to be smaller than 1. Then, according to Eqs.~(\ref{transru-poro}), thermal metamaterials can be designed for use in the conduction-convection process.

In addition to the Darcy law, this methodology can also be applied to the Hele-Shaw flows~\cite{OronRMP97,DaiPRAp22,WangATE21,WangCPB22,YaoIScience22,DaiPRE23,DaiMat23}, which model the Stokes flow between two very close parallel plates. Further research is needed for other cases, such as fluid flows with high Reynolds numbers. Compared to thermal conduction, the experimental demonstration of transformation thermal convection is more complex. In 2019, Park \emph{et al}. fabricated a hydrodynamic cloak by transforming viscosity~\cite{ParkPRL19}. Their microstructures suggest that the same approach can be used to regulate the permeability in porous media. This lead helped Chen \emph{et al}. realize multifunctional metamaterials for fluid flow~\cite{ChenPNAS22}. Moreover, Boyko \emph{et al}. experimentally demonstrated hydrodynamic cloaking and shielding using field-effect electro-osmosis~\cite{BoRPL21}, which might inspire further related research.

\subsection{Applications: Metamaterials and metadevices}

Recently, Jin \emph{et al}.~\cite{JinPNAS23} reported a liquid-solid hybrid
scheme to simultaneously control thermal convection and conduction. Utilizing the advanced convection-conduction transformation theory, they designed,
fabricated, and tested a liquid-solid hybrid thermal metamaterial with remarkable
heat-flux tunability.

\begin{figure}[!ht]
	\includegraphics[width=1.0\linewidth]{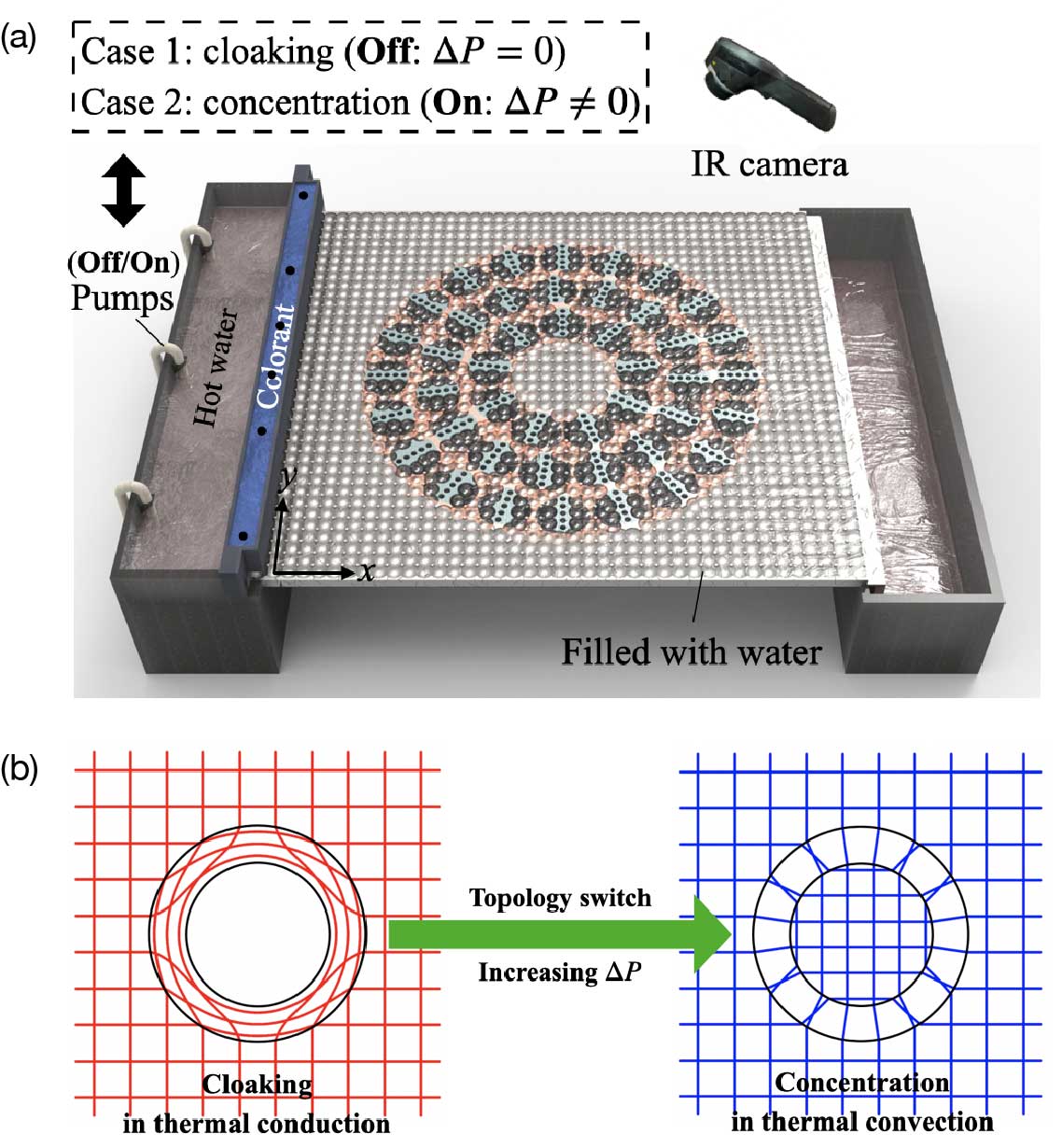}
	\caption{Liquid-solid hybrid thermal metamaterials. (a) Experimental setup. (b) Topological transition occurs in the transformation space. Adapted from~\citealp{JinPNAS23}.}
	\label{4}
\end{figure}

Figure~\ref{4}(a) displays a schematic of the proposed hybrid thermal metamaterial. A steady-state thermal transport should be taken into account,
\begin{equation}\label{Conservation}
	\bm{\nabla}\cdot\bm{J}_{\rm tot}=\bm{\nabla}\cdot\left(\bm{J}_{\rm cond}+\bm{J}_{\rm conv}\right)=0,
\end{equation}
where $\bm{J}_{\rm tot}$, $\bm{J}_{\rm cond}$, and $\bm{J}_{\rm conv}$ are the total, conductive, and convective heat fluxes, respectively. $\bm{J}_{\rm cond}$ and $\bm{J}_{\rm conv}$ are governed by the Fourier law ($\bm{J}_{\rm cond}=-\bm{\kappa}\cdot\bm{\nabla}T$) and the Darcy law ($\bm{J}_{\rm conv}=\rho C\bm{v}T$, $\bm{v}=-\bm{\sigma} / \eta\cdot\bm{\nabla}P$), with $\bm{\kappa}$, $T$, $\rho$, $C$, $\bm{\sigma}$, $\eta$, and $P$ denoting the thermal conductivity, temperature, density, heat capacity, permeability, dynamic viscosity, and hydraulic pressure of the system. Therefore, Eq.~(\ref{Conservation}) can be rewritten as
\begin{equation}\label{invariant}
	\bm{\nabla}\cdot\left[-\bm{\kappa}\cdot\bm{\nabla}T+\rho C\left(-\frac{\bm{\sigma}}{\eta}\cdot\bm{\nabla}P\right)T\right]=0.
\end{equation}
Such an equation is invariant under any coordinate transformations. By means of
appropriate thermotics transformations of the thermal conductivity and
permeability, one obtains both thermal conductive cloaking and convective
concentration, thus ensuring a continuous tunability of the heat flux. By developing an $r$-independent transformation rule, the complexity of the parameter space is significantly reduced. For thermal cloaking, the space transformation rule is given by
\begin{subequations}
	\begin{align}
		r'&=R_1r/R_m, ~~~ 0<r<R_m, \\
		r'&=R_2\left(r/R_2\right)^M, ~~~ R_m<r<R_2, \\
		\theta'&=\theta,
	\end{align}
\end{subequations}
where $M<1$ represents the nonlinear order of the approximate space transformation for cloaking and $R_m = R_2\left(R_1/R_2\right)^{1/M}$ determines the cloaking performance. The transformed thermal conductivity $\bm{\kappa}'$ reads
\begin{subequations}
	\begin{align}
		\bm{\kappa}'&=\left(
		\begin{matrix}
			1 & 0\\
			0 & 1
		\end{matrix}
		\right)\kappa, ~~~ 0<r'<R_1, \\
		\bm{\kappa}'&=\left(
		\begin{matrix}
			M & 0\\
			0 & 1/M
		\end{matrix}
		\right)\kappa. ~~~ R_1<r'<R_2.
	\end{align}
\end{subequations}
When thermal convection dominates, the metamaterial exhibits thermal concentration, and the transformation rule is
\begin{subequations}
	\begin{align}
		r'&=R_1r/R_m, ~~~ 0<r<R_m, \\
		r'&=R_2\left(r/R_2\right)^N, ~~~ R_m<r<R_2. \\
		\theta'&=\theta,
	\end{align}
\end{subequations}
where $N>1$ represents the nonlinear order of the concentration's approximate space transformation and $R_m = R_2\left(R_1/R_2\right)^{1/N}$. The permeability $\bm{\sigma}'$ is obtained,
\begin{subequations}
	\begin{align}
		\bm{\sigma}'&=\left(
		\begin{matrix}
			1 & 0\\
			0 & 1
		\end{matrix}
		\right)\sigma, ~~~ 0<r'<R_1, \\
		\bm{\sigma}'&=\left(
		\begin{matrix}
			N & 0\\
			0 & 1/N
		\end{matrix}
		\right)\sigma. ~~~ R_1<r'<R_2.
	\end{align}
\end{subequations}
Based on the above rule, Jin \emph{et al}. designed a basic unit that can independently and locally regulate thermal convection and conduction. This enabled the liquid-solid hybrid thermal metamaterial to undergo a topology transition in the transformation space, from cloaking to concentration, induced by the external hydraulic pressure difference [Fig.~\ref{4}(b)]. Such property may find extensive
application~\cite{ChenPNAS22,LiuPNAS15,XuCPL20-1,TongPNAS21,OyamaPNAS22,YuCPL23}.

\begin{figure}[!ht]
	\includegraphics[width=1.0\linewidth]{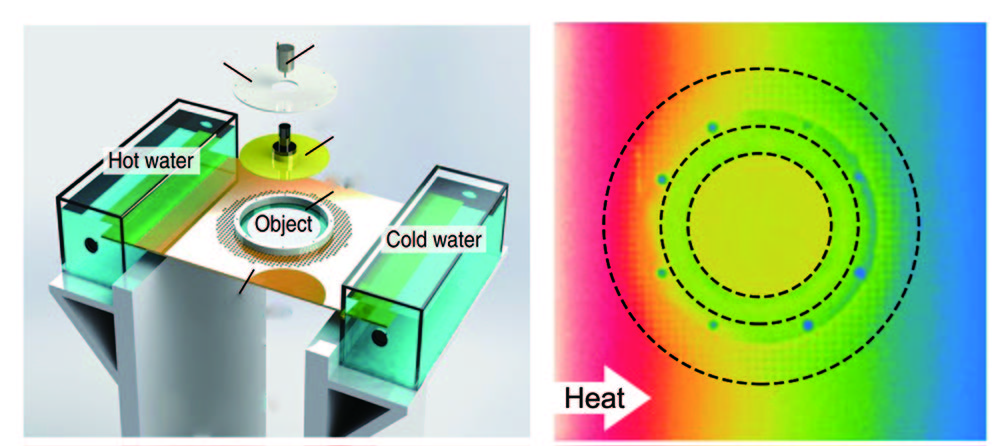}
	\caption{\label{fig-Thermetaconv} Thermal analog to a zero-index photonics metamaterial. Adapted from~\citealp{LiNM19}.}
\end{figure}

Inspired by wave systems, Li \emph{et al}. proposed a thermal analog of zero-index metamaterials with infinite thermal conductivity~\cite{LiNM19} [Fig.~\ref{fig-Thermetaconv}]. The device is a bilayer structure, filling its inner layer with a rotating fluid. The integrated convective part creates an extremely effective thermal conductivity, eliminating the need for a high thermal conductivity in the outer layer. The introduced extreme convection does not violate the original reciprocity of heat transfer, regardless of the detailed fluid motion. Similar rotating
structures have investigated in later works~\cite{LiAM20,ZhuIJHMT21}.

Thermal waves are usually triggered by thermal relaxation~\cite{JosephRMP89}. They have short relaxation times and dissipate quickly, making them difficult to use or control. The emergence of thermal metamaterials helped
circumvent this difficulty. Xu \emph{et al}. demonstrated the control of thermal waves using transformation complex thermotics~\cite{XuIJHMT20}. They proved the form invariance of the complex conduction equation under a coordinate transformation. The real and imaginary parts of the complex thermal conductivity are related respectively to conduction and convection. Similarly, by combining metamaterials and the transformation theory, they designed thermal wave cloaks,
concentrators, and rotators. The form of the complex conduction equation is the same as that of the conduction-convection equation. Therefore, Eq.~(\ref{transru-poro}) is still applicable to this type of thermal waves. In the same line of work,
thermal waveguides have also been proposed~\cite{ZhangTSEP21,XuEPL21}.

\subsection{Nonreciprocal thermal waves, non-Hermitian thermal topology, and other phenomena}

\begin{figure}[!ht]
	\includegraphics[width=0.8\linewidth]{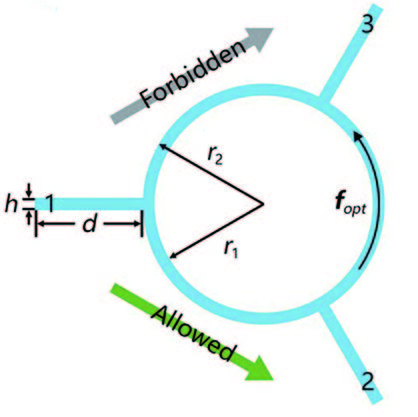}
	\caption{\label{fig-Thermetaconv-2} Schematic of a 3-port angular-momentum-biased ring. Adapted from~\citealp{XuAPL21}.}
\end{figure}

The introduction of convection has resulted in the emergence of numerous remarkable physical phenomena in thermal metamaterials.  An interesting example is the nonreciprocity of heat transport in convective devices [Fig.~\ref{fig-Thermetaconv-2}].
A volume force was imposed in the anticlockwise direction of a three-port ring to generating an angular momentum bias (Fig.~\ref{fig-Thermetaconv-2}). This enabled the observation of nonreciprocity of thermal waves in the angular-momentum-biased ring~\cite{XuAPL21}. The intensity of the volume force was used to manipulate the propagation behavior quantitatively. For a suitable value of the applied volume force, the amplitude of the
oscillating temperature dropped to zero at port 3, but not at port 2, thus
demonstrating the nonreciprocity of thermal wave propagation in the ring. Xu \emph{et al}. further demonstrated thermal edge states by arranging rings in a graphene-like array~\cite{XuEPL21-2}.

\begin{figure}[!ht]
	\includegraphics[width=1.0\linewidth]{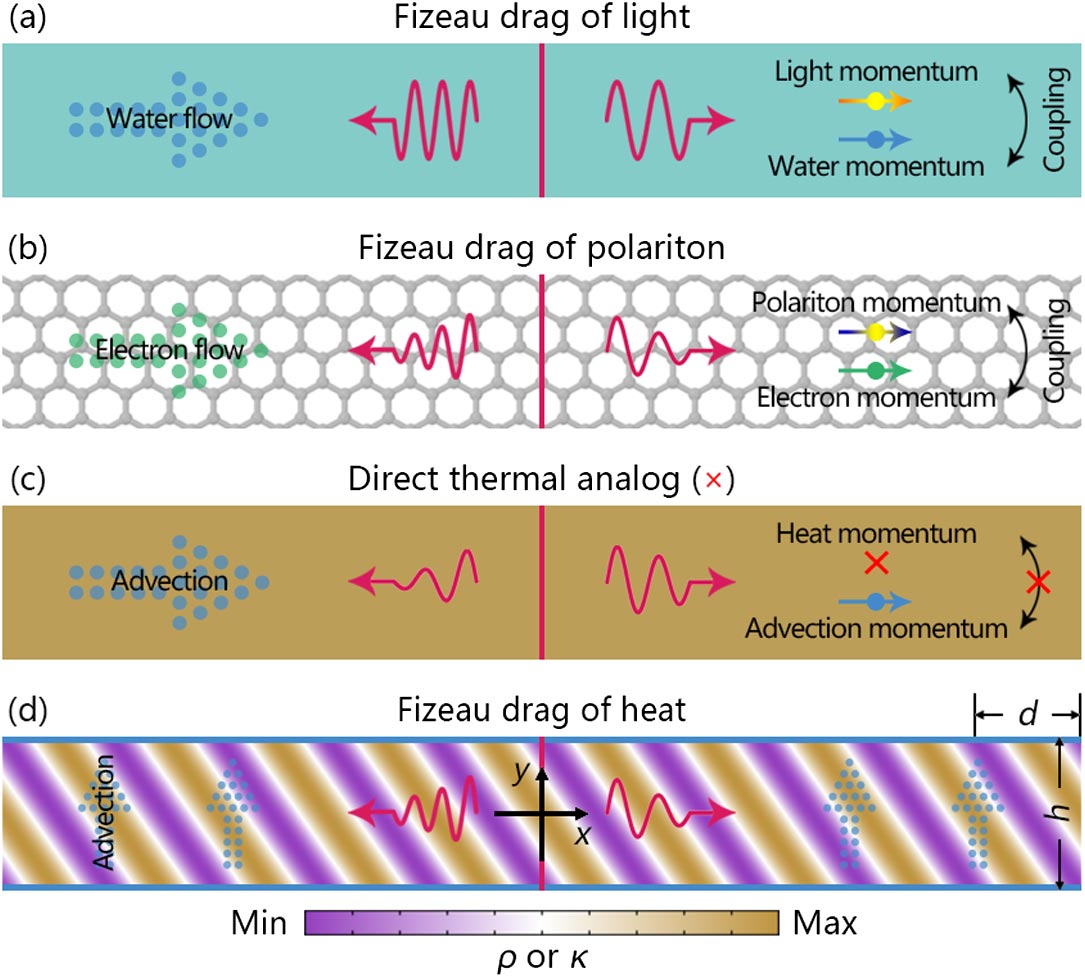}
	\caption{The diffusive Fizeau drag mechanism. (a) Fizeau drag of light in a water flow. (b) Fizeau drag of polaritons in an electron flow. (c) Failure of a direct thermal analogue. (d) Fizeau drag of heat in a porous medium modulated in space and time. Adapted from~\citealp{XuPRL22-1}.}
	\label{Fizeau}
\end{figure}

Diffusive Fizeau drag was also demonstrated with spatiotemporal modulation~\cite{XuPRL22-1}, as illustrated in Fig.~\ref{Fizeau}. Fizeau drag indicates that light travels at different speeds along and against a water flow [Fig.~\ref{Fizeau}(a)]. A similar phenomenon was also revealed for polaritons which propagate at different speeds in an electron flow [Fig.~\ref{Fizeau}(b)]. The next question is whether dragging heat via a similar mechanism is possible. To this end, the concept of propagation speed should be established in heat transfer. Then, Xu and coworkers~\cite{XuPRL22-1} considered a wavelike temperature field, $T=A{e}^{{i}(\beta x-\omega t)}+T_r$, where $A$, $\beta$, $\omega$, and $T_r$ are the amplitude, wave number, angular frequency, and balanced temperature (set as zero for brevity), respectively. In this way, the propagation speed of a wavelike temperature field makes sense, laying a foundation for the Fizeau drag of heat. Common wisdom would suggest that heat advection is the
thermal counterpart of advection in water and electron flows. However, this idea fails because a temperature field does not carry ``heat momentum'' to interact with the advection momentum [Fig.~\ref{Fizeau}(c)]. Specifically, temperature evolution with a horizontal advection is governed by
\begin{equation}\label{FizeauG}
	\rho_0 \dfrac{\partial T}{\partial t}+\bm\nabla\cdot\left(\phi\rho_a \bm{u}T-\kappa_0\cdot\bm\nabla T\right)=0,
\end{equation}
where $\rho_0$ ($\rho_a$) is the product of the mass density and heat capacity of the porous medium (fluid), $\phi$ is the porosity, $\kappa_0$ the thermal conductivity of the porous medium, and $\bm{u}$ the fluid velocity with components $u_x$ and $u_y$. The forward and backward wave numbers (i.e., $\beta_f$ and $\beta_b$) can be derived,
\begin{equation}\label{FizeauT}
	\beta_{f,\,b}=\pm\frac{\sqrt{2}\gamma}{4\kappa_0}+i \frac{-8\phi \rho_a u_x \omega \rho_0 \gamma_0\pm \sqrt{2} \gamma\left(2\phi^2 \tilde{N}_a^2 u_x^2+\gamma{^2}\right)}{16\omega \rho_0 \kappa_0^2},
\end{equation}
with $\gamma=\sqrt{-\phi^2 \rho_a^2 u_x^2+\sqrt{\phi^4 \rho_a^4 u_x^4+16\omega^2 \rho_0^2 \kappa_0^2}}$. A non-zero $u_x$ cannot produce different magnitudes of $|{\rm Re}[\beta]|$ in opposite directions, indicating that no diffusive Fizeau drag is present.

Two crucial factors are taken into account to solve this problem. A vertical advective flow and a spatial distribution of the
porous medium parameters. Neither ingredient alone causes thermal
nonreciprocity in the horizontal direction. Diffusive Fizeau drag results
from their interplay. The underlying mechanism is that time-related advection and space-related inhomogeneity induce spatiotemporal interaction and generate the thermal Willis coupling. The mass density, $\rho(\xi)$, and thermal conductivity, $\kappa(\xi)$, of the porous medium are modulated as
\begin{equation}
	\rho(\xi)=\rho_0(1+\Delta_\rho \cos(G\xi+\theta)),\label{FizeauR}
\end{equation}
\begin{equation}
	\kappa(\xi)=\kappa_0(1+\Delta_\kappa \cos(G\xi)),\label{FizeauK}
\end{equation}
where $\rho_0$ and $\kappa_0$ are two constants, $\Delta_\rho$ and $\Delta_\kappa$ are variation amplitudes, $G$ is the modulation wave number, $\xi$ the generalized coordinate, and $\theta$ the phase difference between these two modulated parameters, Equation~(\ref{FizeauG})
can then be rewritten as
\begin{equation}\label{FizeauG1}
	\overline{\rho}(\xi) \frac{\partial T}{\partial t}+\phi u_y \frac{\partial T}{\partial y}+\frac{\partial}{\partial x} \left(-D_0 \overline{\kappa}(\xi) \frac{\partial T}{\partial x}\right)+\frac{\partial}{\partial y} \left(-D_0 \overline{\kappa}(\xi) \frac{\partial T}{\partial y}\right)=0,
\end{equation}
with $\overline{\rho}(\xi)=\rho(\xi)/\rho_0$, $\overline{\kappa}(\xi)=\kappa(\xi)/\kappa_0$, $\epsilon=\rho_a/\rho_0$, and $D_0=\kappa_0/\rho_0$. The forward and backward propagation speeds in such a metamaterial are different, as shown in Fig.~\ref{Fizeau}(d).

Equation~(\ref{FizeauG1}) is further homogenized to uncover the underlying mechanism. Similarly, the Fourier law should be modified to $\tau\partial_tJ+J=-\kappa_e\partial_xT_0+\sigma_2\partial_tT_0$, where $\tau$, $\kappa_e$, $\sigma_2$, and $T_0$ are the parameters after homogenization. This effect is referred to as the thermal Willis coupling, whereby
the heat flux also depends on the rate of the temperature change. Therefore, whether it is pure conduction~\cite{XuPRL22-2} or conduction-advection~\cite{XuPRL22-1}, the thermal Willis coupling may exist broadly in dynamically inhomogeneous media featuring relative ``motion'' of thermal parameters and mass.

\begin{figure}[!ht]
	\includegraphics[width=0.8\linewidth]{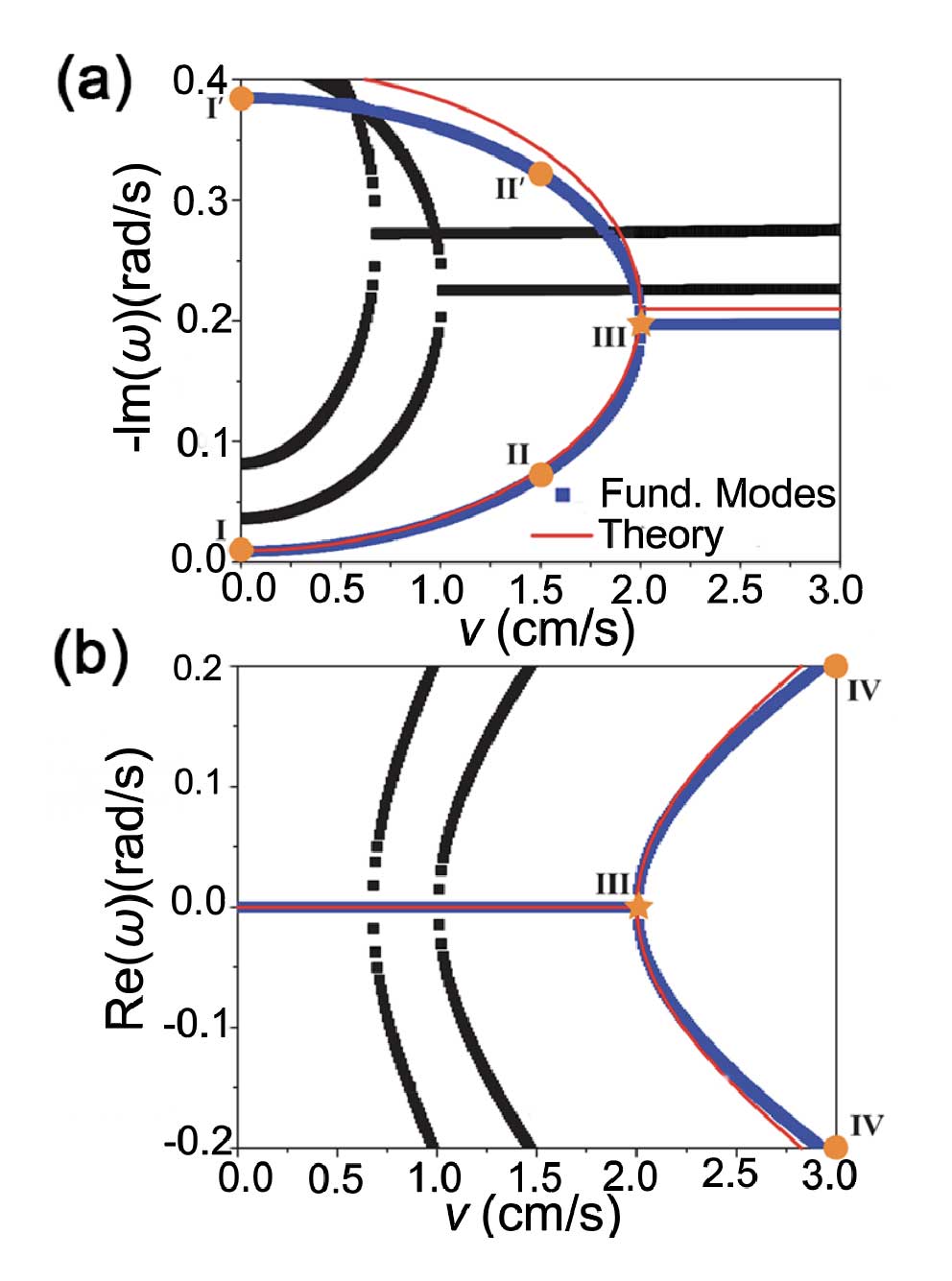}
	\caption{Heat convection-induced APT symmetry in diffusion systems. (a) Decay rates and (b) eigenfrequencies of the double-ring model. The blue squares represent the simulation results of the fundamental mode, and the red lines represent the theoretical eigenvalues of the effective Hamiltonian. Adapted from~\citealp{LiSci19}.}
	\label{LiSci19}
\end{figure}

Intriguing topological and non-Hermitian physics are also revealed in diffusion systems where convective heat transfer is introduced into thermal conduction. Compared with purely conductive systems, conductive-convective systems present three main advantages.

(i) The first is that introducing heat convection allows implementing anti-parity-time (APT) symmetry in diffusive systems. Parity-time (PT) symmetry is an essential symmetry in non-Hermitian physics because the system exhibits a real spectrum~\cite{BenderRPP07, BenderPRL98}. When the eigenstates of the PT-symmetric Hamiltonian preserve (violates) PT symmetry, the system is in the PT unbroken (broken) phase. The critical point between these two phases is the so-called exceptional point (EP). Controlling the PT symmetry in wave systems requires a delicate balance between gain and loss.

However, it is challenging to observe the PT-related phenomenon in diffusion systems that are intrinsically dissipative (anti-Hermitian). This issue was addressed by introducing forward and backward wavelike fields, i.e., heat convection, which act as a Hermitian component. Li {\em et al}.~\cite{LiSci19} considered two coupled rings [Fig.~\ref{CaoCP21}(a)] with
equal but opposite rotating velocities. The governing
equations for such double-ring structure are
\begin{align}	 \frac{{\partial}T_{1}}{{\partial}t}=D\frac{{\partial}^{2}T_{1}}{{\partial}x^2}-v\frac{{\partial}T_{1}}{{\partial}x}+h(T_2-T_1), \\
	 \frac{{\partial}T_{2}}{{\partial}t}=D\frac{{\partial}^{2}T_{2}}{{\partial}x^2}+v\frac{{\partial}T_{2}}{{\partial}x}+h(T_1-T_2),
	\label{double_ring_conv_eq}
\end{align}
where $D$ is the ring diffusivity, $h$ the heat exchange rate between two rings, and $v$ is the angular velocity of rings. By employing the plane-wave solution, an effective APT symmetric Hamiltonian was obtained,
\begin{equation}
	{\hat{H}}=\left(\begin{matrix}
		-i(k^{2}D+h)+kv & ih
		\\
		ih & -i(k^{2}D+h)-kv
	\end{matrix}\right),
	\label{double_ring_conv_Ham}
\end{equation}
where $k$ is the wave number. The eigenvalue is
\begin{align}
	\omega_{\pm}=-i\left[(k^{2}D+h){\pm}\sqrt{h^{2}-k^{2}v^{2}}\right].
	\label{double_ring_conv_eigen}
\end{align}
The EP emerges when $h^{2}=k^{2}v_{\rm{EP}}^2$. Let us focus on the fundamental mode $k=1/R$. The theoretical decay rates and eigenfrequencies of the effective Hamiltonian are shown in Figs.~\ref{LiSci19}(a),(b) to agree with the simulations. In the APT unbroken phase, after a certain time transient the temperature of ring 1 approaches a stationary profile with a localized maximum. Vice versa, in the APT broken phase the temperature profile keeps shifting due to the nonzero eigenfrequency.

By increasing the number of rings, the higher-order EPs of diffusion systems
grow more robust against perturbations and phase oscillations~\cite{CaoES20}. Using the four-ring model, a third-order EP can be obtained at the critical velocity when APT phase transition occurs. Further increasing the number of rings, the device will become a coupled ring chain structure, as mentioned above. Introducing APT symmetry into the coupled ring chain structure will give rise to many exciting phenomena. For example, phase-locking diffusive skin effect in the APT unbroken phase can be induced by heat convection and asymmetric coupling~\cite{CaoCPL22}; dynamical double localization centers can be induced by heat convection and Anderson localization in the APT broken phase~\cite{LiuarXiv22-2}.

\begin{figure}[!ht]
	\includegraphics[width=1.0\linewidth]{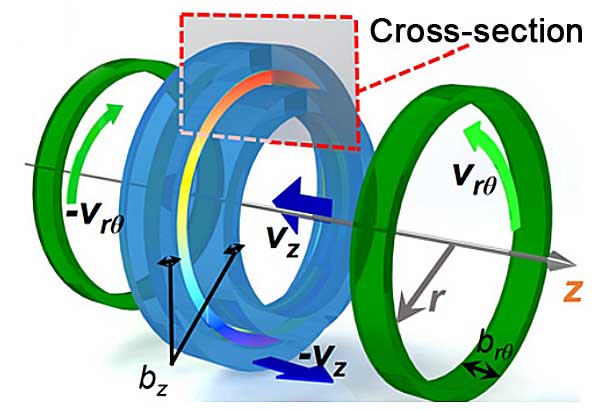}
	\caption{Schematic diagram of a multi-ring system. Adapted from~\citealp{XuPRL21}.}
	\label{XuPRL21}
\end{figure}

(ii) The second is that heat convection acts as an additional highly tunable degree of freedom, which helps one to implement
topological and non-Hermitian effects in diffusion systems.  Recently, the dynamical EP encircling in a non-Hermitian topology has been realized in multiple ring system~\cite{XuPRL21}. This system is composed of two orthogonal pairs of counter-motional convection rings. One is rotating in $r-{\theta}$ space, and the other is translating along the $z$ direction (Fig.~\ref{XuPRL21}). This clever design creates a complex parameter space, providing enough degrees of freedom to encircle an EP. The complex spectrum of the system's effective Hamiltonian exhibits two EPs. The geometric phase generated by a dynamically encircling trajectory determines the topological properties of the EP. It is related to two quantized topological invariants: the eigenstate winding number $\omega$~\cite{LeykamPRL17} and the eigenvalue vorticity $\nu$~\cite{ShenPRL18}. When the adiabatic trajectory of heat convection encircles two EPs, both topological invariants take integer values, which reveals the non-Hermitian thermal topology. Accordingly, the temperature profile of the surface of the central medium exhibits a dynamic-equilibrium distribution without large deviations, and so does the location of the maximum temperature point. In this case, the
system will execute a closed path over a full Riemann sheet and its phase return to its
initial value after one period. By comparison, when the trajectory encircles one EP, both topological invariants are half-integer numbers. The phase of the temperature
profile and the location of the temperature maximun sharply jump by
$\pi$ after one period. The eigenstate will exchange to another state with a $\pi$-phase winding after one period and return to its initial position after two periods.  The geometric phase induced by heat
convection was investigated also by Xu {\em et al}.~\cite{XuIJHMT21}. Based on the double ring model, the temperature field will accumulate a phase during a cyclic evolution of time-varying rotating velocity. When the cyclic trajectory does not contain the EP, the temperature profile will revert to its initial state after one period (zero geometric phase). Vice versa, it gains an extra $\pi$ geometric phase. Moreover, Xu {\em et al}. recently reported a Hall-like heat transfer~\cite{XuPNAS23}. By rotating solid particles, they realized giant thermal chirality without strong magnetic fields or extremely low temperatures.

\begin{figure}[!ht]
	\includegraphics[width=1.0\linewidth]{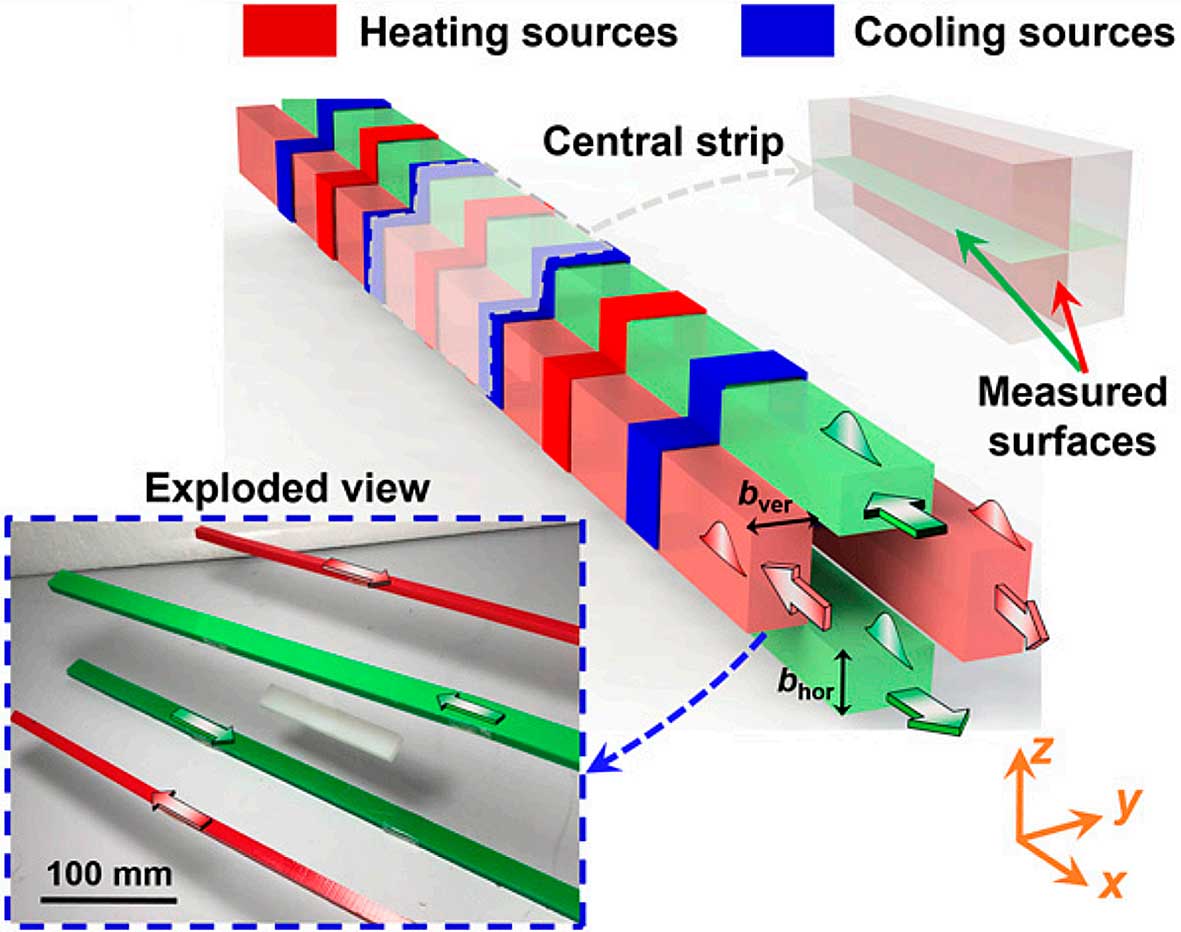}
	\caption{Schematic diagram of experimental setup showing WER. Periodic and alternating heating and cooling sources are applied to the entire system. Green and red strips indicate respectively horizontal and vertical convection, with arrows denoting the initial motion directions. The upper and lower insets are enlarged views of the central strip. Adapted from~\citealp{XuPNAS22}.}
	\label{XuPNAS22}
\end{figure}

Weyl points are band degeneracy points whose dispersion satisfies the Weyl equation in high energy physics. When a non-Hermitian term is introduced into the Weyl Hamiltonian, the Weyl point evolves into a Weyl exceptional ring (WER) composed of exceptional points (EPs)~\cite{XuPRL17}. The WER has been implemented in topological photonics~\cite{CerjanNP19} and topological acoustics~\cite{LiuPRL22}, and, more relevant to the present review, also in diffusive systems~\cite{XuPNAS22}. As shown in Fig.~\ref{XuPNAS22}, the basic model consists of a central strip surrounded by four counter-convection components arranged in orthogonal spaces.

The initial temperature profile of the combined structure was imposed periodically and alternately by the heating and cooling sources to generate a periodic wavelike field [Fig.~\ref{XuPNAS22}]. The synthetic 3D parameter space of this system provides enough degrees of freedom to encompass the WER. If the integration surface encircles the entire WER, the temperature profile and the location of the maximum temperature points in the $xy$ and $xz$ planes remain stationary, which is the signature of a nontrivial topology.

\begin{figure}[!ht]
	\includegraphics[width=\linewidth]{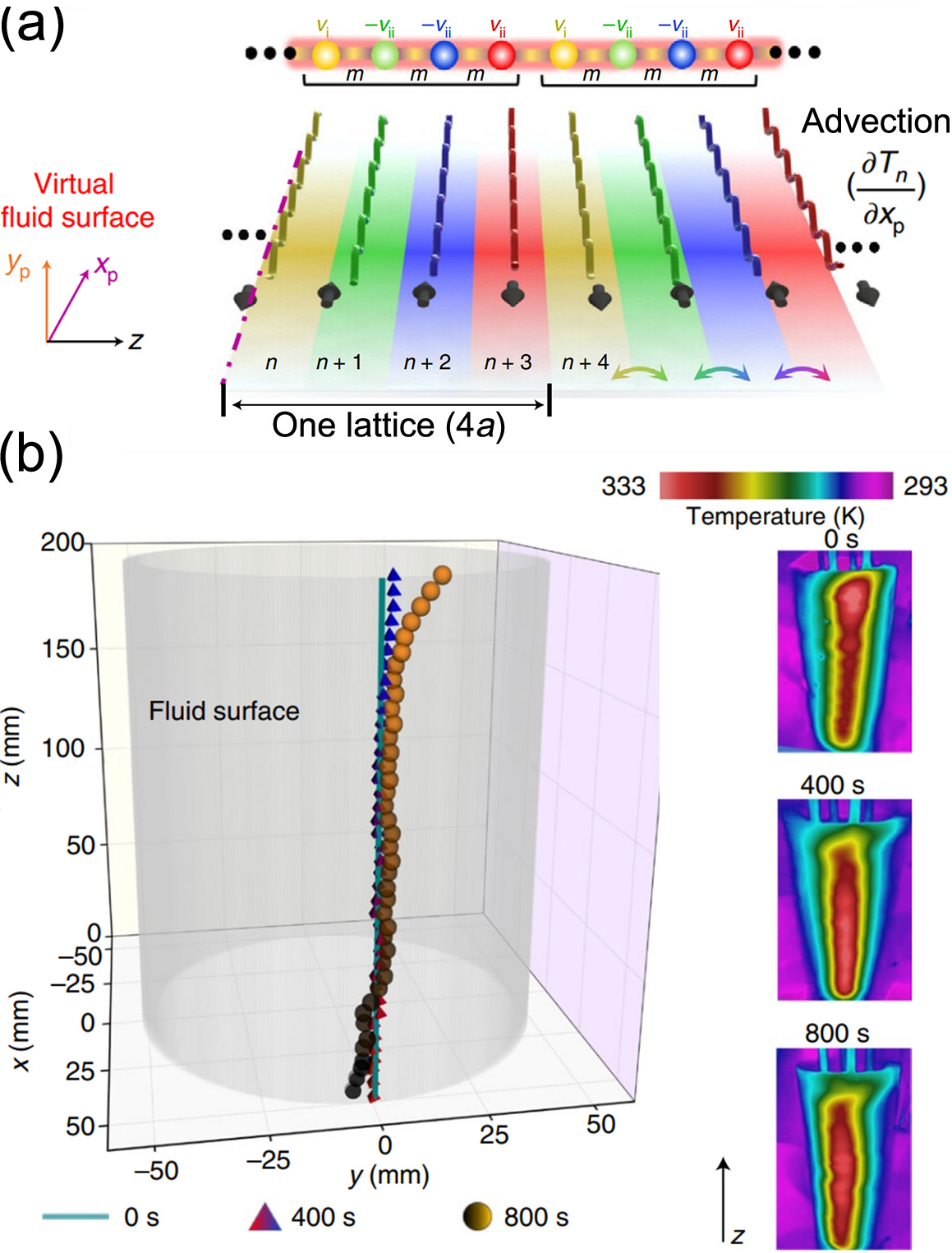}
	\caption{1D topological convective thermal insulators. (a) Planar fluid surface with periodic convection units in the virtual space ($x_{p}, y_{p}, z$). These periodic convection units are oriented in the $x_{p}$ direction and arranged in the $z$ direction to form a lattice of four units. $m$ denotes here the unit heat exchange rate. (b) The experimental locations of the maximum temperature points for each unit at the specific moments with topological edge state. The corresponding temperature profiles are shown aside. Adapted from~\citealp{XuNP22}.}
	\label{XuNP22}
\end{figure}

(iii) The third advantage is that the introduction of heat convection can circumvent the disadvantage of only being able to observe the temperature field of the slowest decaying branch in pure diffusion systems. Heat convection is a wavelike field that induces wave topology rather than diffusive topology. Many interesting and important topological states can be realized in diffusion systems. For instance, the thermal behavior of the topological edge state can be studied by tracking the temperature evolution in pure conduction systems, as mentioned above. However, the temperature field for the edge state decays quickly with no obvious observation, making it difficult to achieve localized heat management. With the help of convection modulation, a recent work has reported the observation of the topological edge state for a 1D system with a robust temperature profile~\cite{XuNP22}.

Following the work of Xu and coworkers, we consider a planar fluid surface in the virtual space ($x_{p}, y_{p}, z$) [Fig.~\ref{XuNP22}(a)]. Periodic heat convection along the $x_{p}$ direction is imposed in regions marked by different colors. A lattice with four units along the $z$ direction are thus generated. Additionally, the two ends for one unit along the $x_{p}$ direction should be connected to form a ring in the transformed space ($x, y, z$) for practical implementation. This new setup is the coupled ring chain structure, as mentioned above. This system has nontrivial topological properties, with a pair of edge states under the convection arrangement ($v_{\romannumeral1}, -v_{\romannumeral2}, -v_{\romannumeral1}, v_{\romannumeral2}$). The locations of maximum temperature points and temperature profiles at the specific moments are presented in Fig.~\ref{XuNP22}(b) for this arrangement. In this case, the thermal profile is robust and stationary with some slight deviations relative to the initial positions at the two system boundaries, which is a signature of topological edge state. 

\begin{figure}[!ht]
	\includegraphics[width=0.8\linewidth]{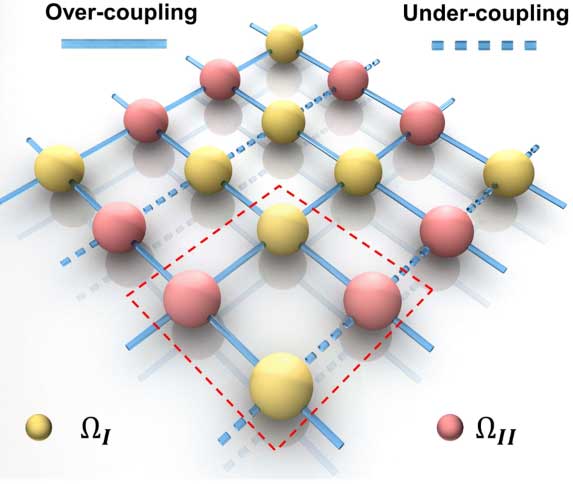}
	\caption{Schematic diagram of a square lattice with 16 sites. The red dashed border denotes a four-site unit cell. Adapted from~\citealp{XuNC23}.}
	\label{XuNC23}
\end{figure}

Recent work has demonstrated the extension of the heat convection-induced topological state to 2D systems~\cite{XuNC23}. As an example, Xu {\em et al}. consider an effective 2D square lattice with sixteen discrete sites sketched in Fig.~\ref{XuNC23}. The corresponding unit cell consists of four sites. The opposite convective velocities of the light-yellow and light-red sites are denoted as $\Omega_{{\rm{\uppercase\expandafter{\romannumeral1}}}}$ and $\Omega_{{\rm{\uppercase\expandafter{\romannumeral2}}}}$, respectively. Channel over- (under-)coupling are indicated by solid (dashed) lines, in
analogy with the positive (negative) hopping amplitudes required for a
quadrupole topological insulator. When tuning the ratio $\Omega_{{\rm{\uppercase\expandafter{\romannumeral2}}}}/\Omega_{{\rm{\uppercase\expandafter{\romannumeral1}}}}$ to a specific value, in-gap corner states can be found in the real spectrum. A Chern insulator with the chiral edge state was demonstrated in diffusion systems in a separate work~\cite{XuEPL21-2}. The basic unit there was the hexagonal
structure, where the vertex regions are solid pumps driving fluids with convective speed $v$. Based on this structure, a honeycomb lattice can be designed to emulate the Haldane model. Then a periodic temperature source is applied at the edge in analogy with an optical pump. Similar to the chiral edge mode, a steady temperature field will propagate unidirectionally along the boundary of the system without backscattering. Moreover, Wettlaufer \emph{et al}.~\cite{WettlauferPRL17,WettlauferJFM21,WettlauferSM20,WettlauferPNAS22,WettlauferJGRO91,WettlauferJGRO92,WettlauferEPL15} made pioneering works in the field of climate science, revealing thermal convection transport mechanism affecting on global warming. Researchers in metamaterial physics notice the necessity of convection in thermal transport.

\section{Thermal conduction and radiation}

Thermal radiation is another major mode of heat transfer. Any object with a non-zero temperature emits thermal radiation. Therefore, like thermal convection, thermal radiation usually occurs in conjunction with thermal conduction. Joint manipulation of thermal conduction and radiation is presently a very active area of metamaterial research.

\subsection{Theory and transformation principles}

Thermal radiation is essentially electromagnetic waves that transformation optics can manipulate in principle. However, in certain cases, where the photon mean free path is far smaller than the system size (i.e., optically thick media), thermal photons have diffusive properties. Recently, Xu \emph{et al.} developed the transformation multithermotics (incorporating conduction and radiation) theory to control thermal radiation and conduction simultaneously~\cite{XuPRAp20}. They employed the Rosseland diffusion approximation to couple the two thermal mechanisms
through the following equation,
\begin{equation}\label{heatradia}
	\rho C\frac{\partial T}{\partial t} + \nabla\cdot\left[-\kappa\nabla T+ \left(-\frac{16}{3}\gamma^{-1}n^2\sigma T^3\nabla T\right)\right] = 0,
\end{equation}
where $\rho$, $C$, and $\kappa$ have been previously mentioned, while $\gamma$, $n$, and $\sigma$ are the Rosseland mean absorption coefficient, relative refractive index, and the Stefan-Boltzmann constant, respectively. This
equation is invariant under coordinate transformations with transformation
\begin{subequations}\label{transru-radia}
	\begin{align}
		\rho' C' &= \frac{\rho C}{\text{det}\bm{J}},\\
		\bm{\gamma}' &= \bm{J}^{-\dag}\gamma \bm{J}^{-1}\text{det}\bm{J},\\
		\bm{\kappa}'&= \frac{\bm{J}\kappa \bm{J}^{\dag}}{\text{det}\bm{J}}.	
	\end{align}
\end{subequations}

\subsection{Applications: Metamaterials and metadevices}

Xu \emph{et al.} applied their transformation theory to design thermal cloaks, concentrators, and rotators. The temperature fields with thermal conduction and radiation exhibit distinct behavior at different temperature ranges compared to pure thermal conduction. In the low-temperature range between 300 and 320~K, thermal conduction is the dominant mode of heat transfer, thus the temperature gradient remains uniform. As the temperature increases, thermal radiation begins to take effect, causing the isotherms to shift to the right. At very high temperatures, thermal radiation dominates and thermal conduction can be disregarded altogether.

\begin{figure}[!ht]
	\includegraphics[width=\linewidth]{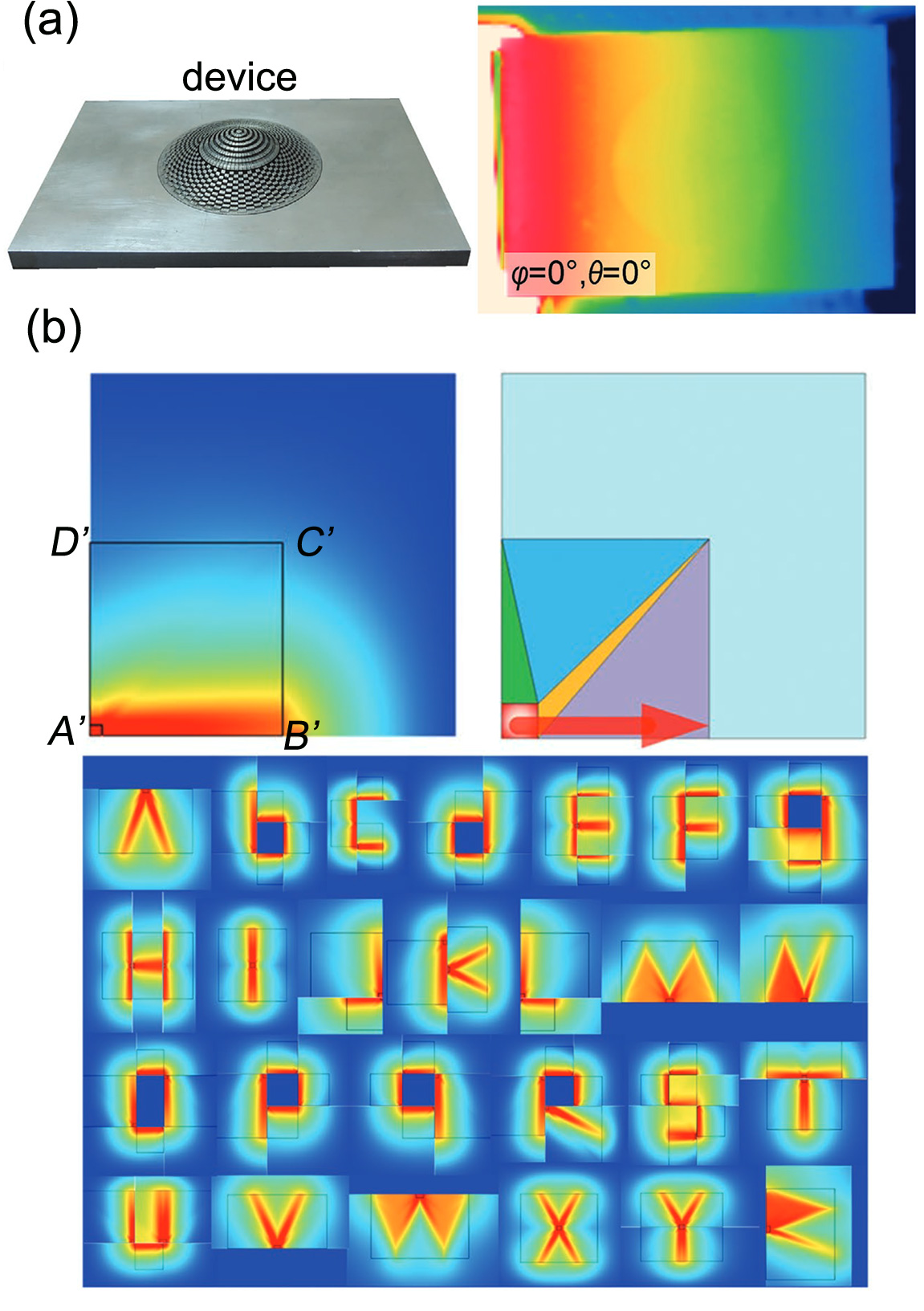}
	\caption{\label{fig-Thermetaconvradia} Joint manipulation thermal conduction and radiation. (a) 3D printed meta-helmet (right) and thermograph of the camouflaged surface (left). Adapted from~\citealp{PengAFM20}. (b) Schematic of the compartmentalized transformation (upper) and heat signatures of the entire alphabet (lower). Adapted from~\citealp{HuAM19}.}
\end{figure}

Radiating thermal metamaterials find application to thermal
imaging. One can control thermal conduction and radiation to achieve the desired radiation pattern~\cite{DedeAPL20}. For example, Li \emph{et al}. realized thermal radiative camouflage based on transformation thermotics~\cite{LiNC18}. They constructed a 2D structured camouflage surface through a two-fold transformation. Thus, the surface temperature field can be adjusted without influencing the background thermal conductivity. Subsequently, 3D thermal camouflage was proposed~\cite{PengAFM20}. They printed a 3D meta-helmet for wide-angle thermal camouflage. The thermograph in Fig.~\ref{fig-Thermetaconvradia}(a) demonstrates the camouflage capabilities of their device. Moreover, Hu \emph{et al}. reported encrypted thermal printing with compartmentalized transformation theory~\cite{HuAM19,ZhouIJHMT18} [Fig.~\ref{fig-Thermetaconvradia}(b)]. A point heat source was applied at point $A'$. By adjusting the thermal conductivities of rectangular regions according to transformation theory, they realized the whole alphabet, which has potential application value in thermal encoding. The heat was conducted directionally as a brush for thermal paintings.

\subsection{Thermal nonlinearities and near-field thermal radiation}

Thermal radiation is closely linked to temperatures, thus producing nonlinear effects. For instance, thermal radiation is much weaker than thermal conduction when the temperature is relatively low. Consequently, thermal radiation can be disregarded, and the system is in a linear region. As the temperature increases, thermal radiation becomes essential and affects the temperature field, and the system enters in a nonlinear region. This nonlinear characteristic was utilized to create a thermostat for flexible temperature control~\cite{OrdPRAP20}.

Compared with far-field thermal radiation~\cite{WangES19}, near-field radiative heat transfer with novel properties has also attracted considerable attention~\cite{SunES19,WangCPL21-2,TianCPL23,FerPRL21,GeCPL23}. Enhancing near-field radiation is an intriguing topic because the enhanced near-field coupling of evanescent waves can surpass the far-field limit predicted by the Stefan-Boltzmann law. In 2017, Fernandez-Hurtado \emph{et al}. achieved this goal by designing Si-based metasurfaces composed of 2D periodic arrays of holes~\cite{Fernandez-HurtadoPRL17}. Their work highlighted the importance of the bandwidth of surface-plasmon polaritons. Later, Du \emph{et al}. reported near-field heat transfer measurements between two hyperbolic metamaterials~\cite{DuNanoE20}. Their work experimentally validated the effective medium theory for near-field radiative metamaterials, stimulating the research of new metamaterials for thermophotonic devices~\cite{WangPRAp19,WangPRAp23,WangAIP22}.

\section{Thermal conduction, convection, and radiation}

In reality, the three fundamental modes of heat transfer always occur together. However, managing them simultaneously remains a challenge due to their distinct characteristics.

\subsection{Theory and transformation principles}

It is beneficial to unify the three fundamental modes within the framework of transformation theory. Xu \emph{et al}. performed this by incorporating the convective flux in Eq.~(\ref{heatradia})~\cite{XuESEE20}. In pure fluids, their omnithermotics equation reads
\begin{equation}\label{heatconvradia}
	\rho_f C_f\frac{\partial T}{\partial t} + \nabla\cdot\left[-\kappa_f\nabla T+ \rho_fC_f\bm{v}T+\left(-\frac{16}{3}\gamma_f^{-1}n_f^2\sigma T^3\nabla T\right)\right] = 0,
\end{equation}
where $\rho_f$, $C_f$, $\kappa_f$, and $\bm{v}$ are the density, heat capacity, thermal conductivity, and fluid velocity, respectively. This equation is form-invariant under the transformation rules
\begin{subequations}\label{transru-convradia}
	\begin{align}
		\rho_f' C_f' &= \frac{\rho_f C_f}{\text{det}\bm{J}},\\
		\bm{\gamma}_f' &= \bm{J}^{-\dag}\gamma_f \bm{J}^{-1}\text{det}\bm{J},\\
		\bm{\kappa}_f'&= \frac{\bm{J}\kappa_f \bm{J}^{\dag}}{\text{det}\bm{J}},\\
		\bm{v}' &= \bm{J}\bm{v}.	
	\end{align}
\end{subequations}
Directed by Eq.~(\ref{transru-convradia}), one can control the total heat flux at their discretion and design various thermal metadevices. Here and in the following, we use the term ``omnithermotics'' to encompass conduction,
convection, and radiation.

The parameters of transformation-based metamaterials are always anisotropic, and omnithermotics is no exception. Fortunately, certain conventional methods still prove effective, such as the layered structure design~\cite{XuPRAp20} and the scattering-cancellation technology~\cite{XuPRAp19-2}.

\subsection{Applications: Metamaterials and metadevices}

Wang \emph{et al.} recently proposed a reconfigurable metasurface~\cite{WangPRAp20}. They theoretically calculated the surface temperature by considering the three fundamental modes of heat transfer. With the radiation-cavity effect, they linked the surface emissivity to the depth of the unit hole. Therefore, the surface temperature can be adjusted by individually tuning the depth of its units. Such a discretized metasurface is reconfigurable for both infrared-light illusion and visible-light similarity.

\begin{figure}[!ht]
	\includegraphics[width=0.8\linewidth]{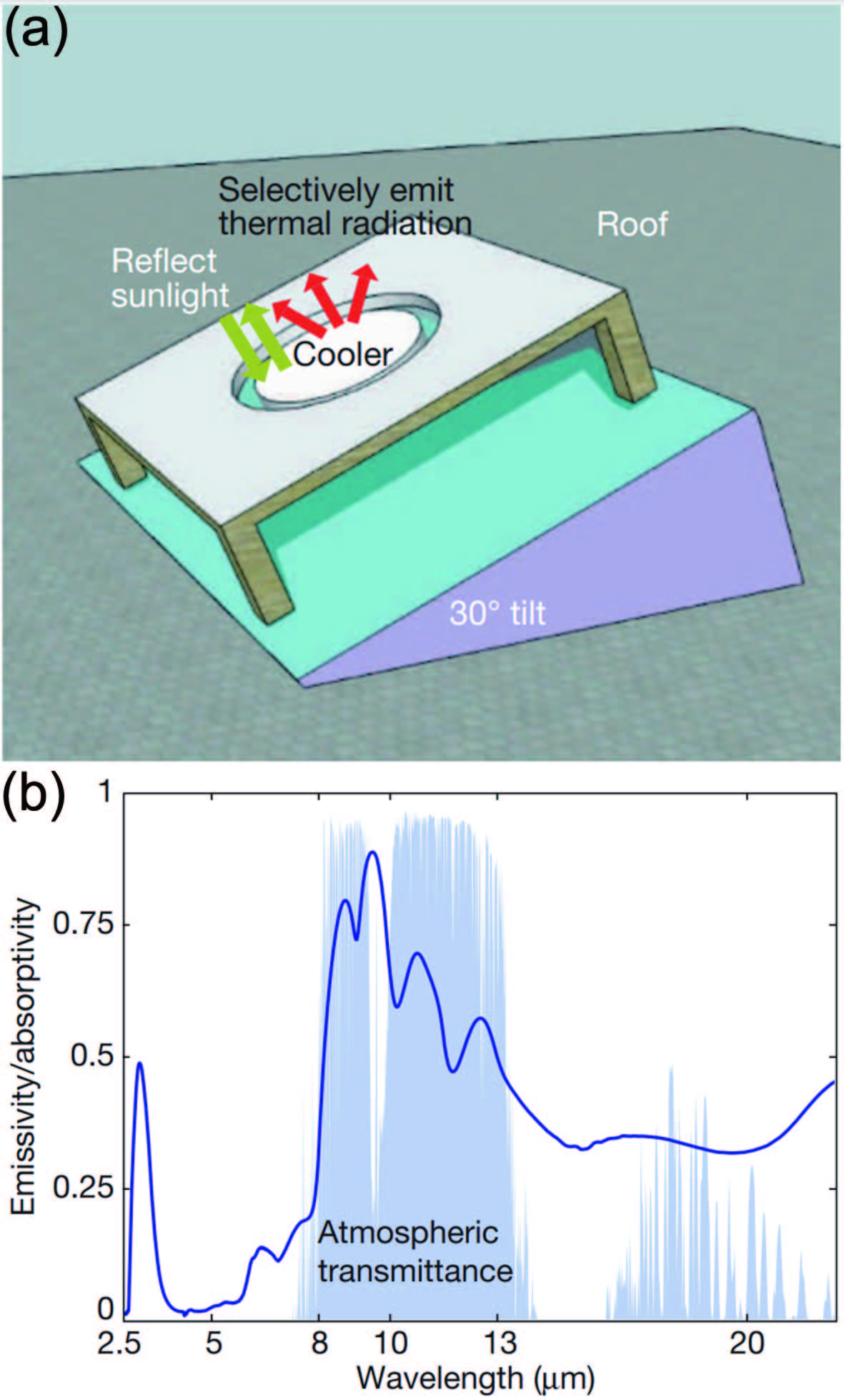}
	\caption{\label{fig-Thermetaconvradia-2} (a) Schematic of a radiative cooling device. (b) Measured emissivity; the shadow area marks the atmospheric transmission window. Adapted from~\citealp{RamanNat14}.}
\end{figure}

Radiative cooling is another key application for omnithermal systems. In 2014, Raman \emph{et al.} proposed a passive radiative cooling system under direct sunlight~\cite{RamanNat14}. The schematic illustration of their setup is shown in Fig.~\ref{fig-Thermetaconvradia-2}(a). To design selective thermal emission in the atmospheric window, they used $\rm{HfO_2}$ and $\rm{SiO_2}$ to construct a multilayered structure. Additionally, a well-sealed air pocket was designed to reduce the influence of thermal convection. The curve in Fig.~\ref{fig-Thermetaconvradia-2}(b) demonstrated the excellent performance of the radiative cooler device.

\subsection{Thermal nonlinearities}

Omnithermotics offers unprecedented possibilities to achieve advanced thermal control and explore novel physics. Following the work of
Xu {\em et al}. ~\cite{XuESEE20}. Yang \emph{et al.} developed an effective
medium theory and designed switchable omnithermal metamaterials that could be
manipulated between transparency and cloaking by controlling three basic
methods of heat transfer~\cite{YangJAP20}. According to the Stefan-Boltzmann law, radiative heat flow is quadratic in the temperature, and its strong nonlinearity is stimulating fresh interest in thermal topology problems. Introducing nonlinearities into classical topological models can lead to a variety of novel phenomena beyond the linear regime~\cite{MorimotoSA16,EzawaJPSJ21,ZhouNJP17,LiuPE17,XiaSci21,SmirnovaAPR20,HangPRA21,KirschNP21,EzawaPRR22,HuLSA21,BhallaPRB21,YuanLPR22,LeykamPRL16,EzawaPRB22,ChenOME14,MaczewskySci20,JurgensenNat21,HaldanePRL83,KosterlitzJPCSSP76}. In a topological heat radiation system, nonlinearity can be realized by converting the higher-order dependence of the radiation coefficient, thus enabling functions that linear thermal systems cannot. Additionally, non-Hermitian physics can be studied in heat radiative systems, due to the their energy exchange with the environment~\cite{Liuelight22}.

\section{Thermal and electric conduction}

Thermoelectric equipment is essential in industry and everyday life. Manipulating both thermal and electric fields in a single device is of critical importance. In this context, the thermal and electric
components of the thermoelectric field can be treated as acting separately
(decoupled) or interacting (coupled). When a device has both a temperature and a voltage difference, the thermal and electrical conductivity of the decoupled thermoelectric field control the thermal and electric fields independently, without affecting each other. On the other hand, the thermal and electric fields of a coupled thermoelectric field interact with each other and cannot be controlled separately. In other words, when the thermoelectric effect is taken into account, the coupling terms cannot be ignored, significantly increasing the complexity of the problem. 

\subsection{Theory and transformation principles}

The steady-state heat and electric conduction equations without heat or power sources can be formulated as
\begin{subequations}\label{Lei1}
	\begin{align}
	\nabla \cdot (\kappa \nabla T)=&0,\\
	\nabla \cdot (\sigma \nabla \mu)=&0,
	\end{align}
\end{subequations}
where $\kappa$ and $\sigma$ are thermal and electrical conductivities, respectively, $T$ the temperature, and $\mu$ the electric potential. Equations~(\ref{Lei1}) satisfy form invariance with transformation rules~\cite{LiJAP10}.
\begin{subequations}\label{Lei2}
	\begin{align}
	\bm{\kappa}'&=\frac{\bm{A}\kappa\bm{A}^{\dag}}{\det \bm{A}},\\ \bm{\sigma}'&=\frac{\bm{A}\sigma\bm{A}^{\dag}}{\det \bm{A}},
	\end{align}
\end{subequations}
where $\bm{\kappa}'$ and $\bm{\sigma}'$ are the transformed thermal and electrical conductivity, respectively. Here $\bm{A}$ is the relevant Jacobian transformation matrix. The physical parameters in the real space can be obtained by substituting the transformed coordinates into the Jacobian matrix.

In coupled thermoelectric fields, electricity and heat can be converted into each other. The heat flow generated by the potential difference is known as the Peltier effect. If the charge carriers flow from high to low energy levels,
an excess heat is released into the material. Conversely, if the charge
carriers move in the opposite direction, heat is absorbed from the
environment (i.e. the material is refrigerated). The current generated by a
temperature difference is known as Seebeck effect. Indeed, under a
temperature gradient, the charge carriers in the conductor move from a hot
to  cold end and accumulate there, thus causing an electric potential
difference, which, in turn, induces an electric current in the opposite
direction. In the case of the thermoelectric effect, the thermal and electric fields are coupled by the Seebeck coefficient, which is used to characterize the strength of the Seebeck effect. When a temperature and a voltage difference are applied simultaneously to a medium, coupled heat and current are separately induced by each other in addition to their independent transport. In the linear regime, the coupling between the electric, $\bm{J}_E$, and the heat
current density, $\bm{J}_Q$, is expressed as
\begin{subequations}\label{Lei3}
	\begin{align}
		{\boldsymbol J}_E&=-\sigma\nabla\mu - \sigma S \nabla T,\\
		{\boldsymbol J}_Q&=-\kappa\nabla T+T S^{\dag}{\boldsymbol J}_E,
	\end{align}
\end{subequations}
where $S$ is the Seebeck coefficient, and $S^T$ its transpose. Accordingly, Eqs. (\ref{Lei1}) can be rewritten as
\begin{subequations}\label{Lei4}
	\begin{align}
		\nabla \cdot \boldsymbol J_E&=0,\\
		\nabla \cdot \boldsymbol J_Q&=-\nabla\mu \cdot \boldsymbol J_E.
	\end{align}
\end{subequations}
These two equation express the local conservation in the steady state
of charge and energy, respectively. The thermoelectric coupling transport produces a heat source term, $-\nabla\mu \cdot {\boldsymbol J}_E$, corresponding
to the Joule effect. In view of Onsager reciprocal relations~\cite{OnsPR1931}, electric and thermal conductivity matrices should be symmetric, $\sigma=\sigma^\dag$ and  $\kappa= \kappa^{\dag}$. Substituting Eqs.~(\ref{Lei4}) into Eqs.~(\ref{Lei3}), the complete governing equations for the thermoelectric effect can be formulated as
\begin{subequations}\label{Lei5}
	\begin{align}
		\nabla \cdot \left(\sigma\nabla\mu +\sigma S\nabla T\right)&=0,\\
		-\nabla \cdot \left(\kappa\nabla T+ T S^{\dag}\sigma S\nabla T
		+ T S^{\dag} \sigma\nabla\mu\right)
		&=\nabla\mu \cdot \left(\sigma\nabla\mu +\sigma S\nabla T\right).
	\end{align}
\end{subequations}

This equation satisfies formal invariance under coordinate transformation~\cite{StedmanSR17,ZhuangIJMSD23}, and the relevant transformation rules can be written as
\begin{equation}\label{Lei7}
	\begin{split}
		\bm{\kappa}'(r')&=\frac{\bm{A}\kappa{\bm{A}}^{\dag}}{\det{\bm{A}}},\\
		\bm{\sigma}'(r')&=\frac{\bm{A}\sigma{\bm{A}}^{\dag}}{\det{\bm{A}}},\\
		S'(r')&={\bm{A}}^{-\dag}S{\bm{A}}^{\dag},
	\end{split}
\end{equation}
where $\kappa'(r')$, $\sigma'(r')$, and $S'(r')$ are the transformed thermal conductivity, electrical conductivity, and Seebeck coefficient, respectively. Compared with the decoupled thermoelectric field, the transformation of the Seebeck coefficient should be taken into account in the coupled thermoelectric field.

\subsection{Applications: Metamaterials and metadevices}

\begin{figure}[!ht]
	\centering
	\includegraphics[width=\linewidth]{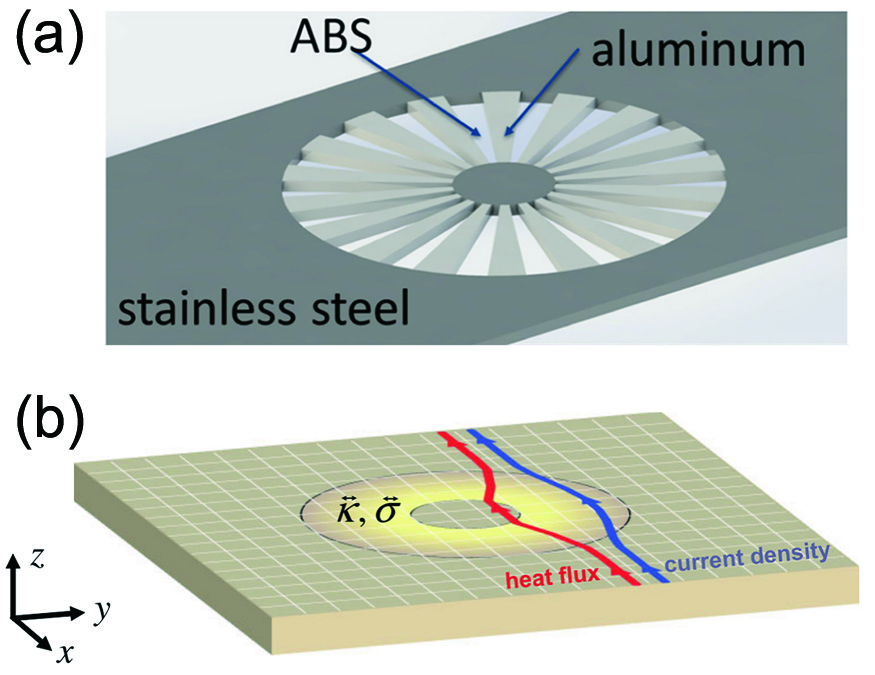}
	\caption{Multifunctional devices for decoupled thermoelectric fields. (a) Experimental realization of a thermal and electric concentrator in a single device. The concentrator shell alternately consists of 18 wedges of material A (acrylonitrile butadiene styrene) and 18 wedges of material B (aluminum). Adapted from~\citealp{LanOE15}. (b) Schematic of implementing thermal concentration and electric cloaking in a single device. Adapted from~\citealp{MocciaPRX14}.}
	\label{FigL2}
\end{figure}

For practical purposes, finding the anisotropic thermoelectric materials required by the transformation theory is challenging. Therefore, in thermoelectric metamaterials, the effective medium theory and scattering-cancellation theory are often used. We first discuss the function of a decoupled thermoelectric field~\cite{LanAPL16,LanOE16-2,ZhangCS20,ZhangOME18}. A thermal and electric concentrator was first realized
experimentally by returning to the effective medium
theory~\cite{LanOE15}. In the proposed geometry, the concentrator shell
consists of wedges of two different materials that are alternately arranged to form
an annulus [Figs.~\ref{FigL2}(a)]. Electric and thermal fields can also achieve different functions. Thermal concentration and
electric cloaking were simultaneously realized by embedding a mixture of different shapes and materials in the host medium~\cite{MocciaPRX14} [Figs.~\ref{FigL2}(b)]. An invisible sensor that can simultaneously sense and hide thermal and electric fields is proposed and realized~\cite{YangAM15}. Li \emph{et al}. theoretically formed a composite cloak shell by distributing non-spherical nanoparticles with different shapes and volume fractions along the cloak radius. Based on the effective medium theory, the effective electric and thermal conductivity were shown to meet the cloaking requirement calculated through the transformation theory~\cite{LiJAP10}. Thermal and electric cloaks were realized experimentally in a bilayer structure designed by scattering-cancellation theory~\cite{MaPRL14}.

In addition to temperature-dependent transformation theory, functional
switching can be obtained through {\em ad hoc} geometries of the
thermoelectric metamaterials. Adjusting the functions of multiple physical fields requires fine tuning many parameters simultaneously, which is
experimentally very difficult. Therefore,  Lei {\em et al.} brought forward the
notion of spatiotemporal multiphysics metamaterial to expand the functional
parameter space~\cite{Lei23-1}. Their spatiotemporal metadevice consists of a
rotating checkerboard structure, where the even layers rotate to modulate its
geometry and, therefore, continuously vary its functions.

When temperature and electric potential differences are applied to the two ends of the structure,
background thermal and electric currents flow through it. Their direction in the central region determines the function of thermal and electric fields, as can
be predicted by calculating the effective thermal and electric conductivity. Using the effective medium theory layer by layer, the effective thermal and electric conductivity of the checkerboard structure can be obtained. The chessboard structure of two (four) materials makes the thermal and electric fields switch over time with the same (different) function. Two-material (four-material) spatiotemporal metamaterials allow multiphysics fields to achieve three (five) functional combinations in a single device. The simulation further verifies the theoretical prediction. The multifunctional combination of multiple physical fields can also be adjusted by controlling the checkerboard material composition, geometry, and rotation time.

Transformation theory also suggests how the basic functions of
cloaking, concentration, and rotation apply to a coupled thermoelectric field~\cite{LeiEPL21,ShiJPE19}. A bilayer structure based on the generalized scattering-cancellation method can effectively reduce the difficulty of preparing thermoelectric metamaterials with the parameters
required by transformation theory~\cite{QuEPL21}. Two practical applications in coupled thermoelectric fields are proposed based on the temperature-dependent transformation theory~\cite{LeiEPL21}: a switchable thermoelectric device, which adapts to
the ambient temperature by switching between cloak and concentrator mode; an improved thermoelectric cloak that keeps the internal temperature nearly constant. In the coupled thermoelectric field, the thermal and electric fields are difficult to control independently due to coupling terms. This conundrum can be solved within a general computational scheme for designing composite materials that follow the simple circuit theory of in-parallel and in-series connected transmission properties~\cite{ShiJAP20}.

\subsection{Nonlinear thermoelectric effects and effective thermoelectric metamaterials toward high-energy conversion efficiency}

To improve the adaptability and versatility of the actual complex scene, multifield transformation theory can be generalized to
account for nonlinear effects. Nonlinear transformation theory refers to temperature-dependent or temperature-independent coordinate transformations of temperature-dependent background parameters. For the thermoelectric effect coupled by the Seebeck coefficient, the governing equations coincide with Eq.~(\ref{Lei5}), considering the temperature-dependent thermal conductivity $\kappa(T)$, electrical conductivity $\sigma(T)$, and Seebeck coefficient $S(T)$. According to this theory, the transformation rule can be expressed as~\cite{LeiEPL21}
\begin{subequations}\label{Lei8}
	\begin{align}
		\bm{\kappa}'(r,T)&=\frac{\bm{A}\kappa(T){\bm{A}}^{\dag}}{\det{\bm{A}}},\\
		\bm{\sigma}'(r,T)&=\frac{\bm{A}\sigma(T){\bm{A}}^{\dag}}{\det{\bm{A}}},\\
		S'(r,T)&={\bm{A}}^{-\dag}S(T){\bm{A}}^{\dag},
	\end{align}
\end{subequations}
where $\kappa(T)$, $\sigma(T)$, and $S(T)$ are the temperature-dependent background parameters, the transformed parameters $\bm{\kappa}'(r,T)$, $\bm{\sigma}'(r,T)$, and $S'(r,T)$ are related to temperature and the transformed coordinates. The temperature-dependent transformation theory applies to the decoupled thermoelectric field, too, upon setting $S(T)=0$~\cite{JiaCry19}.


In summary, both temperature-independent and temperature-dependent transformation theories can be used to analyze decoupling and coupling thermoelectric fields. By designing appropriate coordinate transformations, the thermal and electric fields can be manipulated to achieve desired functions.

Thermoelectric conversion technology utilizes the Seebeck effect (thermoelectric generation) and the Peltier effect (electrified refrigeration) to convert electricity into heat. This technology is widely used, from waste heat recovery in motor vehicles to satellite powering
in deep space exploration. The thermoelectric conversion efficiency is an important parameter to measure the performance of thermoelectric materials, which is determined by the thermoelectric performance value $zT=\alpha^2\sigma T/\kappa$. Thermoelectric materials with excellent performance should have a large thermoelectric force $\alpha$, high conductivity $\sigma$, and low thermal conductivity $\kappa$. Research on thermoelectric materials focuses on improving efficiency, reducing cost, and expanding the application range of thermoelectric devices. The thermoelectric parameters are intertwined in the coupled thermoelectric field, and one parameter cannot be changed unilaterally. Effectively regulating coupled thermoelectric parameters is key to improving the $zT$ value and conversion efficiency. Tin sulfide (SnS) crystal materials have the essential characteristics of effective thermoelectric materials due to their abundant reserves, low cost, and friendly environment. He \emph{et al}. used the evolution law of multiple energy bands of SnS with temperature to optimize the contradictions between effective mass and mobility, thereby achieving high thermoelectric properties (the maximum $zT$ of SnS$_{0.91}$Se$_{0.09}$ crystal at 873~K is close to 1.6)~\cite{HeSCI19}. The conversion efficiency can be further improved by doping. Bromine doping in tin selenide (SnSe) can maintain low thermal conductivity in the out-of-plane direction of the layered material, making it another promising thermoelectric material~\cite{ChangSCI18}. Pore-doped tin selenide (SnSe) crystals have reached $zT$ values ranging between 0.7 and 2.0 for temperatures from 300 to 773~K~\cite{ZhaoSCI16}. The improved conversion performance of thermoelectric materials has brought new vitality to the research of thermoelectric metamaterials. It is relatively
simple to couple thermal and electric fields in order to have them perform
the same function; vice versa, much harder is realizing simultaneously
different field functions, such as thermal cloaking and electric
concentration.

Finally, the Seebeck-driven transverse thermoelectric generation has recently gained attention. Most existing thermoelectric applications are based on the Seebeck effect alone, which limits the flexibility and durability of the devices. Zhou \emph{et al.} addressed this issue by combining Seebeck and anomalous Hall effect~\cite{ZhouNM21,ZhouNC23,YamamotoJAP21}. They designed a closed circuit with thermoelectric and magnetic materials and used the Seebeck-induced charge current to drive transverse thermoelectric generation. Their approach demonstrated a significant increase in transverse thermopower.

\section{Particle diffusion}

Particle diffusion is a ubiquitous phenomenon in systems driven by a concentration gradient and is typically described by the Fick equations~\cite{PeterRMP09}. Although the transformation theory is not perfectly applicable to the Fick second equation, mass diffusion cloaking can be achieved by employing the low-diffusivity approximation~\cite{GuenneauJRSI13}. This work sparked an increased interest in the research of transformation particle diffusion. In 2014, Schittny \emph{et al}. reported a water-based invisibility cloak for light diffusion propagation~\cite{SchittnyScience14}. By using the scattering-cancellation method, they fabricated cylindrical invisibility cloaks with thin shells made of polydimethylsiloxane doped with melamine-resin microparticles. Subsequently, they discussed the transient case and experimentally demonstrated the robustness of their diffusive light cloaking in static and quasi-static regimes~\cite{SchittnyOpt15}. This light diffusion cloaking overcomes one of the limitations of conventional electromagnetic cloaking and provides a new direction for designing electromagnetic cloaking~\cite{SchittnyLPR16}.

\subsection{Theory and transformation principles}

We introduce now the transformation theory of particle diffusion. The Fick Second Equation is used to describe the diffusion process in the absence of external sources,
\begin{equation}\label{particlediff}
	\frac{\partial c}{\partial t} = \nabla\cdot(D\nabla c - \bm{v}c),	
\end{equation}
where $c$, $D$, and $\bm{v}$ are the particle concentration, diffusivity, and advection velocity, respectively. The transformed equation is expressed as
\begin{equation}\label{particlediff'}
	\frac{1}{\text{det}\bm{J}}\frac{\partial c}{\partial t} = \nabla'\cdot(D''\nabla' c - \bm{v}''c),	
\end{equation}
where $\bm{D}'' = \bm{J}D\bm{J}^\dag/\text{det}\bm{J}$ and $\bm{v}'' = \bm{J}\bm{v}/\text{det}\bm{J}$. The additional metric term $1/\text{det}\bm{J}$ breaks the transformation invariance of Eq.~(\ref{particlediff}). Therefore, transformation particle metadevices do not work ideally. To address this issue, an optimized theory was proposed to control transient particle diffusion~\cite{ZhangATS22}. This theory transformed the starting Fick equation as
\begin{equation}\label{particlediff-opti}
	\frac{\partial c}{\partial t} = \nabla'\cdot(D'\nabla' c - \bm{v}'c),	
\end{equation}
with $\bm{D}' = \bm{J}D\bm{J}^{\dag}$ and $\bm{v}' = J\bm{v}$. Equation~(\ref{particlediff-opti}) has the same form as Eq.~(\ref{particlediff}) under low diffusivity and velocity approximation. Consequently, they designed four typical devices for separating, cloaking, concentrating, and rotating chemical waves. 

\subsection{Applications: Metamaterials and metadevices}

Particle diffusion plays a key role in chemical and biological systems. With diffusion metamaterials, Restrepo-Fl$\acute{\rm o}$rez \emph{et al}. reported cloaking and focusing of particle diffusion~\cite{RestrepoAPL17,RestrepoJPDAP17} in the delivery of water-soluble drugs by concentric liposomes. Furthermore, Zhou \emph{et al}. achieved binary mixture manipulation through a bilayer metadevice~\cite{ZhouCS21}. They separated $\rm{O_2}$ and $\rm{N_2}$ in the central region by exploiting the different diffusivities of these two particles in the same medium.

One can realize particle separation by selecting suitable materials to cloak one kind of particle and concentrate the other. Particle separation can also be achieved by metamaterial membranes~\cite{RestrepoJPDAP17,LiuJPCL20,ChenACIE21}. Recently, Li \emph{et al}. reported a ``Plug and Switch'' modulus to design metamaterials for controlling ion diffusion~\cite{LiAS22}. They employed the scattering-cancellation theory to obtain the parameters of the cloak, concentrator, and selector modes. Then they resorted to the effective medium theory to calculate the diffusivity of resins and filling media (the materials used). To experimentally obtain the switchable particle devices, they adopted fixed backgrounds and ``plug and switch'' central region. All parts were fabricated by 3D printing. The resin was chosen as zero-index material, and copper sulfate solution was the filling media. For the selector, they adopted fan-shaped structures filled with ion-exchange resin to achieve selection through selective particle penetration.

Particle diffusion metamaterials have various applications. For instance, a monolayer cloak based on scattering-cancellation was designed~\cite{LiAFM20} to improve the electrical conductivity by eliminating electron scattering. Furthermore, numerical algorithms are also employed to design diffusion metamaterials, providing a useful approach to facilitate the realization of experiments~\cite{Khodayi-mehrIEEE20,AvanziniPRE20}.

Recently, there has been progress in the development of particle diffusion-based intelligent devices. Zhang \emph{et al}. reported a chameleon-like metashell based on transformation-invariant metamaterials~\cite{ZhangPRAp23}. They assumed chemical components with the
same solubility and neglected effects due to different chemical potentials. As a result, the passively stable mass diffusion process is governed by the Fick law~\cite{FickJMS95},
\begin{equation}\label{fick}
	\nabla\cdot\left(-D\nabla c\right)=0,
\end{equation}
where $c$ is the volume concentration and $D$ is the diffusivity. Consider an arbitrary 2D transformation expressed in cylindrical coordinates~\cite{PendryScience06,LeonhardtScience06},
\begin{subequations}\label{act}
	\begin{align}
		r' &=R(r,\theta),\\
		\theta' &=\Theta(r,\theta),
	\end{align}
\end{subequations}
where $R(r,\theta)$ and $\Theta(r,\theta)$ are certain functions of radius $r$ and angle $\theta$, the transformed diffusivity $\bm{D}'$ is $\bm{D}'=\bm{J}D\bm{J}^{\dag}/\det \bm{J}$, with $\bm{J}$ being the Jacobian transformation matrix. Or more explicitly,
\begin{equation}\label{d}
	\begin{split}
	\bm{D}'&=\frac{1}{\det \bm{J}}\times\\
	&\left[
	\begin{matrix}
		\begin{smallmatrix}
			D_{rr}\left(\dfrac{\partial R}{\partial r}\right)^2+D_{\theta\theta}\left(\dfrac{\partial R}{r\partial \theta}\right)^2 & D_{rr}\dfrac{\partial R}{\partial r}\dfrac{r'\partial \Theta}{\partial r}+D_{\theta\theta}\dfrac{\partial R}{r\partial \theta}\dfrac{r'\partial \Theta}{r\partial \theta}\\
			D_{rr}\dfrac{\partial R}{\partial r}\dfrac{r'\partial \Theta}{\partial r}+D_{\theta\theta}\dfrac{\partial R}{r\partial \theta}\dfrac{r'\partial \Theta}{r\partial \theta} & D_{rr}\left(\dfrac{r'\partial \Theta}{\partial r}\right)^2+D_{\theta\theta}\left(\dfrac{r'\partial \Theta}{r\partial \theta}\right)^2
		\end{smallmatrix}
	\end{matrix}
	\right],
	\end{split}
\end{equation}
where $D_{rr}$ and $D_{\theta\theta}$ are the radial and tangential components of diffusivity, respectively. Consequently, the eigenvalues of $\bm{D}'$ can be obtained as
\begin{subequations}\label{eigen}
	\begin{align}
		\lambda_1 &\approx \frac{D_{rr}}{\det \bm{J}}\left[\left(\dfrac{\partial R}{\partial r}\right)^2+\left(\dfrac{r'\partial \Theta}{\partial r}\right)^2\right],\\
		\lambda_2 &\approx \frac{D_{\theta\theta}}{\det \bm{J}}.
	\end{align}
\end{subequations}
For transformation-invariant metamaterials with $D_{rr}\to \infty$ and $D_{\theta\theta}\to 0$ for transformation-invariant metamaterials, Eq.~(\ref{eigen}) can be further simplified to
\begin{subequations}\label{eigen2}
	\begin{align}
		\lambda_1 &\approx \infty,\\
		\lambda_2 &\approx 0.
	\end{align}
\end{subequations}
This result indicates that the eigenvalues are invariant under any coordinate transformation, meaning that the transparency of transformation-invariant metamaterials is robust under coordinate transformations.

Zhang \emph{et al}. further demonstrated the chameleon-like behavior of transformation-invariant metamaterials for particle diffusion. They divided the system into three regions with diffusivities $\bm{D}_1$, $\bm{D}_2$, and $\bm{D}_3$. To this purpose, they rewrote Eq.~(\ref{fick}) directly in
cylindrical coordinates,
\begin{equation}\label{cc}
	\frac{1}{r}\dfrac{\partial}{\partial r}\left(rD_{rr}\dfrac{\partial c}{\partial r}\right)+\frac{1}{r}\dfrac{\partial}{\partial \theta}\left(\frac{D_{\theta\theta}}{r}\dfrac{\partial c}{\partial \theta}\right)=0.
\end{equation}
For the transformation-invariant metashell, ${D}_1={D}_3={\rm constant}$, and $\bm{D}_2=\text{diag}\left(D_{rr}, D_{\theta\theta}\right)$. Hence, the effective diffusivity of the core-shell structure can be calculated as~\cite{XuEPJB19,XuPRAp19-2,XuSC20}.
\begin{equation}\label{effd}
	D_e=\eta D_{rr}\frac{\left(1+\xi^{\eta}\right)D_3+\left(1-\xi^{\eta}\right)\eta D_{rr}}{\left(1-\xi^{\eta}\right)D_3+\left(1+\xi^{\eta}\right)\eta D_{rr}},
\end{equation}
with $\xi=r_1^2/r_2^2$ and $\eta=\sqrt{D_{\theta\theta}/D_{rr}}$. The effective diffusivity can be approximatd as $D_e\approx D_3$ since $D_{rr}\approx \infty$ and $D_{\theta\theta}\approx 0$, which indicates that it can adapt to the environment. Zhang \emph{et al}. referred to this as chameleon-like behavior. Furthermore, it should be noticed that this metashell does not require external energy, making it suitable for advanced intelligent devices. Furthermore, Zhang \emph{et al}. designed an irregular-shaped chameleonlike concentrator and a circular chameleonlike rotator, to demonstrate the advantages of transformation-invariant metashells over conventional devices. To design the irregular-shaped chameleonlike concentrator, they twisted the annulus region
into an irregular shape in the virtual space by means of the coordinate transformation
\begin{subequations}\label{cont}
	\begin{align}
		r' &=r\Gamma(\theta),~~~r_1<r<r_2\\
		\theta' &=\theta,
	\end{align}
\end{subequations}
where $\Gamma(\theta)=1.2+0.5\sin(\theta)+0.5\cos(2\theta)$. $r_1$ and $r_2$ are the inner and outer radii of the annulus, respectively. Then the transformed diffusivity of the metashell can be obtained by referring to Eqs.~(\ref{d}) and~(\ref{cont}).

To compare the capabilities of standard devices and chameleon-like devices, they calculated the coordinate transformation for the standard concentrator with the same shape,
\begin{subequations}\label{ntc}
	\begin{align}
		r' &=\frac{r_1}{r_m}r,~~~r'<r_1\Gamma(\theta),\\
		r' &=\alpha r+\beta\Gamma(\theta),~~~r_1\Gamma(\theta)<r'<r_2\Gamma(\theta),\\
		\theta' &=\theta ,
	\end{align}
\end{subequations}
where $r_m$ is a constant, $\alpha=(r_2-r_1)/(r_2-r_m)$, and $\beta=(r_1-r_m)r_2/(r_2-r_m)$. Thus, the transformed diffusivity of the normal concentrator is given by~\cite{BaratiSedehPRAp20}
\begin{equation}\label{dntc}
	\bm{D}'=D_3
	\left[
	\begin{matrix}
		D_{11} & D_{12}\\
		D_{21} & D_{22}
	\end{matrix}
	\right],
\end{equation}
where the components $D_{ij}$ are expressed as
\begin{subequations}\label{ntc-2}
	\begin{align}
		D_{11} &=\frac{(r_2-r_m)r'+r_2(r_m-r_1)\Gamma(\theta)}{(r_2-r_m)r'}\nonumber\\
		 &+\frac{r_2^2(r_m-r_1)^2\left(d\Gamma(\theta)/d\theta\right)^2}{\left[(r_2-r_m)^2r'+r_2(r_m-r_1)(r_2-r_m)\Gamma(\theta)\right]r'},\\
		D_{12} &=D_{21} =\frac{r_2(r_1-r_m)(d\Gamma(\theta)/d\theta)}{(r_2-r_m)r'+r_2(r_m-r_1)\Gamma(\theta)},\\
		D_{22} &=\frac{(r_2-r_m)r'}{(r_2-r_m)r'+r_2(r_m-r_1)\Gamma(\theta)}.
	\end{align}
\end{subequations}
A larger ratio $r_m/r_1$ indicates a higher convergence effect. For the
transformation-invariant concentrator, the ratio of $r_m/r_1$ approaches the
maximum value possible, $r_2/r_1$, which implies a larger concentration
gradient in the core region compared to the normal concentrator.

For the circular chameleon-like rotator, the coordinate transformations for both the chameleon-like rotator and the regular rotator are identical,
\begin{subequations}\label{rot}
	\begin{align}
		r' &=r,\\
		\theta' &=\theta+\theta_0\dfrac{r-r_2}{r_1-r_2},~~~r_1<r<r_2,
	\end{align}
\end{subequations}
where $\theta_0$ is the rotation angle, this coordinate transformation can be explained by rotating the space in which the annulus is located in the virtual space. Similarly, they derived the transformed diffusivity of the chameleon-like rotator according to Eqs.~(\ref{d}) and (\ref{rot}). The designed rotator has both converging and rotating mass flow functions. For the normal rotator, the transformed diffusivity is, instead,
\begin{equation}\label{dntr}
	\bm{D'}=D_3
	\left[
	\begin{matrix}
		1 & \dfrac{r'\theta_0}{r_1-r_2}\\
		\dfrac{r'\theta_0}{r_1-r_2} & 1+\left(\dfrac{r'\theta_0}{r_1-r_2}\right)^2
	\end{matrix}
	\right].
\end{equation}

\begin{figure}[!ht]
	\includegraphics[width=1.0\linewidth]{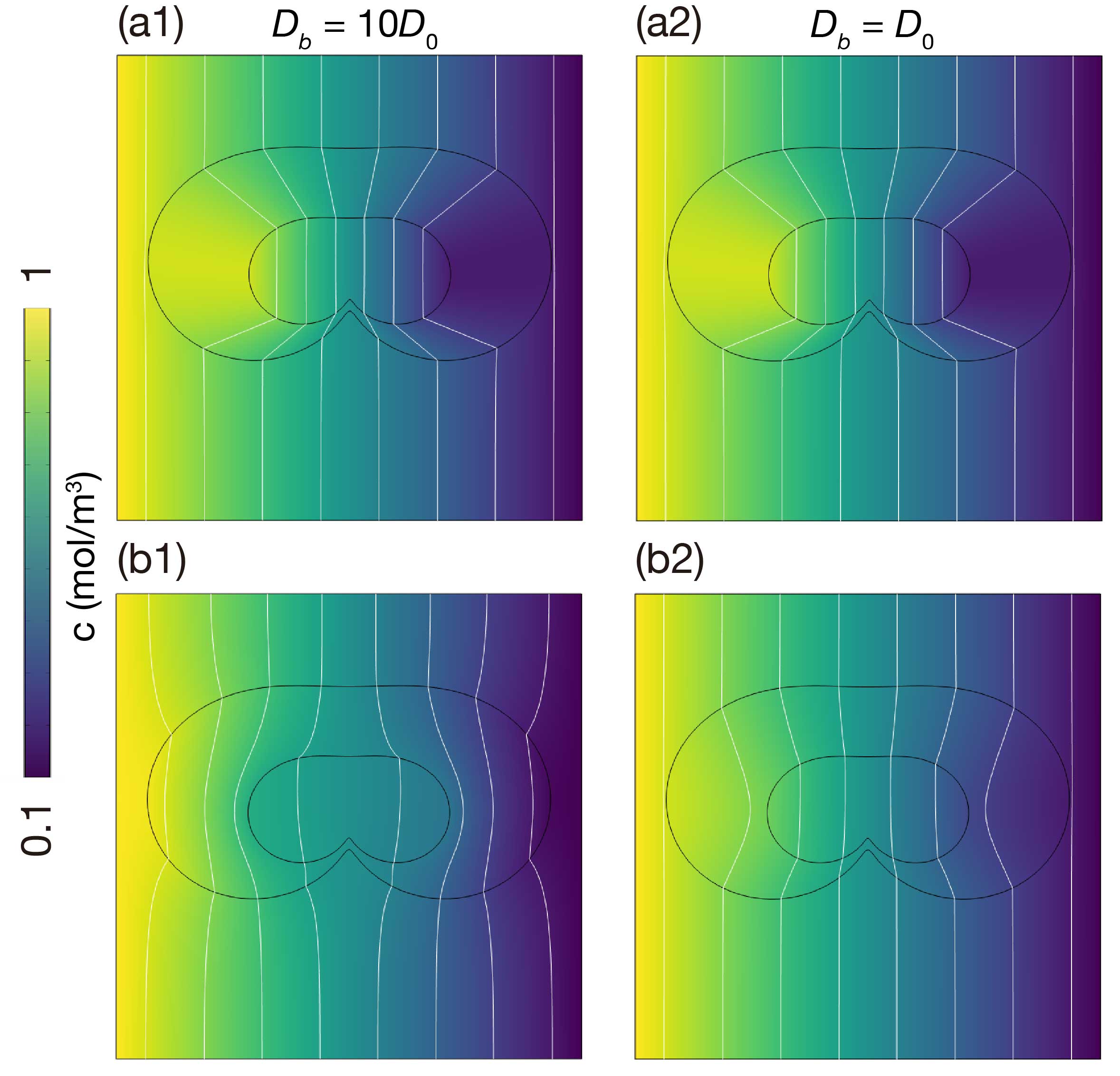}
	\caption{(a1)-(a2) Simulations of the transformation-invariant concentrator. (b1)-(b2) Simulations of the normal-transforming concentrator. The background diffusivity in the first and second column is $10D_0$ and $D_0$, respectively. Adapted from~\citealp{ZhangPRAp23}.}
	\label{tifig2}
\end{figure}

Simulations to validate the theory prediction are presented in Fig.~\ref{tifig2}. Figure~\ref{tifig2} compares the performance of transformation-invariant and normal-transforming metashells for different values of the background diffusivity, $D_b$. In particular simulated mass flows along the $x$
axis of a transformation-invariant (first row) and a
normal-transforming concentrator (second row) are compared. Concentration distributions
are displayed for different values of the background diffusivity, namely, $D_b=10D_0$ and $D_0$, respectively in the first and second. In Fig.~\ref{tifig2}(b1), the concentration in the device inner
region is affected by the increase or decrease of $D_b$, pointing to the
ineffectiveness of the normal concentrator. Conversely, the
transformation-invariant concentrator remains effective regardless of whether
the background diffusivity is varied.

Although materials with extreme anisotropy are rare, two isotropic materials with different diffusivities can be used to achieve a similar effect. Zhang \emph{et al}. used this method to design experimentally more accessible chameleon-like concentrators and rotators with a multilayered structure. As an example, the concentrator is composed of a background medium with a chessboard structure and a composite metashell, both made of materials with high, $D_h$, and low, $D_l$, diffusivity. Their experimental setup consisted of two
parts~\cite{LiAS22}. One part provides particle sources with two tanks positioned on two sides, each filled with a solution to produce the desired concentration gradient. The other part is the device itself, which comprises the background medium and the metashell. The effective diffusivity of the background medium can be calculated using the Maxwell-Garnett formula~\cite{XuPRAp19-2},
\begin{equation}\label{mg}
	D_e=\frac{(1+\delta)D_{i}+(1-\delta)D_{h}}{(1-\delta)D_{i}+(1+\delta)D_{h}}D_{h},
\end{equation}
where $\delta$ and $D_{i}$ are the inclusion volume fraction and diffusivity,
and $D_{h}$ is diffusivity of the host. The background diffusivity can be thus
varied by tuning the volume fraction, $\delta$. 

\subsection{Geometric phase, odd diffusivity, and asymmetric diffusion}

\begin{figure}[!ht]
	\includegraphics[width=1.0\linewidth]{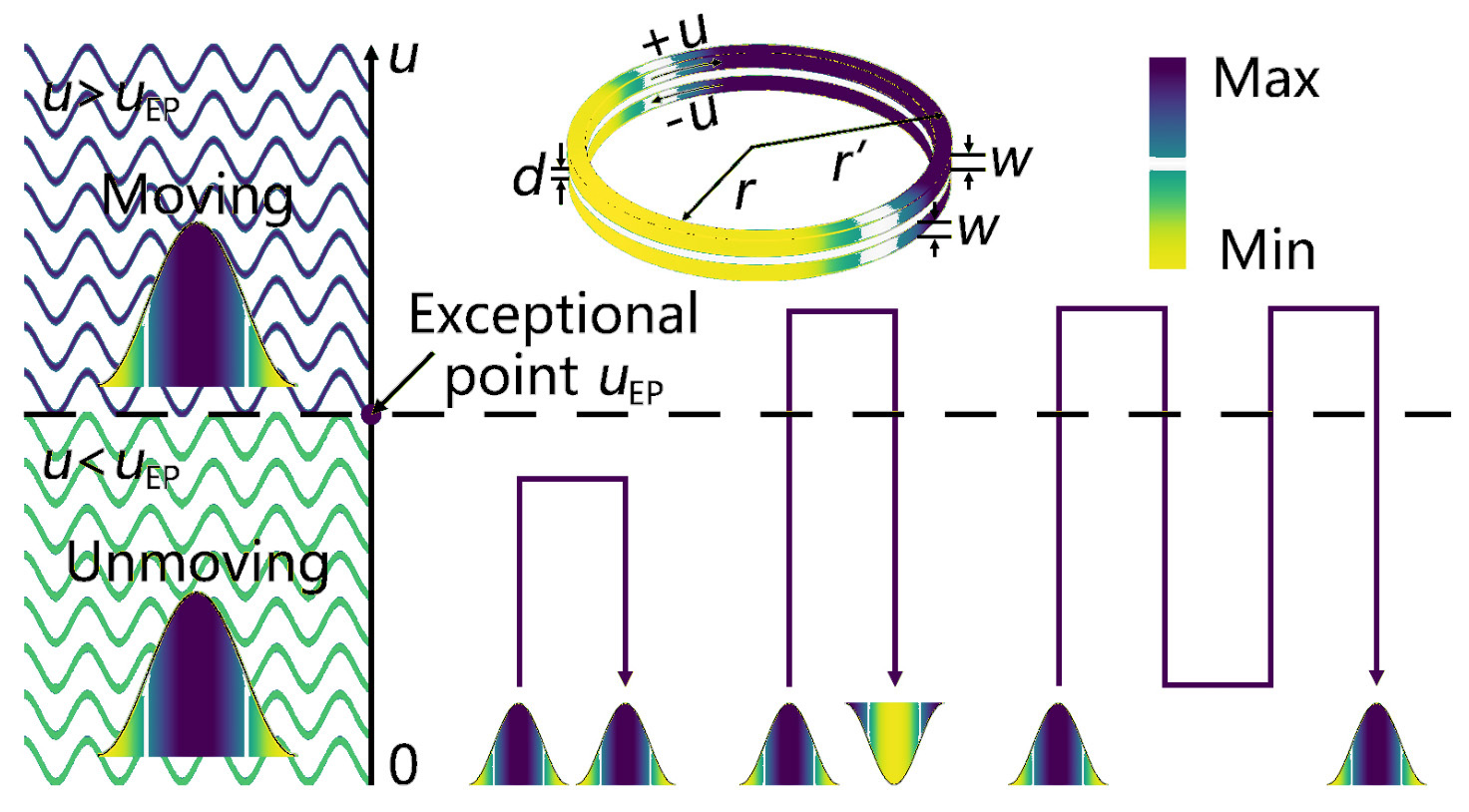}
	\caption{Schematic diagrams of the EP and geometric phase in a particle diffusion system. Adapted from~\citealp{XuPRE20}.}
	\label{particlegeometric}
\end{figure}

As discussed in Sec. III.C, wave propagation usually involves energy
conservation and is therefore described by a Hermitian Hamiltonian; non-Hermitian
wave propagation requires a fine balancing of gain and loss. Particle diffusion,
instead, is inherently dissipative, thus offering a natural playground for novel
non-Hermitian and topological physics. In this context, Xu \emph{et al}.
investigated the exceptional point (EP) and geometric phase of particle
diffusion systems~\cite{XuPRE20}.

To this purpose, these authors considered a three-layer ring structure
[Fig.~\ref{particlegeometric}]. The top and bottom rings were made to rotate at
the same speed, but in opposite directions. The layer in the middle was kept
at rest and assumed to allow particle exchange with the rotating rings,
coupling them together. The equation describing particle diffusion in such a
system is
\begin{subequations}\label{PDG}
	\begin{align}
		\frac{\partial C_1}{\partial t}&=D\left(\frac{\partial^2C_1}{\partial x^2}+\frac{\partial^2C_1}{\partial z^2}\right)-u\frac{\partial C_1}{\partial x}, ~~~\dfrac{d}{2}\leq z\leq w+\dfrac{d}{2}\\
		\frac{\partial C_m}{\partial t}&=D_m\left(\frac{\partial^2C_m}{\partial x^2}+\frac{\partial^2C_m}{\partial z^2}\right),~~~\dfrac{-d}{2}< z< \dfrac{d}{2}\\
		\frac{\partial C_2}{\partial t}&=D\left(\frac{\partial^2C_2}{\partial x^2}+\frac{\partial^2C_2}{\partial z^2}\right)+u\frac{\partial C_2}{\partial x},~~~-w-\dfrac{d}{2}\leq z\leq \dfrac{-d}{2}
	\end{align}
\end{subequations}
where $C_1$, $C_2$, and $C_m$ are the concentrations in the top, bottom, and
central layers, respectively. $D$ is the diffusivity of the top and bottom
layers, $D_m$ the diffusivity of the middle layer, and $u$ the ring rotation
speed. Since the central layer allows particle exchange, it acts on
both rotating rings as an effective particle source, which couple them. Therefore, Eq.~(\ref{PDG}) can be reformulated as
\begin{subequations}\label{PDGS1}
	\begin{align}
		\frac{\partial C_1}{\partial t}&=D\frac{\partial^2C_1}{\partial x^2}-u\frac{\partial C_1}{\partial x}+\frac{D_m\left(C_2-C_1\right)}{wd}, \dfrac{d}{2}\leq z\leq w+\dfrac{d}{2}\\
		\frac{\partial C_2}{\partial t}&=D\frac{\partial^2C_2}{\partial x^2}+u\frac{\partial C_2}{\partial x}+\frac{D_m\left(C_1-C_2\right)}{wd},-w-\dfrac{d}{2}\leq z\leq \dfrac{-d}{2}
	\end{align}
\end{subequations}
Since the thickness, $w$, was assumed to be small, here we set
$\partial^2C_1/\partial z^2=\partial^2C_2/\partial z^2=0$. Consider next a
concentration wave represented by the real part of $A{e}^{{i}\left(kx-\omega t\right)}+B$, where $A$, $B$, $k$, and $\omega$ are the wave amplitude,
reference concentration, wave number, and wave frequency, respectively.
Consequently, Eq.~(\ref{PDGS1}) can be further reduced to the eigenvalue
equation,
\begin{equation}\label{EE}
	\bm{H}\bm{\psi}=\omega \bm{\psi},
\end{equation}
with Hamiltonian
\begin{equation}\label{ParticleH}
	\bm{H}=
	\left(
	\begin{matrix}
		-{i}\left(k^2D+h\right)+ku & {i}h\\
		{i}h & -{i}\left(k^2D+h\right)-ku
	\end{matrix}
	\right),
\end{equation}
where $h=D_m/(wd)$ is the particle exchange rate between the upper and lower layers. The eigenvalues of $\bm{H}$ are
\begin{equation}\label{ParticleEV}
	\omega=-{i}\left[\left(k^2D+h\right) \pm \sqrt{h^2-k^2u^2}\right].
\end{equation}
Then, an intriguing property related to $u$ appears. When $u<u_{\rm EP}\left(=h/k\right)$, the eigenvalue is purely imaginary (the
concentration wave does not move), and the system is in the APT unbroken state. When $u=u_{\rm EP}\left(=h/k\right)$, the two eigenvalues
merge and the corresponding eigenvectors are degenerate. When $u>u_{\rm EP}\left(=h/k\right)$, the real part of the eigenvalue appears (demonstrating that the concentration wave moves), and the system is in the APT broken state. Therefore, $u_{\rm EP}=h/k$ acts as system's EP. We know that encircling an EP is an effective technique to reveal its topological properties. To this end, the rotation speed was
made to execute a closed loop variation over time
[Fig.~\ref{particlegeometric}]. When the loop did not enclose the EP, the
system returned to its initial state, no phase difference was observed. Vice
versa, when the EP was encircled, a phase difference of $\pi$ was detected,
the system acquired a geometric phase.


\begin{figure}[!ht]
	\includegraphics[width=1.0\linewidth]{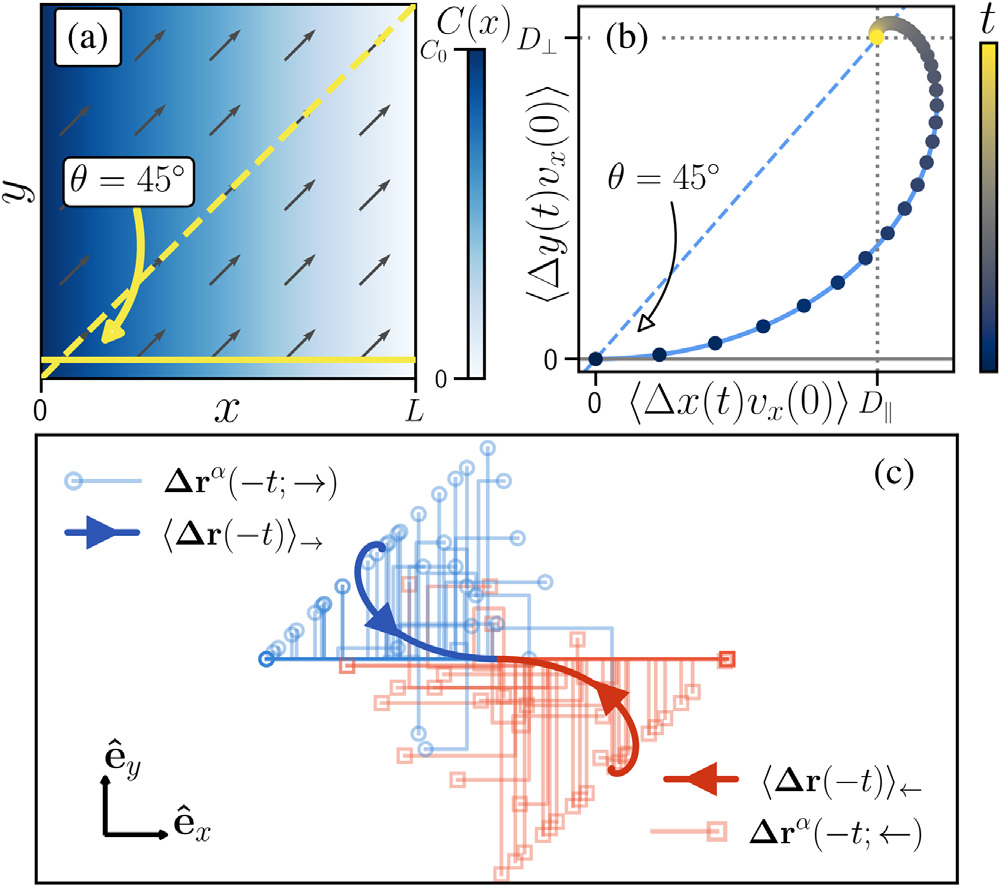}
	\caption{Odd diffusivity: (a) concentration profile; (b) position-velocity correlation functions; (c) 50 time-reversed trajectories. Adapted from~\citealp{CoryPRL21}.}
	\label{odddiffusivity}
\end{figure}

Besides non-Hermitian and topological physics, particle diffusion systems
also exhibit the intriguing property of odd
diffusivity~\cite{CoryPRL21,ErikPRL22}. Odd diffusivity is distinctly
different from the known isotropic and anisotropic diffusivity. Its tensor is
characterized by opposite off-diagonal components, that is
\begin{equation}\label{OddDiffusivity}
	\bm{D}=
	\left[
	\begin{matrix}
		D_\parallel & -D_\perp\\
		D_\perp & D_\parallel
	\end{matrix}
	\right],
\end{equation}
where $D_\parallel$ is the common isotropic diffusivity driving the particle flux from areas of high to low concentrations, and $D_\perp$ is the antisymmetric odd diffusivity inducing the transverse particle flux, a longitudinal concentration gradient can induce a transverse particle flux; see Fig.~\ref{odddiffusivity}(a). Hargus and coworkers investigated such a remarkable diffusivity tensor for chiral random walk, where chiral active matter breaks parity and time-reversal symmetry. They used the Green-Kubo relations to derive the position-velocity correlation functions [Fig.~\ref{odddiffusivity}(b)] and analyzed a random sample of 50 time-reversed trajectories to elucidate the microscopic origin of odd diffusivity [Fig.~\ref{odddiffusivity}(c)].

\begin{figure}[!ht]
	\includegraphics[width=\linewidth]{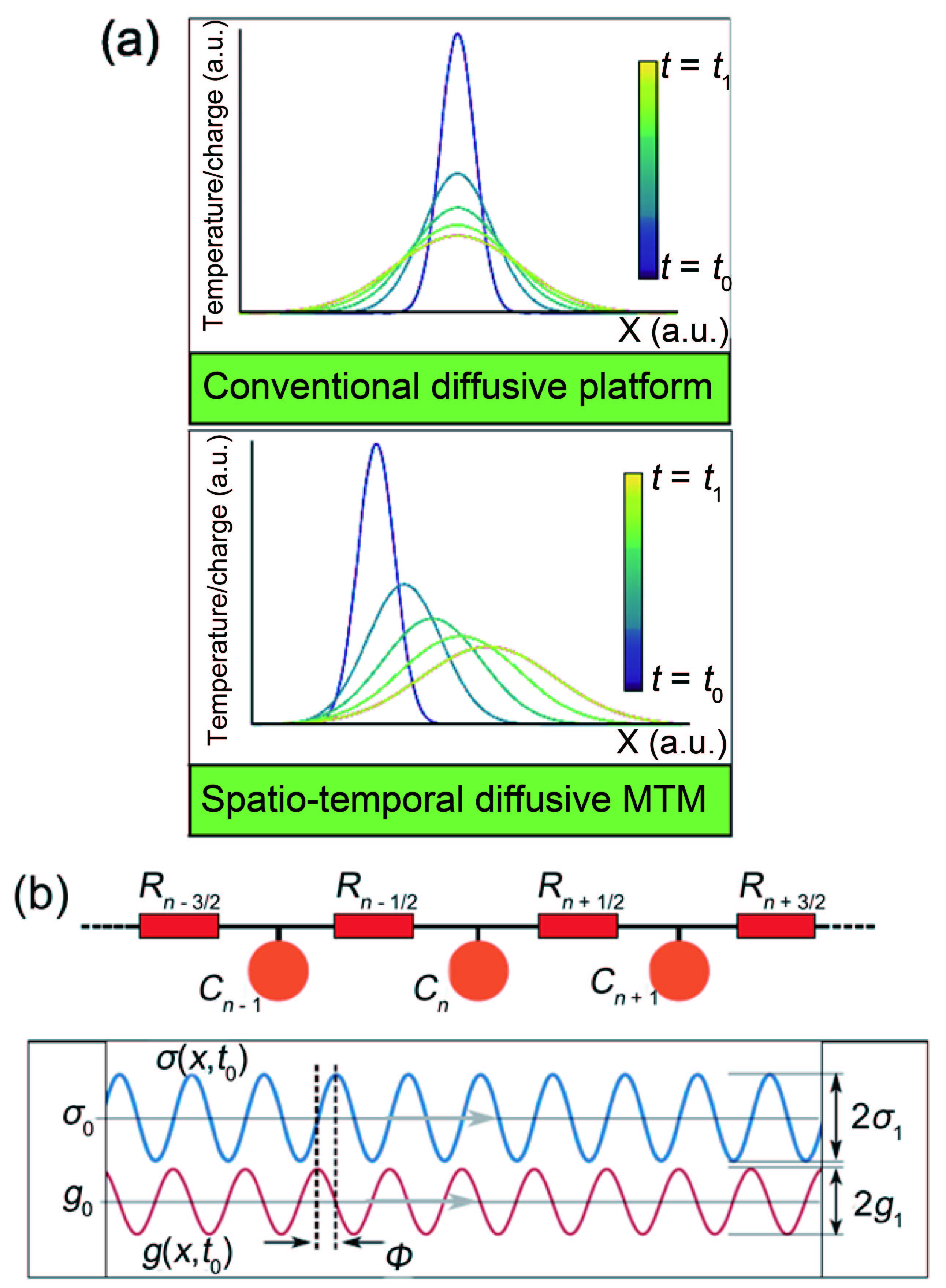}
	\caption{Asymmetric particle diffusion under spatiotemporal modulation: (a) comparison between a conventional material and a spatiotemporally modulated metamaterial; (b) schematic illustration of spatiotemporal parameters. Adapted from~\citealp{CamachoNC20}.}
	\label{chargediffusion}
\end{figure}

Asymmetric diffusion is an important physical concept, indicating that the diffusion properties vary in opposite directions~\cite{BuraCPC09}. However, due to the inherent space inversion symmetry of the underlying microscopic mechanisms, to demonstrate asymmetric diffusion proves a challenging task. To address this issue, Camacho \emph{et al}. proposed a novel spatiotemporal modulation mechanism to experimentally realize asymmetric charge diffusion~\cite{CamachoNC20}.

The Fick Law describes a diffusion process, and the governing equation for 1D diffusion can be written as
\begin{equation}\label{chargediffusionG}
	\dfrac{\partial q\left(x,t\right)}{\partial t}=\dfrac{\partial}{\partial x}\left(\sigma\dfrac{\partial}{\partial x}\left[gq\left(x,t\right)\right]\right),
\end{equation}
where $\sigma$ denotes the conductivity, $g$ represents the inverse of the capacity, and $q\left(x,t\right)$ is the charge density varying in space and time. Generally, if a conventional diffusion process occurs in a static and uniform material, $\sigma$ and $g$ can be combined to form the diffusivity $D=\sigma g$ for brevity. In this case, Eq.~(\ref{chargediffusionG}) would take the standard form
\begin{equation}\label{chargediffusionGR}
	\dfrac{\partial q\left(x,t\right)}{\partial t}=D\dfrac{\partial^2}{\partial x^2}q\left(x,t\right).
\end{equation}
A diffusion process described by Eq.~(\ref{chargediffusionGR}) exhibits space
inversion symmetry. Therefore, an initial symmetric pulse-like profile, $q(x,0)$, would
not drift in either direction, but rather symmetrically spread out over time
[Fig.~\ref{chargediffusion}(a)].

To break symmetric particle  diffusion, Camacho and coworkers modulated $\sigma$ and
$g$ periodically in space and time with the same wave number, $k$, and angular
frequency, $\omega$ [Fig.~\ref{chargediffusion}(b)],
\begin{equation}\label{sigma}
	\sigma\left(x,t\right)=\sigma_0+\sigma_m\sin\left(kx-\omega t\right),
\end{equation}
\begin{equation}\label{g}
	g\left(x,t\right)=g_0+g_m\sin\left(kx-\omega t+\phi\right),
\end{equation}
where $\sigma_0$ and $g_0$ are the parameter reference values, $\sigma_m$ and $g_m$ are the modulation amplitudes (with $0<\sigma_m<\sigma_0$ and $0<g_m<g_0$), and $\phi$ is the phase difference between them. Substituting
Eqs.~(\ref{sigma}) and~(\ref{g}) into Eq.~(\ref{chargediffusionG}) yields a
diffusion process with dynamically inhomogeneous parameters, making a full
analytical treatment complicated. However, homogenization simplifies the
mathematical description. Indeed, the homogenized equation can be
conveniently formulated as~\cite{CamachoNC20}
\begin{equation}\label{chargediffusionGRH}
	\dfrac{\partial \tilde{q}\left(x,t\right)}{\partial t}=\tilde{D}\dfrac{\partial^2}{\partial x^2}\tilde{q}\left(x,t\right)-v\dfrac{\partial \tilde{q}\left(x,t\right)}{\partial x}.
\end{equation}
where $\tilde{q}\left(x,t\right)$ is the time-averaged charge density, $\tilde{D}$ is the homogenized diffusivity, and $v$ can be treated as an advection velocity. The explicit expression for $\tilde{D}$ and $v$ are
\begin{equation}\label{ChargeD}
	\tilde{D}=\sigma_0g_0+\frac{1}{2}\sigma_mg_m\cos\phi,
\end{equation}
\begin{equation}\label{chargeg}
	v=\frac{1}{2}k\sigma_mg_m\sin\phi.
\end{equation}
Since the advection term proportional to $v$ involves a first-order derivative with respect to $x$, the space inversion symmetry of the charge diffusion process is broken. For the advection velocity to be non-zero, the amplitudes of both spatiotemporal parameters must be nonzero, $\sigma_m\neq 0$ and $g_m\neq 0$,
and the phase difference between them chosen so as $\sin\phi\neq 0$, i.e., $\phi \neq 0$ and $\phi \neq \pi$. Under these conditions, a pulse not only decays but also moves in a specific direction due to an effective advection effect [Fig.~\ref{chargediffusion}(a)].

When it comes to experimental realizations, square and triangular parameter
wave profiles are easier to implement. For instance, one can chose
\begin{equation}\label{sigma1}
	\sigma\left(x,t\right)=\sigma\Pi\left(\beta x-\omega t\right),
\end{equation}
\begin{equation}\label{g1}
	g\left(x,t\right)=\dfrac{1}{C\left(x,t\right)}=\dfrac{1}{c_1\Lambda\left(\beta x-\omega t+\phi\right)+c_0},
\end{equation}
where $\Pi$ denotes a square function, $\Lambda$ a triangular function, $C$ is the capacity, $c_0$ its reference value, and $c_1$ its modulation amplitude. Equations~(\ref{sigma1}) and~(\ref{g1}) can be realized by rotating disks, thus making an experimental demonstration possible. These results can also be extended to two or three dimensions. The above effect is the basis for diffusion trapping, whereby a pulse can be trapped to an interface where the effective advection points in opposite directions, thus providing a flexible tool to control charge diffusion.

\begin{figure}[!ht]
	\includegraphics[width=\linewidth]{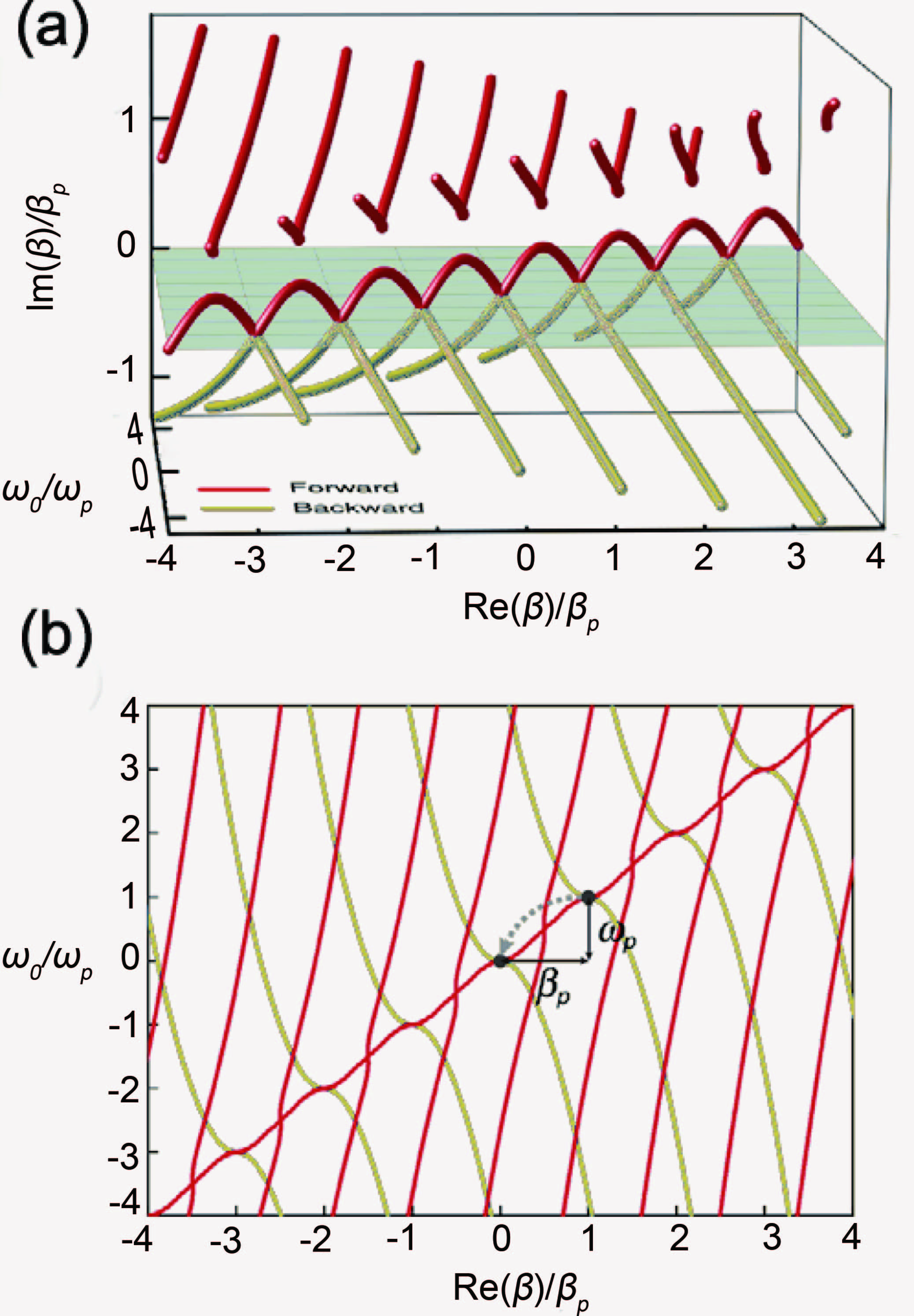}
	\caption{(a) Dispersion diagrams of the spatiotemporally modulated medium. (b) Projection of the dispersion diagrams onto the ${\rm Re}\left(\beta\right)-\omega_0$ plane. Adapted from~\citealp{LiPRL22}.}
	\label{diffusionwave}
\end{figure}

The asymmetric particle diffusion investigated by Ref.~\cite{CamachoNC20} assumed
time-independent boundary conditions. The next question is what happens when
the boundary potential is periodically modulated, thus exciting a diffusion wave. Changing boundary conditions may bring about novel nonreciprocal mechanisms. Based on the diffusion-wave framework, Li \emph{et al}. uncovered a different mechanism to rectify a diffusion wave in a medium, where the electrical conductivity alone is spatiotemporally modulated~\cite{LiPRL22}. In particular, they showed that for a forward diffusion wave, the average output is higher than for the corresponding backward diffusion wave.

The diffusion process is governed once again by the Fick law, $\partial\left(c\varPhi\right)/\partial
t=\nabla\cdot\left(\sigma\nabla\varPhi\right)$, where $\varPhi$ represents
the potential field, $c$ the capacity, and $\sigma$ the conductivity. Here,
however, the authors imposed oscillating boundary conditions with frequency $\omega_0$. The capacity of the medium is set to a constant, $c_0$, and the
conductivity, $\sigma\left(x,t\right)$, is assumed to vary periodically in space and time. The corresponding 1D diffusion equation was rewritten as
\begin{equation}\label{dwG}
	c_0\dfrac{\partial \varPhi\left(x,t\right)}{\partial t}=\dfrac{\partial}{\partial x}\left[\sigma\left(x,t\right)\dfrac{\partial\varPhi\left(x,t\right)}{\partial x}\right],
\end{equation}
where conductivity has the profile of a traveling wave,
\begin{equation}\label{dwS}
	\sigma\left(x,t\right)=\sigma\left(\beta_px-\omega_pt\right),
\end{equation}
with wave number and angular frequency $\beta_p$ and $\omega_p$, respectively.
In view of the spatiotemporal periodicity, the Bloch-Floquet theorem can be
applied to expand the potential field, $\varPhi\left(x,t\right)$, as
\begin{equation}\label{dwBF}
	\varPhi\left(x,t\right)={e}^{{i}\left(\beta x-\omega_0t\right)}\sum_mF_m{e}^{{i}m\left(\beta_p x-\omega_pt\right)},
\end{equation}
where $\beta$ is the wave number of the diffusion wave and $F_m$ is the
Bloch-Floquet modulation coefficients. Truncating the Fourier expansions to
the tenth order yields  accurate solutions of the dispersion relation
[Fig.~\ref{diffusionwave}(a)]. Particle diffusion waves differ from their electromagnetic counterpart, in that, due to their inherently dissipative nature, the corresponding dispersion relations usually involve complex wave numbers [Fig.~\ref{diffusionwave}(a)], the imaginary part of the wave number being responsible for the spatial decay of the diffusion wave amplitude. The
projection of the dispersion relations on the ${\rm Re}\left(\beta\right)-\omega_0$ plane are also suggestive [Fig.~\ref{diffusionwave}(b)]. The shifts of the $\Phi(x,t)$ wave parameters are due to the interplay of all applied spatiotemporal modulations, similar to the mode transition in diffusion systems. When the angular frequency $\omega_p$ satisfies the condition $\omega_0=n_0\omega_p$, with $n_0$ an integer, $\Phi(x,t)$ admits of a nonnull zero-frequency component, which implies particle rectification. Both $\beta_p\neq0$ and $\omega_p\neq0$ are necessary conditions for the rectification effect to occur, as spatial or temporal modulation alone do not suffice.

Finally, Camacho and coworkers also designed a 1D equivalent circuit for experimental demonstration. The spatiotemporal modulation of the electrical conductivity was obtained by an appropriate combination of standard capacitors and resistors varied periodically in time by means of step motors. The rectification of diffusive charge waves was experimentally reported, in close agreement with simulation.


\section{Plasma transport}

Plasma transport can also be viewed as a diffusion process. Plasma is a highly ionized gaseous substance with collective properties so unique to be regarded as the fourth state of matter together with solids, liquids, and gases. Its high conductivity allows it to be captured, moved, and accelerated by electromagnetic forces; hence the possibility to control its transport by carefully designed external electromagnetic fields. Plasma can be created artificially, for instance, by charging gases with currents, radio-frequency waves, or microwave sources~\cite{LiebermanWiley05}. Unlike other diffusive systems, plasma is strongly coupled to electromagnetic fields, making it an intrinsically nonlinear medium~\cite{ZhengPSST21,CuiJAP19,HuangIEEE15}. The analytical treatment of plasma transport is therefore an extremely difficult problem, which lies outside the scopes of the present review. Here, we will focus, instead, on the most common numerical approaches employed in the literature, such as the particle-in-cell or Monte Carlo collision models.

\subsection{Theory and transformation principles}

Recently, Zhang \emph{et al}. applied transformation theory to plasma physics~\cite{ZhangCPL22}. Plasma transport was described by a simplified diffusion-migration model, ignoring convection and magnetic field, and assuming a constant temperature of charged particles. The governing equation was written according to the Einstein relation, which was used to define mobility in terms of diffusivity as
\begin{equation}\label{plasma}
	\frac{\partial n_i}{\partial t} - \nabla\cdot(D\nabla n_i) + \nabla\cdot\left[\left(\frac{D\bm{E}}{T_i}\right)n_i\right] = 0,
\end{equation}
where $n_i$, $D$, and $\bm{E}$ are the density, diffusivity, and electric field, respectively. $T_i = \frac{Tk_B}{q}$ is the reduced temperature, where $T$ is the temperature, $k_B$ is the Boltzmann constant, and $q$ is the unit charge. Additionally, under an optimized transformation theory, which is used in Eqs.~(\ref{particlediff}) and~(\ref{particlediff-opti}), the transformed form of Eq.~(\ref{plasma}) is
\begin{equation}\label{plasma'}
	\frac{\partial n_i}{\partial t} - \nabla'\cdot(D'\nabla' n_i) + \nabla'\cdot\left[\left(\frac{D'\bm{E}'}{T_i}\right)n_i\right] = 0,
\end{equation}
with $\bm{D}' = \bm{J}D\bm{J}^\dag$ and $\bm{E}' = \bm{J}^{-\dag}\bm{E}$. The
two key parameters of plasma control in this model, namely, the diffusivity
and the electric field, can thus be calculated according to the relevant
transformation rules. Following this procedure, Zhang and coworkers designed
plasma metadevices performing the fundamental functions of cloaking,
concentration, and rotation, and thus numerically validated their theory.

\subsection{Applications: Metamaterials and metadevices}

The advancement of plasma physics has enabled the emergence of new technologies
and processes with innovative applications in biomedicine, crystal industry,
and materials science~\cite{LiangAEM18}. The surface of a material can be uniquely modified by plasmas filled with high-energy particles. This makes plasma essential for the development of advanced processes, such as etching, curing, and depositing. Additionally, plasmas can generate chemical reactions at the plasma-matter interface due to the highly reactive radicals, which are effectively used in catalyst preparation and greenhouse gas conversion. In the medical field, cold plasma is also used to treat infected wounds and cancers. Moreover, plasma is beneficial in the aerospace industry, as it can break down high-carbon molecules into low-carbon molecules, thus improving the ignition and combustion performance of aerospace engines. Notably, the transformation theory provides a new perspective for plasma regulation~\cite{ZhangCPL22}. For example, plasma cloaks can be designed in other diffusion systems to prevent healthy tissue from being damaged, while treating infected wounds. Plasma concentrators can also be used to increase plasma concentration in local areas, thus improving catalytic efficiency. Other metadevices, such as rotators, separators, or waveguides, can be realized with appropriate coordinate transformations. According to the theory, diffusivity and electric field are two key parameters for manipulating plasma transport. Although it is challenging to realize the transformed parameters in practice, there are still possible ways to achieve the same effect, such as scattering-cancellation technology. One can fabricate a bilayer structure using two homogeneous materials to achieve the same effect predicted by the transformation theory.

Notably, the term ``plasma metamaterials'' has already been used~\cite{RodriguezPRAp21,SakaiPSST12,NavarroJAP19,ZengIEEE20}. They refer to plasma-based metamaterials in the field of electromagnetic waves. Indeed, transformation-based plasma metamaterials could open up a new research direction in plasma physics. People can extend the simplified diffusion-migration model to more complex cases. For example, the effects of a magnetic field, gas-phase reaction, and changing temperature fields can be gradually taken into account. Despite the inevitable challenges, further exploration of transformation plasma physics is warranted.


\subsection{Potential nonreciprocal and topological impacts}

Since the field of plasma transport transformation is still in its
infancy, much novel physics remains to be explored. Promising directions
may include the following:

{\em (i)}Nonreciprocal plasma transport surely is a first challenging topic. As diffusion
equations always exhibit space-inversion symmetry, uncovering nonreciprocity
mechanisms in plasma transport can prove difficult. To this purpose, one might
have to resort to spatiotemporally modulated or nonlinear parameters, as
reported for heat and particle diffusion.

{\em (ii)}Topological plasma transport is another intriguing topic. Inspired by the
known topological particle and heat diffusion mechanisms, one could try
different options, like devising a plasma counterpart of the Su-Schrieffer-Heeger model to uncover bulk-boundary correspondence and edge
states~\cite{LiuCPL23}. The existence of exceptional points and geometric phases is also worth
investigating~\cite{ParkerJPP21}. In linear approximation, Chern numbers can be defined to characterize nontrivial plasma topologies~\cite{ParkerPRL20}. On the other hand, a flow shear or a density gradient might cause non-Hermiticity, thus making plasma an excellent platform to investigate non-Hermitian physics. 

Finally, plasma control may be a potential bridge connecting diffusion and wave metamaterials. Despite the many distinctions between diffusion and wave metamaterials, it is a more meaningful advancement to control diffusion and wave propagation simultaneously. On the one hand, plasma transport is a diffusion process. On the other hand, plasmas are usually utilized to control electromagnetic waves due to their unique permittivity. Consequently, novel physics could be anticipated when diffusion and wave propagation occur together in one system.

\section{Manipulating diffusion beyond metamaterials}

\begin{figure}[!ht]
	\centering
	\includegraphics[width=1.0\linewidth]{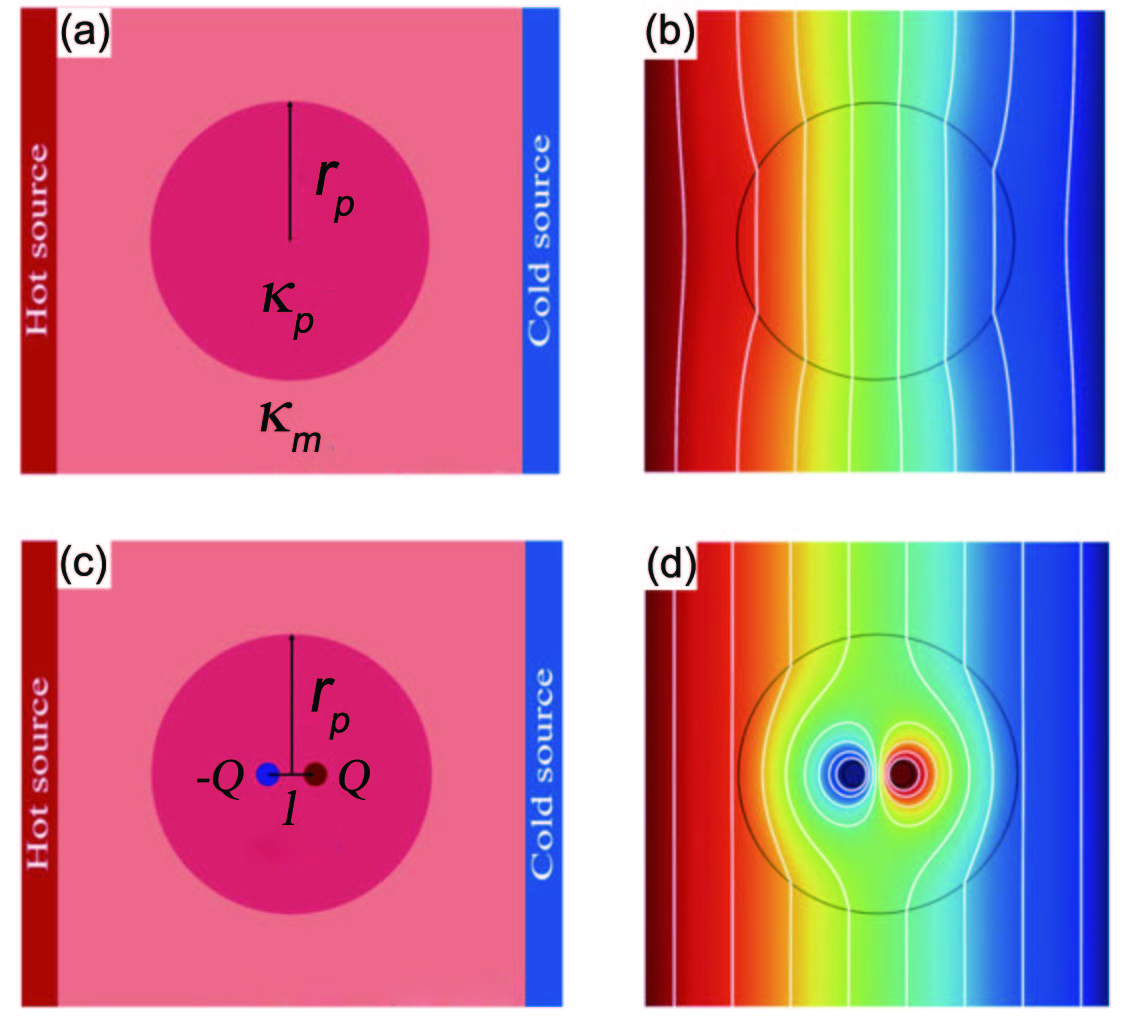}
	\caption{(a) and (c) Thermal control by means of a thermal dipole: system without dipole (top panels) and with dipole (bottom panels). The relevant simulated temperature distributions are reported on the right column, with white lines representing isotherms. Adapted from~\citealp{XuPRE19}.}
	\label{beyondmeta1}
\end{figure}

In addition to metamaterials, there exist other approaches and techniques to control thermal diffusion. Hansson \emph{et al}.~\cite{HanssonPRD99,HanssonPLB99} provided the thermal dipole model in nuclear physics, providing inspiration for a classical thermal dipole model~\cite{XuPRE19}. To explain it, let us consider first a thermal matrix with uniform thermal conductivity, $\kappa_m$, with a particle of radius $r_p$ and thermal conductivity $\kappa_p$ embedded at the center. When a hot and cold heat sources are applied to the opposite sides of the matrix, its
temperature distribution, $T_{me}$, can be expressed in terms of polar coordinates, $(r, \theta)$, with origin located at the center~\cite{XuPRE19-2}, that is
\begin{equation}\label{ndip}
	T_{me}=-G_0r\cos{\theta}-\frac{\kappa_m-\kappa_p}{\kappa_m+\kappa_p}r_p^2G_0r^{-1}\cos{\theta}+T_0.
\end{equation}
Here, $G_0$ is the thermal field of the matrix without the embedded particle
and $T_0$ is the temperature in the middle of the matrix. As illustrated in
Figs.~\ref{beyondmeta1}(a) and~\ref{beyondmeta1}(b), the particle distorts
the matrix thermal field.

The thermal dipole is composed of two adjacent heat sources, each with power $Q$, separated by a distance $l$, as shown in Fig.~\ref{beyondmeta1}(c). When placed at the center of a particle, the thermal field generated by it is given by
\begin{equation}\label{dip}
	T_{md}=\frac{M}{\pi(\kappa_m+\kappa_p)}r^{-1}\cos{\theta}+T_0,
\end{equation}
where $M$ is the thermal dipole moment, defined as $M = Ql$.

According to the superposition principle, the total thermal field of the dipole and the external heat source turns out to be
\begin{equation}\label{tdip}
	T_{s}=-G_0r\cos{\theta}-\left[\frac{\kappa_m-\kappa_p}{\kappa_m+\kappa_p}r_p^2G_0-\frac{M}{\pi(\kappa_m+\kappa_p)}\right]r^{-1}\cos{\theta}+T_0.
\end{equation}
This results tells us how to design thermal dipoles with specific thermal
functions. For instance, thermal transparency would require a uniform matrix thermal field, which can be achieved imposing
\begin{equation}\label{dipm}
	M=\left(\kappa_m-\kappa_p\right)\Gamma G_0,
\end{equation}
where $\Gamma=\pi r_p^2$ denotes the area of the particle. Then the heat sources in the particle can be determined to achieve thermal transparency, whose simulation result is shown in Fig.~\ref{beyondmeta1}. This thermal-dipole-based scheme eliminates the need for singular thermal conductivities, facilitating further progress in active thermal control.

Inspired by negative refraction in wave systems, a complex thermal conductivity, $\kappa$, was proposed as the analogous of the complex refractive index, to explore a new path toward negative thermal transport~\cite{XuCPL20}. Considering a typical thermal conduction-advection process and applying a wavelike solution, $T=A_0e^{i\left(\bm{\beta\cdot r}-\omega t\right)}+T_0$, to the conduction-advection equation, a dispersion relation was derived,
\begin{equation}\label{disp}
	\omega=\bm{\beta\cdot v}-i\frac{\sigma \beta^2}{\rho C},
\end{equation}
where $\omega$, $\bm{v}$, $\bm{\beta}$, $\sigma$, $\rho$ and $C$ are the
angular frequency, advection velocity, wave vector, thermal conductivity,
density, and heat capacity, respectively. $A_0$ and $T_0$ are two constants.
The resulting equation naturally defines a complex thermal conductivity,
$\kappa=\sigma+i\rho C\bm{v\cdot \beta}/\beta^2$). Indeed, for the above
wavelike solution, the dispersion relation in terms of $\kappa$ would read
\begin{equation}\label{comdisp}
	\omega=-i\frac{\kappa \beta^2}{\rho C}=\bm{\beta\cdot v}-i\frac{\sigma \beta^2}{\rho C},
\end{equation}
consistently with Eq.~(\ref{disp}).


The physical representation of the complex frequency can be elucidated by rewriting the waveform solution as
\begin{equation}\label{ws}
	T=A_0e^{\text{Im}(\omega)t}e^{i\left[\bm{\beta\cdot r}-\text{Re}(\omega) t\right]}+T_0,
\end{equation}
Im$(\kappa)$ and Re$(\kappa)$ determine the propagation and dissipation of the wavelike heat diffusion according to Eq.~(\ref{ws}) and Eq.~(\ref{comdisp}). Im$(\kappa)$ indicates the direction of propagation of the wavelike temperature field. Positive (or negative) Re$(\kappa)$ indicates the amplitude decrement (or increment) of the wavelike temperature field. Because energy flow is only determined by thermal advection, $\bm{J}=\rho C\bm{v}T_0$, negative thermal transport can be achieved when $\text{Im}(\kappa)<0$. To realize negative thermal transport, solid ring structures, an approach widely used in thermal topology~\cite{LiSci19}, may offer a solution.

In chemistry, researchers have found that controlling the pH value of the environment, chemical potential gradients, and concentration gradients can affect the reaction-diffusion process~\cite{CeraAM18}. For instance, the growth of polycrystalline nanorods can hierarchically self-organize into intriguing structures such as conical funnels, leaf-like sheets, and worm-like ropes, which
are seemingly generated by pH fluctuations~\cite{KellermeierEJIC12,KaplanScience17}. Furthermore, particle diffusion can be controlled by mechanisms based on acoustic~\cite{CetinBEJ14} or magnetic effects~\cite{CetinBEJ14}. All the aforementioned approaches provide a wide range of possibilities for controlling diffusion through specially designed metamaterials.

Besides energy and mass diffusion, information diffusion is another particular type. Unlike energy and mass diffusion, which are physical processes in a strict sense, information diffusion is not. Nevertheless, some physical equations can still be used to describe it~\cite{ZhangPR16,IribarrenPRL09,HuangFSS97,NematzadehPRL14,CuiJMCC2017,YurevichSIM18,DorsognaPLR15}. For instance, information can be treated as molecular gas, which flows between two locations in the presence of a gradient. This suggests the possibility to control information diffusion by physical mechanisms, such as transformation theory.

\section{Conclusion and future research prospects}

\subsection{Conclusion}

Transformation theory has flourished over the past decade, bringing a wealth of new physics and phenomena to diffusion systems. In this review, we provided a comprehensive overview of transformation theory in diffusive processes and the most recent advances in diffusion metamaterials, with focus on heat diffusion, particle dynamics, and plasma transport. We discussed theoretical foundations, practical applications, and novel physics of diffusion metamaterials. Unlike metamaterials in wave systems, diffusion metamaterials are not limited by working bandwidth due to their frequency independence. This property alone makes diffusion metamaterials attractive for many practical applications. In conclusion, diffusion metamaterials is a booming field with a huge scientific and technological potential still to be fully explored.


\subsection{Future research prospects}

Although significant progress has been made since 2008, many challenges and issues remain to be addressed. The transformed parameters are often anisotropic and singular due to the inherent issues of transformation theory, making experimental realization difficult. Extremely high parameters are also inaccessible, thus impacting the performance of designed intelligent metamaterials. Additionally, research on multifunctional metadevices involving multiple physical fields and intelligent diffusive devices is still inadequate. Furthermore, most diffusion metamaterials concentrate on macroscopic physics, while studies at the micro- and nano-scale are still rare~\cite{YunCPL23,ChenCPL23,WeiCPL23,RenAPP23,JianCPL23,ChoeNanoLett2019,YePCCP16,ShirzadkhaniIJHMT23}. To fill this gap, investigating phonon diffusion could be a viable option~\cite{GandolfiPRL20,HuNSR19,WuCPL23,SeifNC18,GuRMP18}.

Transformation thermotics and scattering-cancellation are typically considered two distinct mechanisms for designing diffusion metamaterials. However, a recent theory of diffusive pseudoconformal mapping has revealed the geometric connection between them~\cite{DaiarXiv23}. By constructing a specific angle-preserving mapping, a unified perspective of transformation thermotics yields the same parameters as those derived from scattering-cancellation for designing bilayer cloaking. In other words, the intrinsic anisotropic parameters from transformation thermotics can be eliminated without having to solve the diffusion equations. This insight could revolutionize the paradigm of diffusion metamaterials and have a profound impact on the field of wave manipulation by metamaterials.

The interfacial thermal resistance between two materials is an essential factor that affects heat transfer~\cite{CanbazogluAPL15,LiuDNL2014,ChenRMP22,YangCPL21,LinM23}. It is often caused by two key factors: the presence of air and lattice mismatch. However, this effect is often overlooked at the macro-scales due to the lack of air between materials~\cite{NarayanaPRL12,SchittnyPRL13}. This phenomenon is currently under investigation~\cite{LiJAP10,ZhengPRAp20}, especially at the micro- or nano-scale~\cite{LiuDNL2014,ChoeNanoLett2019,YePCCP16}, where lattice mismatch plays a major role. Moreover, for mass diffusion, the issue is that the transformation theory is not perfectly applicable to transient cases. Fortunately, the optimized transformation theory performs well. Therefore, more  devices similar to those proposed in thermal systems can be realized, such as invisible sensors~\cite{XuEPL20,WangEPL21}.

To date, only a small number of exotic phases of matter have been realized in diffusion systems. The implementation of various topological states is an emerging trend in this field. Topological wave metamaterials have inspired the development of topological thermal metamaterials~\cite{FengNP17,LuNP14}. Besides the non-Hermitian formalism, Floquet thermal metamaterials can be created by spatiotemporally modulating the material parameters~\cite{Yinelight22}.  Furthermore, higher-dimensional systems can bring about new interesting topological diffusive phenomena. Various peculiar configurations, such as fractal systems~\cite{BiesenthalSci22}, Moiré lattices~\cite{WangNat20}, and disclinations~\cite{LiuNat21}, can be engineered in thermal lattices. Higher-dimensional non-Hermitian topology may also be implemented in diffusion systems, such as geometry-dependent skin effect~\cite{ZhangNC22} and higher-order skin effect~\cite{ZhangNC21}. In addition, diffusion metamaterials may also contribute to the study of protein diffusion inside bacteria~\cite{SadoonPRL22,IrbackPSFB06}, non-Gaussian diffusion near surfaces~\cite{ArthurPRL23}, metastatic diffusion of cancer~\cite{KenJCO10,StackerNM01,GundemNat15} and the study of thermal superconductivity~\cite{HavilandPBCM94,HavilandPRL89, HavilandPRB89}.

Several interesting phenomena can be more intuitively realized in diffusion than in wave systems. One typical example is the implementation of APT symmetry in a diffusion system due to its natural dissipation~\cite{LeggettRMP87,ParisiPRL13,ParisiPRE19}. Additionally, the flexible adjustment of heat convection can help to realize the braiding statistics in non-Hermitian topological bands~\cite{WangNat21, ZhangPRL23}. Furthermore, temperature-dependent thermal transport can add nonlinearity to topological thermal metamaterials, in addition to heat radiation.

Emerging machine learning techniques~\cite{PilozziCP18,LongPRL20, ZhangPRL18} can be used to reveal topological effects in diffusion systems, a technique which promises to further broaden the research in topological diffusion metamaterials. The potential applications of topological thermal metamaterials are also worthy our attention, as the topological properties of the edge state may be beneficial for localized heat management and robust heat transport. Additionally, diffusion metamaterials have unique properties that make them useful in many applications, such as radiative cooling~\cite{BaranovNM19} and crystal growth control~\cite{GaoNM21,ChenNP23}. In conclusion, research into the fundamental theory and practical applications of diffusion metamaterials is still in its infancy.


\begin{acknowledgments}
	We are grateful to Dr. Gaole Dai and  Dr. Jun Wang for the stimulating conversations regarding the geometric connection between transformation thermotics and scattering-cancellation theory. J.H. is grateful for the financial support from the National Natural Science Foundation of China under Grants No.~12035004, No.~11725521 and No.~12320101004, from the Innovation Program of Shanghai Municipal Education Commission under Grant No.~2023ZKZD06, and from the Science and Technology Commission of Shanghai Municipality under Grant No.~20JC1414700. J.-H.J. is also thankful for the financial support from the National Natural Science Foundation of China under Grants No.~12125504 and No.~12074281. F. M. expresses gratitude for the financial support from the National Natural Science Foundation of China under Grant No. 12350710786. L. X. also extends appreciation for the financial support from the National Natural Science Foundation of China under Grants No. 12375040, No. 12088101, and No. U2330401.
\end{acknowledgments}

\bibliographystyle{apsrmp4-2}
\bibliography{rmp}

\end{document}